\documentclass[sigconf]{acmart} 

\AtBeginDocument{%
  }

\copyrightyear{2024}
\acmYear{2024}
\setcopyright{rightsretained}

\acmConference[CHI '24]{Proceedings of the CHI Conference on Human Factors in Computing Systems}{May 11--16, 2024}{Honolulu, HI, USA}
\acmBooktitle{Proceedings of the CHI Conference on Human Factors in Computing Systems (CHI '24), May 11--16, 2024, Honolulu, HI, USA}
\acmDOI{10.1145/3613904.3642700}
\acmISBN{979-8-4007-0330-0/24/05}

\usepackage{colortbl}
\usepackage{tabularx}
\usepackage{enumitem}
\usepackage{changepage}
\usepackage{afterpage}
\newcommand\aff[1]{{\emph{--#1}}}
\newcommand\affpre[1]{{\emph{#1}}}

\newenvironment{listquote}{ 
  \setlength{\partopsep}{2pt}
  
  \begin{itemize}[leftmargin=0.5cm,topsep=4pt,itemsep=1pt,partopsep=1ex,parsep=1ex]
    \setlength{\itemsep}{3pt}
    \setlength{\parskip}{3pt}
    \setlength{\rightskip}{15pt}
    \setlength{\parsep}{3pt}    
}{ \end{itemize} }

 \def\Hype/{{\fontfamily{lmss}\selectfont\textbf{Sensationalized}}} \def\Academic/{{\fontfamily{lmss}\selectfont\textbf{Transformative}}}
 \def\Tool/{{\fontfamily{lmss}\selectfont\textbf{Effort-Saving Tool}}}
 \def\Narrative/{{\fontfamily{lmss}\selectfont\textbf{Narrative}}}
 
\newcommand\pid[1]{\emph{#1}}

\newcommand{\eg}{e.g., }

\newcommand{\ie}{i.e., }

\definecolor{asparagus}{rgb}{0.53, 0.66, 0.42}
\definecolor{royalazure}{rgb}{0.0, 0.22, 0.66}
\definecolor{brickred}{rgb}{0.75, 0.25, 0.33}
\definecolor{bluegray}{rgb}{0.4, 0.6, 0.8}
\definecolor{darkmagenta}{rgb}{0.55, 0.0, 0.55}
\definecolor{darkslateblue}{rgb}{0.28, 0.24, 0.55}

\newcommand{\tagNarr}[1]{    
  \newline\sffamily{\textbf{\textcolor{royalazure}{#1}}} }
\newcommand{\tagForc}[1]{
  \newline\sffamily{\textbf{\textcolor{brickred}{#1}}} }
  
\newcommand{\TopThree}[1]{\textbf{#1}}
\newcommand{\notTopThree}[1]{{#1}}

\newcommand{\agendaG}[1]{\textcolor{bluegray}{#1}}

\newcommand{\rqf}[1]{\vspace{1.0em}\noindent{\textcolor{darkslateblue}{\textbf{\sffamily{#1}}}}\vspace{0.4em}}
\newcommand{\rqft}[1]{\begin{adjustwidth}{1.4em}{}{#1}\end{adjustwidth}}

\begin{document}

\title{How Knowledge Workers Think Generative AI Will (Not) Transform Their Industries}


\author{Allison Woodruff}\email{woodruff@acm.org}
\affiliation{\institution{Google}\country{USA}}

\author{Renee Shelby}\email{reneeshelby@google.com}
\affiliation{\institution{Google}\country{USA}}

\author{Patrick Gage Kelley}\email{patrickgage@acm.org}
\affiliation{\institution{Google}\country{USA}}

\author{Steven Rousso-Schindler}\email{stevenrs@gmail.com}
\affiliation{\institution{
  \fontsize{9.5}{12}\selectfont{California State University, Long Beach}}
  \fontsize{10.5}{12}\selectfont\country{USA}}

\author{Jamila Smith-Loud}\email{jsmithloud@google.com}
\affiliation{\institution{Google}\country{USA}}

\author{Lauren Wilcox}\email{lgw231@acm.org}
\affiliation{\institution{Google*}\authornote{This work was conducted while the author was at Google.}
\country{USA}}

\renewcommand{\shortauthors}{Woodruff et al.}

\begin{abstract}
Generative AI is expected to have transformative effects in multiple knowledge industries. To better understand how knowledge workers expect generative AI may affect their industries in the future, we conducted participatory research workshops for seven different industries, with a total of 54 participants across three US cities. We describe participants' expectations of generative AI's impact, including a dominant narrative that cut across the groups' discourse: participants largely envision generative AI as a tool to perform menial work, under human review. Participants do not generally anticipate the disruptive changes to knowledge industries currently projected in common media and academic narratives. Participants do however envision generative AI may amplify four social forces currently shaping their industries: deskilling, dehumanization, disconnection, and disinformation. We describe these forces, and then we provide additional detail regarding attitudes in specific knowledge industries. We conclude with a discussion of implications and research challenges for the HCI community.
\end{abstract}




\sloppy
\maketitle

\fussy
\section{Introduction}

On November 30, 2022, OpenAI released a demo of ChatGPT, a chatbot powered by a large language model (LLM)\footnote{An LLM is a machine learning model trained on large amounts of data (\eg at the terabyte or petabyte scale), with model parameter estimates in the many billions, capable of generating well-structured and stylized natural language output in response to natural language input.}. The chatbot's impressive ability to converse and provide information immediately drew international attention, attracting over one million users in just a few days~\cite{marr2023}.
While LLMs had been under active development for several years, and build on technologies that have been researched and used for decades, ChatGPT's release was heralded as a disruptive moment in which the general population, as well as many technologists, became aware of a significant leap in AI's capabilities.

Since then, the rapid uptake of generative AI systems such as ChatGPT~\cite{chatgpt}, Bard~\cite{bard}, DALL$\cdot$E~\cite{dalle2}, Imagen~\cite{imagen}, and Midjourney~\cite{midjourney} has been accompanied by striking narratives. These include discussion of its increased ability to perform complex tasks and predictions about how generative AI will disrupt knowledge industries (industries such as law, journalism, software development, and more, in which workers apply knowledge gained through specialized training to engage in non-routine problem solving and develop products and services~\cite{janzknowledge}). These narratives imagine generative AI as a resource that can automate much of the knowledge work currently done by humans, thus having detrimental effects on labor such as eliminating significant numbers of jobs across multiple industries~\cite{BBC-jobs}. However, while initial evidence points to productivity gains when generative AI is used for particular tasks \cite{brynjolfsson2023, weisz2021}, much remains unknown regarding its future impact.

This unique moment in the uptake of generative AI offers a timely opportunity to deeply consider expectations of its future use. To better understand how people anticipate generative AI may affect knowledge work in the future, we conducted three-hour participatory research workshops for seven different knowledge industries, with a total of 54 participants across three US cities. Our contributions are as follows:

\begin{itemize}
\item We present a novel qualitative study of how knowledge workers in seven professional fields\textemdash advertising, business communications, education, journalism, law, mental health, and software development\textemdash expect generative AI may affect their fields in the future. In particular, we explore perceptions of knowledge workers not only regarding how specific tasks or jobs might be automated or otherwise change, but also how and whether generative AI might more broadly shift the nature and structure of work in their industries.
\item We describe a dominant narrative regarding generative AI that participants across multiple knowledge industries envisioned, of generative AI as a tool to perform menial work under human review, and we also highlight unique perspectives from each industry.
\item We describe how specific social forces that participants are experiencing in their industries\textemdash in parallel to the development of generative AI\textemdash create a unique confluence of circumstances that frame participants' expectations of generative AI's impact as it intersects with social trends surrounding its deployment and use. These forces include deskilling, dehumanization, disconnection, and disinformation.
\item Building on our participants' insights, we identify five HCI research challenges at the intersection of generative AI and knowledge work.
\end{itemize}

In the remainder of the paper, we review relevant background, describe our methodology, present our findings, discuss implications and research challenges for the HCI community, and conclude.
\section{Background}

Our research examines knowledge workers’ perceptions of how generative AI will impact their fields. In this section, we provide background on generative AI technologies and knowledge work, as well as research at their intersection, and provide an overview of generative AI narratives as well as public perception of AI.

\subsection{Generative AI Models}

Generative AI is a type of machine learning system that generates realistic and credibly human-like content (\eg text, images, code, audio) in response to an input. These systems are trained on web-scale databases and differ from conventional machine learning (ML) systems designed to function as expert systems, marking a shift in the purpose of ML from ``problem solving to problem finding''~\cite[p. 1]{muller2022}. The concept of generative AI dates to the 1950s when researchers began exploring the possibility of using computers to produce new content \cite{knill1997hidden, reynolds2009gaussian}. However, it was not until the 1990s and 2000s with further investment in neural networks, interconnected nodes that process and identify patterns from large datasets \cite{murphy2022probabilistic}, that the field of machine learning gained momentum. The beginning of the  ``deep learning revolution'' \cite[p. xxvii]{murphy2022probabilistic} occurred in 2014 when Alex Krizhevsky and colleagues \cite{krizhevsky2012imagenet} used deep neural networks to win an ImageNet image classification challenge.  Since then, researchers developed new kinds of neural networks that can generate novel content\textemdash including Generative Adversarial Networks \cite{goodfellow2014generative}, Recurrent Neural Networks (RNNs) \cite{mikolov2010recurrent}, Variational Autoencoders (VAEs) \cite{kingma2013auto}, and diffusion generation models \cite{song2019generative}\textemdash marking technical advances that set the stage for generative AI's growth. 

The next significant milestone occurred in 2017 with the development of the modern transformer architecture \cite{vaswani2017attention} that allows for \textit{parallelization} in which computation is divided into smaller tasks that run simultaneously. Transformer architectures enable generation of detailed and realistic content, and remain the state-of–the-art for a range of generative tasks, including text \cite{ramesh2021zero, lewis2019bart}, image \cite{dosovitskiy2020image, liu2021swin}, and code generation \cite{codex}; and underpin ``foundation models'' trained on massive datasets, such as LaMDA \cite{thoppilan2022lamda}, GPT \cite{brown2020language}, and DALL$\cdot$E \cite{ramesh2022hierarchical}. The field of generative AI is rapidly evolving; however, large language models (\eg ChatGPT \cite{chatgpt}, Bard \cite{bard}) and text-to-image models (\eg Midjourney \cite{midjourney}, DALL$\cdot$E 2 \cite{dalle2}, Stable Diffusion \cite{stablediffusion}) are today's popular modalities. Large language models process and generate text that can be used for a variety of applications, including question and answer interactions, language translation, chatbots, code and text generation, summarization, and language analysis. Text-to-image (T2I) systems allow users to generate photorealistic images from free-form and open-ended text prompts. While modalities differ, generative AI systems are marked by three key characteristics: (1) applicability to generalized rather than specialized use cases; (2) production of original content that is often indistinguishable from human content; and (3) intuitive and accessible interfaces~\cite{briggs2023}.
\subsection{Knowledge Work}

Knowledge work\footnote{The term ``knowledge worker'' was coined by Peter Drucker~\cite{druckerlandmarks} in 1959.} concerns specialized labor, the main capital of which is knowledge~\cite{galbraith2007new}; knowledge workers apply theoretical and analytical knowledge gained through specialized training to engage in non-routine problem solving and develop products and services~\cite[p. 878]{janzknowledge}. Knowledge work is a characteristic of post-industrial societies~\cite{castells2011rise} in which capitalist production shifts towards technical knowledge \cite{bellpostindustrial, pyoria2005concept}. In this way, knowledge work reflects the reconcentration of socioeconomic and political power into institutions that translate information resources into marketable commodities and services~\cite{moscoknowledge}. Knowledge work thus reframes workers as a ``capital asset'' and locates the responsibility for productivity on individual workers who have autonomy to determine how a particular task should be approached within particular boundaries~\cite[p. 87]{druckerknowledge}. The most defining characteristic of a knowledge worker is their possession of ``knowledge,'' not necessarily completion of an education program~\cite{dove1998}, although certain roles or fields may require licensure or credentials (\eg attorneys, medicine). As such, knowledge work comprises professions that use information, data, or ideas as ``raw material'' for ``planning, analyzing, interpreting, developing, and creating products and services''~\cite[p. 511]{heerwagencollaborative}.
\subsection{Generative AI and the Future of Knowledge Work}
\label{sec:fow}

Generative AI's ability to enable users to easily generate new content across generalized use cases raises questions about potential macroeconomic impacts on knowledge labor~\cite{eloundou2023}. Unlike many previous innovations that have impacted physical work such as manufacturing or transportation, generative AI directly implicates professionalized knowledge work~\cite{briggs2023}. Discourse on this topic forms a \Academic/ \Narrative/---we discuss narratives further in Section~\ref{sec:narrative}---that suggests that generative AI will have substantial impact on knowledge industries. For example, Briggs et al.~\cite{briggs2023} identify thirteen kinds of knowledge work activities vulnerable to automation (\eg gathering, organizing, analyzing, and interpreting information), estimating that approximately two-thirds of US occupations may experience disruption in how workers do their work. While more research is needed given the rapid pace of change in the field of generative AI, nascent work suggests potential disruption in fields such as creative work~\cite{davenport2022}, customer service~\cite{shaji2023}, education~\cite{felten2023}, healthcare~\cite{shaji2023,varghese2023}, journalism~\cite{jones2023}, legal services~\cite{felten2023},  marketing~\cite{chui2023}, management consulting~\cite{dell2023}, and software engineering~\cite{welsh2023}. Novice workers are likely to disproportionately benefit from efficiency increases for particular work tasks~\cite{noy2023}. For instance, introduction of generative AI conversational assistants in customer service roles enhanced worker productivity, with disproportionate increases in the performance of novice customer service agents~\cite{brynjolfsson2023}. Economists anticipate that occupations experiencing significant automation will be offset by new job creation~\cite{briggs2023}, such as new roles to review and manage content created by generative AI systems~\cite{davenport2022}. However, appropriate investments in occupational change management are required~\cite{chui2023}, including understanding worker perspectives, a goal that directly guided our research. 

\subsubsection{Generative AI Impacts on Worker Tasks}
HCI scholars are beginning to examine the impacts of generative AI on different kinds of work tasks. A dominant thread of research focuses on different writing tasks~\cite{gero2023, roemmele2015creative, shakeri2021, singh2022, yang2022ai, yuan2022}, including marketing slogans~\cite{clark2018} and storytelling~\cite{roemmele2018}. Generative AI systems can significantly influence what topics users write about and how they are framed~\cite{jakesch2023}. For instance, when people use a generative AI writing assistant (\eg ChatGPT), they write more frequently about topics suggested by the system~\cite{poddar2023}. Writing suggestions made by generative AI systems also influence the tone and sentiment of communications~\cite{arnold2018sentiment, bhat2023}, including length and how generic the text is~\cite{arnold2020}. However, user experience of generative AI systems is not simply passive, but also involves cognitive work to negotiate machine-in-the-loop writing approaches~\cite{singh2022, dang2023}. As creative work is not necessarily outcome-focused, but process-focused, there remain gaps in our understanding of the impacts of generative AI in explicitly commodified labor contexts. 

An emerging thread of HCI scholarship has focused on potential impacts of generative AI on workers in professionalized settings, examining potential professional impacts of generative AI, including how it may improve knowledge workers’ task efficiency~\cite{zheng2022, arakawa2023}. Some HCI research has examined impacts on specific professional sectors, notably creative~\cite{inie2023, guzdial2019, davis2023} and educational~\cite{ali2021, lu2023, kazemitabaar2023} domains. While research suggests that, overall, creative professionals are not yet worried about generative AI-related job displacement, they did flag concerns related to worsening work quality, an erosion of the creative process, and training data copyright concerns~\cite{inie2023}. Within the education domain, research has focused on how generative AI can support classroom teaching in terms of preparing course materials~\cite{sarsa2022, lu2023} and student learning outcomes~\cite{kazemitabaar2023}. However, there remains a dearth of HCI research on generative AI’s impacts on professional industries, particularly outside of creative contexts.

\subsection{Sensationalized Generative AI Narratives}
\label{sec:hype}

The rapid uptake of generative AI systems has been accompanied by a \Hype/ \Narrative/ about how it will disrupt society and work. Building on previous AI narratives, it contains dramatic or even hyperbolic claims (see critiques:~\cite{bartholomew2023, chomsky2023}) about topics such as AI sentience (\eg~\cite{roose2023}) and labor displacement (\eg~\cite{corfield2022}). Leading figures in AI have framed expectations in evocative terms (\eg~\cite{metz2023}), for example, likening AI to ``new electricity'' to emphasize the expectation that it will disrupt most or all industries~\cite{ng2017}. The media, futurists, and others have also explored generative AI's potential, pondering for example whether AI could ``replace humans''~\cite{lock2022}. Exaggerated claims about machine intelligence envision a global underclass of human workers supplanted by generative AI~\cite{jaupi2023}, with one headline musing: ``Could ChatGPT Write My Book and Feed My Kids?''~\cite{heritage2022}. While such narratives may draw on claims made by AI developers to predict a worst case scenario for knowledge work (\eg~\cite{meyer2023, barrabi2023}), they often incorporate evidence in misleading ways (see critiques:~\cite{marcus2019,omaar2023}). Moreover, they rarely engage the perspectives of knowledge workers with expertise in their field. 

While some argue generative AI is at the peak of inflated expectations~\cite{gartner2023, vynck2023}, sensationalized AI narratives can narrow and close public debate~\cite{katz2020}. In response to the hype cycle~\cite{jesuthasan2023}, there have been calls for more measured narratives that create space for publics to meaningfully consider and grapple with the labor questions of generative AI~\cite{bartholomew2023, siegel2023}). This includes consideration of the specific ways generative AI may interact with different kinds of professional work~\cite{joshi2023} and also broader knowledge industry dynamics.
\subsection{Public Perception of AI}

Because the public's awareness of generative AI is quite recent, little research has been done on public perception of it. However, a fair amount of research has been done on public perception of AI more generally, much of it survey-based~\cite{rainie2022,bao2022,blumberg2019, cave2019,edelman2019,zhang2019artificial,northeastern2018,kapania2022,selwyn2020,yougov2021,european2017,arm2017,funk2020,lloyds2020,neudert2020}. Respondents typically expect AI will have a significant impact on the future, and often anticipate that beneficial effects are possible, with the most favorable impressions in emerging and/or Asian markets and more negative impressions (particularly recently) in Western countries such as the US~\cite{arm2017,edelman2019,funk2020,northeastern2018,european2017,lloyds2020,neudert2020,yougov2021}. At the same time, AI is neither interpreted as exclusively beneficial nor exclusively disadvantageous, and public response often indicates contradictory emotions~\cite{blumberg2019,kelley2021,kelley2021b,mozilla2019}. Job loss (especially related to robotics and manufacturing), increased social isolation, privacy and other social topics have been highlighted as key concerns~\cite{arm2020,edelman2019,kelley2023,kelley2021,kelley2021b,smith2018,rainie2022}. In fact, a recent Pew survey found that, among the 37\% of US respondents who indicated they were ``more concerned than excited'' about the increased presence of AI in daily life, about one in five explicitly mentioned job loss as the main reason for their concern~\cite{rainie2022}.
\section{Methodology}

To explore the expected impact of generative AI applications in knowledge industries, we conducted participatory research workshops with 54 participants from seven knowledge industries in three US cities. While our institution does not have an IRB, we adhere to similarly strict standards. Our research objectives were to learn more about the following:

\begin{itemize}
\item{How do knowledge workers expect generative AI will affect their industries?}
\item{How do knowledge workers view generative AI in relationship to other changes they anticipate in their industry?}
\end{itemize}

\subsection{Participants}
We recruited 7-8 workshop participants from each of seven industries (for 54 in total) through Gemic, who worked with partners to arrange professional recruiting firms in each city. Our participants played a co-research role in the activities and discussion; we refer to them as ``participants'' throughout the paper not to suggest a passive role in the research, but to avoid confusion about formal researcher contributions and workshop co-researcher contributions.
We chose a wide range of industries\textemdash advertising, business communications, education, journalism, law, mental health care, and software development\textemdash to elicit diverse perspectives across fields which are likely to be substantially affected by generative AI, according to economic analyses and industry perspectives~\cite{briggs2023,chui2023,davenport2022,felten2023,shaji2023,welsh2023}. Approximately half of the participants in each workshop held central roles which we expected would be well-positioned to speak broadly about each industry, \eg teachers for education, reporters for journalism, and attorneys for law. The other half were more specialized roles in each profession to introduce a greater diversity of perspectives. See Appendix A, Table~\ref{table:industry_overview}. Screening also aimed to recruit a balanced mix of genders and half non-White participants in each workshop. All participants were compensated the same amount, at industry standard. For a summary of participant information, see Table~\ref{table:participants}. 

\begin{table*}[ht]
\centering
\begin{tabular}{ l l l l l  p{8.5em}  }
                &               &   \multicolumn{3}{l}{\textbf{Participants}} \\
    Industries &        Location &  $n$     &   Age Range  & Gender   & Time in Industry  \\
    \midrule
    Advertising &       New York City &  8   &   26 -- 38     & Women: 4   & 2-5 years: 2        \\
                                                              & & & & Men: 4   &  6-10 years: 1       \\
                                                              & & & & &  10+ years: 5           \\
    \midrule
    Business  &          Columbus & 8 & 24 -- 54  & Women: 4 &    2-5 years: 3        \\
    Communications                              & & & & Men: 4 &  6-10 years: 1       \\
                                                & & & & & 10+ years: 4           \\ 

        \midrule
    Education  &        Oakland &        7  &  29 -- 51   & Women: 6 &   2-5 years: 2        \\
                                                & & & & Men: 1 &  6-10 years: 1       \\
                                                & & & & & 10+ years: 4           \\
    \midrule
    Journalism &        New York City &  8  &     25 -- 42     & Women: 5      &  2-5 years: 3        \\
                                                & & & & Men: 3 &  6-10 years: 2       \\
                                                & & & & &  10+ years: 3           \\
    \midrule
    Law &               Columbus &       8  &    30 -- 64         & Women: 4 &  2-5 years: 3        \\
                                                & & & & Men: 4 &  6-10 years: 2       \\
                                                & & & & & 10+ years: 3           \\
    \midrule
    Mental Health &     Oakland &        8  & 32 -- 50          & Women: 5 &  2-5 years: 2        \\
                                                        & & & & Men: 3 &  6-10 years: 1       \\
                                                        & & & & & 10+ years: 5           \\
    \midrule
    Software  &   New York City &  7  & 25 -- 36    &  Women: 3 & 2-5 years: 3        \\
    Development                                         & & & & Men: 4 &  6-10 years: 1       \\
                                                        & & & & & 10+ years: 3           \\
     \bottomrule
\end{tabular}
  \vspace{0.4em}
  \caption{Details about the location and participant makeup of each industry group. Gender was self-reported by participants from a range of options, including the option to self-describe. Recruitment was limited to participants who had been in their industry for two or more years. }
  \label{table:participants}
\end{table*}

\subsection{Workshops}

Participatory workshops have a long methodological history in HCI, used frequently in participatory design engagements~\cite{till2022, harrington2019, ledantec2015, rosner2016}, living labs~\cite{dell2014}, hackerspaces~\cite{fox2015}, and more. The participatory workshop as a research site and method has roots in participatory action research~\cite{cooper2022,hayes2011}. Inspired by participatory action research, we used participatory workshops to engage with specific communities of practice representing knowledge industries. We followed this approach as it is a practical complement to the application of critical theoretical research traditions in HCI (\eg feminist~\cite{rode2011,bardzell2010}, queer~\cite{light2011,spiel2019}, intersectional~\cite{kumar2019,rankin2020,schlesinger2017,wong2018,scheuerman2019}, and critical-race-theory-informed~\cite{ogbonnaya2020,schlesinger2018} work), in that it: (1) centers lived experiences of those impacted by technology; (2) is concerned with understanding social change\textemdash in our case, understanding stages of generative AI adoption in knowledge work to explore participatory understandings of the forms that sociotechnical and organizational interventions might take; and (3) enacts research practices that strive to mitigate asymmetrical power relationships in research. We thus structured our workshops to incorporate a range of activities, including probes ~\cite{boehner2007,gaver1999,bray2022,graham2008,beignon2020} and provocations, in order to collaboratively envision and reflect on expected impacts based on participants’ lived experiences and points of view, and to engage in collaborative discourse and activities to inform and shape future agendas and policies around use of generative AI.

\begin{figure*}[htp]
    \section{Artifacts}
    \vspace{2.0em}
    \includegraphics[width=6.2in]{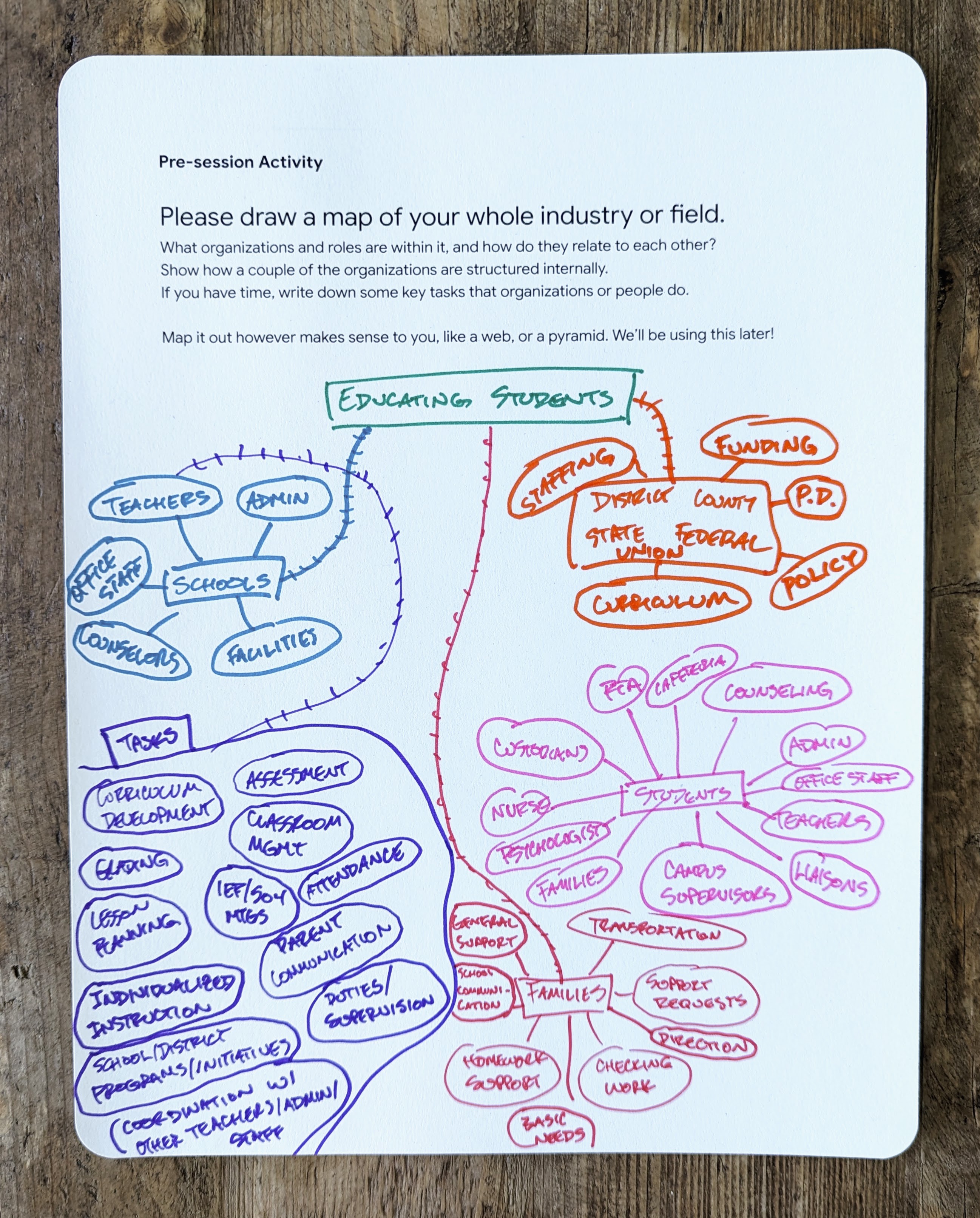}
    \caption{\pid{E6}'s industry map for education. During check-in participants were invited to draw a map of their industry or field.}
    \label{fig:industry_map1}
\end{figure*}

We held one workshop per industry in July 2023\footnote{Workshops took place several months after the March 2023 release of OpenAI's GPT-4 model, which is broadly considered to be a significant improvement over the already impressive original November 2022 release.} in Columbus, Ohio; New York City, New York; and Oakland, California (for industries in each city, see Table~\ref{table:participants}). These cities were selected as centers of activity for the respective industries, as well to represent a socio-geographic range. Each workshop was held in person for three hours at a third-party facility. Three to four researchers from Google and Gemic moderated each workshop to facilitate and share information about generative AI. Participants and researchers sat together around a large table to facilitate group discussion. Two videographers also attended each workshop and participants were aware that several researchers and staff were in an observation room or viewing remotely. Participants were aware of Google's involvement in the study. We followed COVID-19 precautions (\eg all participants and moderators took tests during check-in), and refreshments were available throughout the workshops. As a short pre-work activity before the workshops began, we asked participants to draw a map of their industry to establish a reference for discussion of potential changes 
(Appendix C: Figure~\ref{fig:industry_map1}).

\begin{table*}[ht]
    \centering
    \begin{tabular}{l l l}
         \toprule
         \emph{Activity} & \emph{Time} \\
         \midrule
         Arrive, sign-in, COVID-19 testing, Industry Map & \\
         \arrayrulecolor{bluegray}
         \midrule
         \arrayrulecolor{black}
         Welcome, Introductions, Prior Experience with Generative AI   & \emph{15 minutes}     \\
         Generative AI Teaching and Q/A      & \emph{40 minutes}    \\
         \agendaG{Meal Break}                & \agendaG{\emph{20 minutes}}    \\
         Change Cards Activity               & \emph{45 minutes}    \\
         \agendaG{Break}                     & \agendaG{\emph{10 minutes}}    \\
         Policy Activity and Voting          & \emph{40 minutes}    \\
         Personal Impact and Wrap-Up         & \emph{10 minutes}    \\
         \bottomrule
    \end{tabular}
    \vspace{0.4em}
    \caption{Agenda for participatory research workshops. Times are approximate. }
    \label{table:agenda}
\end{table*}

Workshops ran as shown in Table~\ref{table:agenda}. We began with introductions and expectations for the workshop, and then invited participants to share any prior experiences or impressions of generative AI. We then gave a presentation to provide participants with a common working understanding of generative AI to ground the subsequent activities, allowing a generous amount of time for questions and discussion. The presentation included a brief overview of how generative AI models learn and generate content, how they differ from other AI models, and key characteristics and risks of generative AI (Appendix B).

\begin{figure*}[htp]
    \centering
    \includegraphics[width=5in]{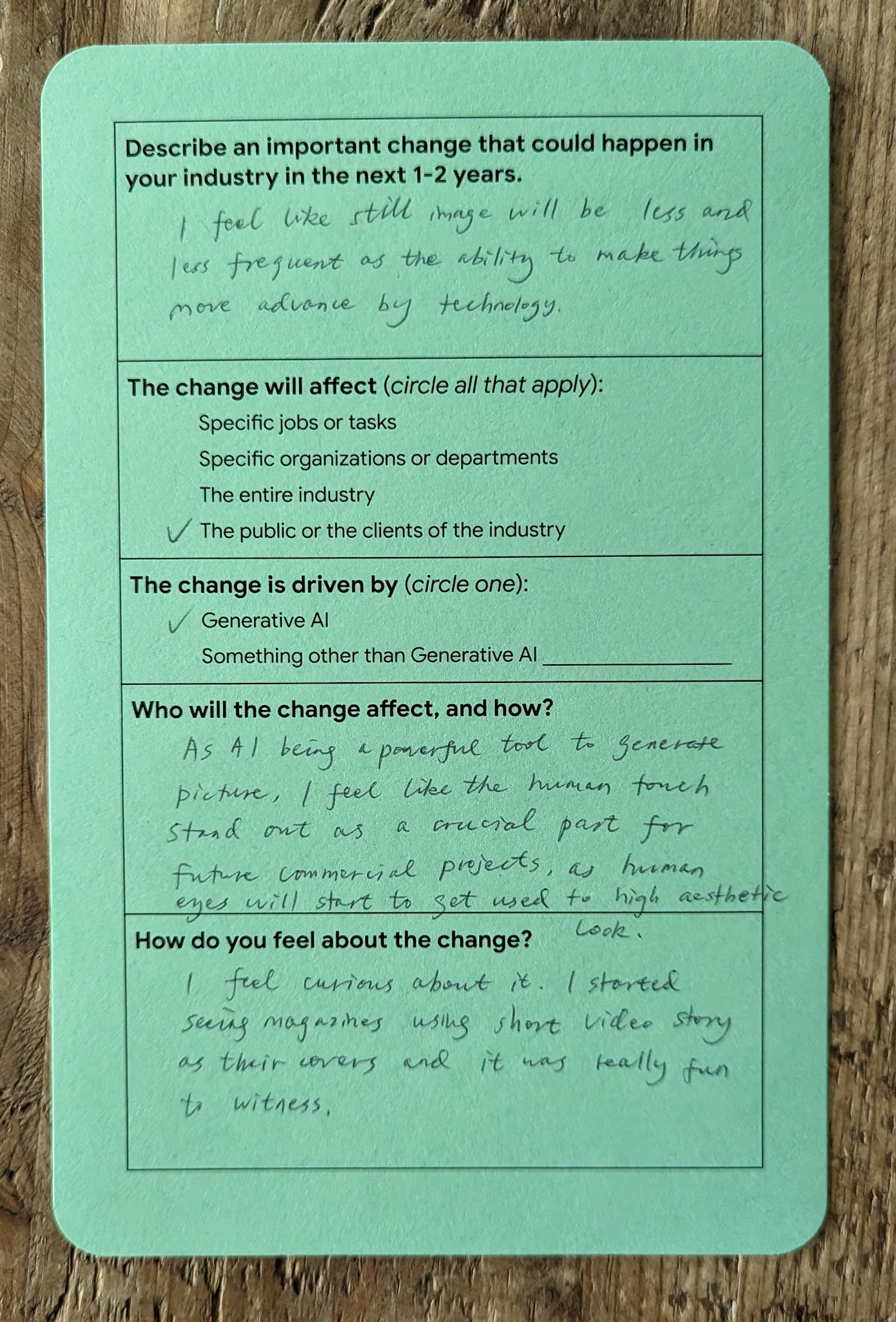}
    \caption{One of \pid{A4}'s change cards, discussing their expectations regarding generative AI's ability to generate images. Participants completed change cards individually and then shared them in a facilitated discussion.}
    \label{fig:changecard1}
\end{figure*}

After a meal break, we facilitated a discussion of potential industry changes. We began by introducing participants to a probe: a large, physical card that we term a \textit{change card} that encourages participants to reflect on important changes that could happen in their industry in the next one to two years; changes did not need to be related to generative AI 
(Appendix C: Figure~\ref{fig:changecard1}). 
Change cards enabled us to: (1) scaffold an envisioning process for individual participants about industry futures, through both open- and closed-ended questions about changes they expect; (2) elicit reflection on the participant's feelings about those changes, through an open-ended prompt; and (3) capture thoughts that participants could revisit and build on in collaborative discourse. 
Each participant spent about 10 minutes filling out their individual change card(s). We then facilitated an open-ended group discussion in which participants shared one or more of their change card(s) and responded to each other's ideas. These discussions, which built on the cards, contributed to the larger set of possibilities we subsequently analyzed together with participants, which we report on in Findings.

\begin{figure*}[htp]
    \centering
    \includegraphics[width=6.3in]{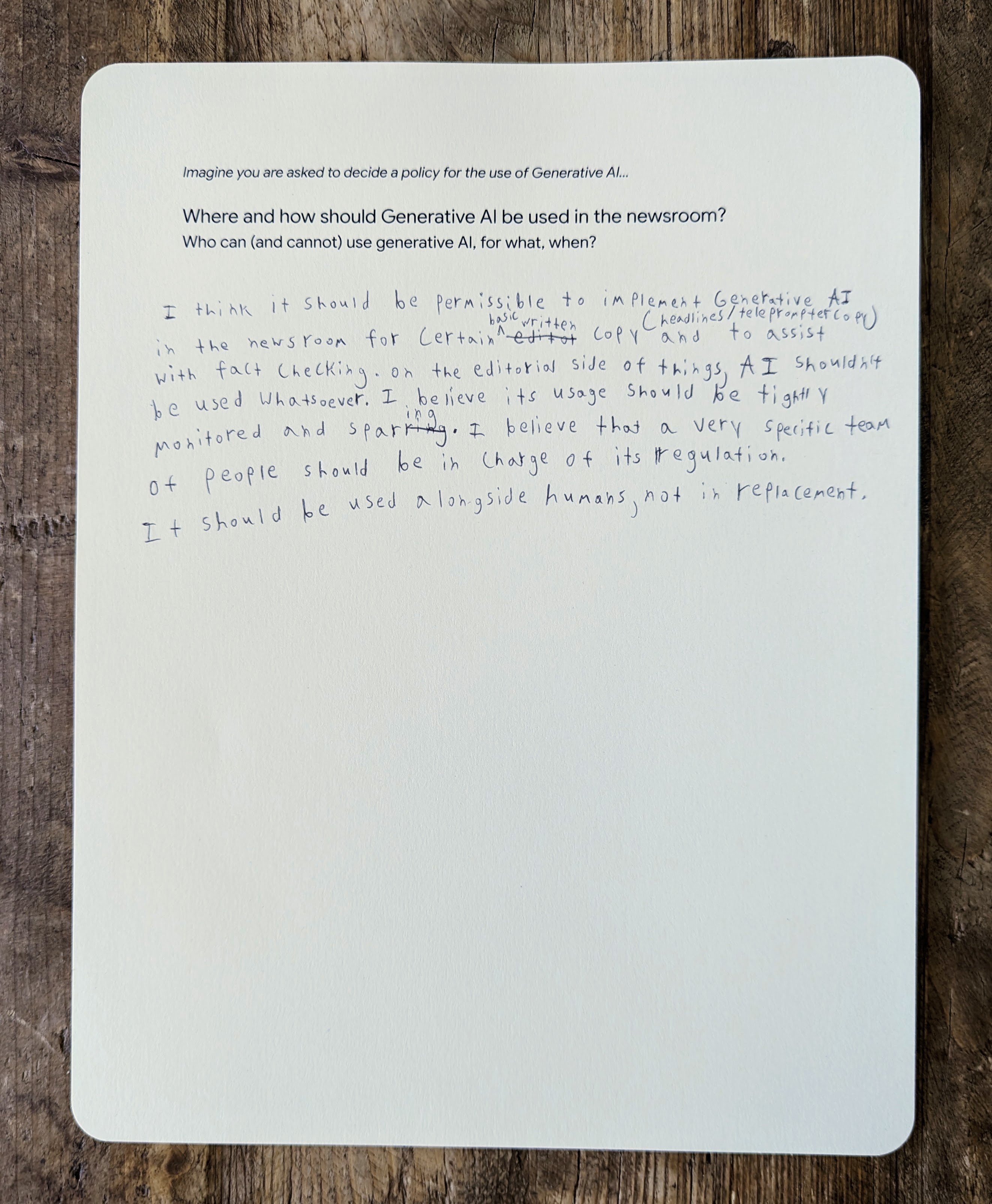}
    \caption{\pid{J5}'s policy suggests constraints on the use of generative AI in the newsroom. We lightly customized the policy handout to each group's industry. Participants completed their policy handouts individually and then shared them in a facilitated discussion.}
    \label{fig:policy1}
\end{figure*}

\sloppy
Following another short break, participants discussed and sketched out the broad contours of a policy to guide the use of generative AI in their industry 
(Appendix C: Figure~\ref{fig:policy1}), 
during which we further explored their expectations and attitudes. Participants were again given about 10 minutes to work individually before taking turns presenting their policies in a facilitated discussion. We followed the discussion with an exercise in which participants voted on which aspects of generative AI they find most helpful and most detrimental for their industry. After a brief discussion of this voting activity, in the final minutes of the workshop, participants were encouraged to speak about the potential impact of generative AI outside of their work. Throughout the workshops, we took care to encourage collaborative interpretation, problem-solving, and discussion among participants and moderators, and to make space for all participants to share their ideas and opinions. 

\fussy
\subsection{Analysis}
All sessions were recorded and transcribed verbatim with an automated speech-to-text service, and we then manually corrected the transcripts against the original recordings. All artifacts (54 industry maps, 140 change cards, and 54 policies) were collected and archived. We analyzed data from the corrected transcripts and artifacts inductively. Drawing on reflexive thematic analysis approaches~\cite{braun2019, braun2020}, four authors reviewed transcripts, in multiple configurations of paired and independent review, and one of these authors conducted comparisons with corresponding artifacts. These four authors engaged in deep and prolonged data immersion and discussion, independent open coding followed by memo writing to generate themes~\cite{birks2008}, and collaborative discussions in multiple rounds to compare their interpretations of the data reflexively and finalize themes.

\subsection{Limitations}
Several limitations of our study methodology should be considered when interpreting this work. First, it carries with it the standard issues attendant with qualitative methodologies and group interviews. We conducted the research in only three US cities with only one group per industry, and our small sample was not statistically representative of the roles or demographics of the professional fields we explored. Our findings should be viewed as a deep exploration of our participants' perspectives on their industries, but should not be taken as generalizing to their industries as a whole. Second, our choice of teaching content and activities, while appropriate to our research objectives, may have influenced participants, although we tried to minimize any effects through participant-led discussion. Finally, while we confirmed our understanding of participants' comments during the sessions, and one or more of the authors have experience in most of the industries represented, our interpretations may lack context or nuance that would have been more readily available to members of the same professional categories.
\section{Findings}
\label{sec:findings}

In this section, we present our findings. We begin by describing participants' expectations of generative AI's impact, including a dominant narrative that emerged across the groups. We then turn to four current social forces that shaped our participants' perspectives on how generative AI will affect their industries: deskilling, dehumanization, disconnection, and disinformation. Finally, we describe some of the unique perspectives within particular knowledge industries.

\subsection{Expected Impact of Generative AI}
\label{sec:narrative}

In this subsection, we give an overview of participants' perspectives on how generative AI may impact their industries.\footnote{We note their perspectives were informed by content in the workshop as well as their prior exposure. Almost all participants had heard of generative AI or a specific generative AI system, \eg in the press or from colleagues, friends, or family. Many participants also had experience using an app, especially ChatGPT or Midjourney, at least lightly, for either personal purposes (\eg writing a birthday card) or professional purposes (\eg drafting a memo).} We begin by briefly summarizing the participants' dominant narrative regarding generative AI, which we term the \Tool/ \Narrative/ (see Table~\ref{table:narrative}). For ease of comparison, we also briefly summarize the \Academic/ \Narrative/ and the \Hype/ \Narrative/ introduced in Sections~\ref{sec:fow} and \ref{sec:hype}, respectively. We introduce these narratives to situate participant data against extant meaning-making discourses, in alignment with reflexive thematic analysis \cite[p.211]{braun2022}. In the remainder of this subsection, we describe the \Tool/ \Narrative/ in more detail.

\begin{table}[hp!]
\begin{tabular}{>{\raggedright\arraybackslash}p{0.95\columnwidth}}
\specialrule{1pt}{0pt}{7pt}
{\large \Tool/ \Narrative/} --- Participants largely envision that it will be realistic and desirable to use generative AI as a tool to perform menial work, subject to human review. Further, they believe that existing guardrails in their industries can be leveraged and augmented to perform such review. They expect certain roles will be impacted, but for most industries they do not anticipate substantial transformation or elimination of a wide range of jobs. In most cases, participants do not anticipate change as broad as that predicted in the \Academic/ \Narrative/ or the \Hype/ \Narrative/. 
\\\specialrule{0.5pt}{7pt}{7pt}
{\large \Academic/ \Narrative/} --- Typically shared in technical reports\footnotemark[5] or peer-reviewed papers by think tanks, consulting firms, or academic groups, and echoed in some rigorous news reports, this narrative takes an analytic approach to arguing that generative AI will have broad, substantial impact across industries, jobs, and tasks. Work in this vein often includes projections (\eg \% of tasks that can be automated by generative AI, \# of jobs affected), outlining a range of possibilities, but usually including an exceedingly high upper bound for such estimates. While this narrative is not as catastrophic or exaggerated as the \Hype/ \Narrative/, it usually predicts transformative change. 
\\\specialrule{0.5pt}{7pt}{7pt}
{\large \Hype/ \Narrative/} --- A predominant discourse in social media and many news articles, this narrative makes dramatic or even hyperbolic claims about how generative AI will change/replace human labor in the future. It often draws comparisons between generative AI and previous historic innovations or even ``the big bang'' (\eg~\cite{sahai2023}). In some cases this narrative originates from leading figures in AI~\cite{marcus2019} who are arguably framing evocative messages for the lay public. However, it often misrepresents expert opinion or incorporates evidence from sources such as the \Academic/ \Narrative/ in misleading ways and emphasizes anxiety-inducing worst case scenarios~\cite{omaar2023}. 
\\\specialrule{1pt}{7pt}{7pt}

\end{tabular}
\vspace{0.6em}
\caption{An overview of three differing narratives regarding generative AI. The Effort-Saving Tool narrative emerged from our analysis of study data. We synthesized the Transformative and Sensationalized Narratives from extant literature as described in Sections~\ref{sec:fow} and \ref{sec:hype}. Overall, participants did not embrace broader narratives of disruption, and the \Tool/ \Narrative/ is a fairly limited view of expected labor impacts compared with current expectations in the \Academic/ \Narrative/ and the \Hype/ \Narrative/. }
\vspace{1.3em}
\label{table:narrative}
\begin{flushleft}
  \footnotesize{
    \footnotemark[5]Due to the recency of the topic, much of the relevant literature appears in pre-prints or white papers.}
\end{flushleft}
\end{table}

\subsubsection{Generative AI can automate menial work: ``We need to just churn it out.''}

While a wide range of possible use cases were discussed, our participants' experience, imagination, and preference centered on the use of generative AI for menial tasks. They characterized these as routine and/or formulaic tasks that they thought could easily be automated, or which they had actually tried in apps like ChatGPT, such as drafting social media content, troubleshooting tickets, leading guided breathwork sessions, or creating lesson plans.

\begin{listquote}
\item ``For the tasks that I find to be tedious or taking away from the bigger picture or [taking away from] what's actually on my plate, those to-do items that I don't really care about, but we need to just churn it out, ChatGPT all the way.''~\aff{A1}\footnote[6]{Throughout the paper, we assign participants pseudonyms that begin with the letter associated with their industry, \ie \textit{A} (Advertising), \textit{B} (Business Communications), \textit{E} (Education), \textit{J} (Journalism), \textit{L} (Law), \textit{M} (Mental Health), and \textit{S} (Software Development). In some cases, we have lightly cleaned the quotes for readability, \eg to remove inconsequential false starts or to remove filler words such as ``like'' or ``um''.}
\item ``I've used it to help write release notes... I'll have this list of really boring updates or features that aren't that sexy, and then they're like, ‘Just make it sound fun.’ And that is really time consuming actually. And generative AI is extremely helpful for that.''~\aff{S5}
\item ``It does seem useful when you have to sort of pound out outlines that are just very robotic.''~\aff{J7}
\item ``Within law firms there's some work product that you produce that takes a lot of thought and decision making and some creativity to really draft something that is unique in the way that you put it together. There's other tasks that I do on a daily basis that are a lot more formulaic... If you're comparing [the first] to say Citizen Kane, [the second] is more like an SVU episode where you can throw it together in 10 minutes.''~\aff{L1}
\end{listquote}

Participants were typically enthusiastic about offloading these tasks to generative AI, both because they were tedious and because they felt the time saved would allow them to focus on more meaningful, human aspects of their work. For example, mental health professionals eagerly proposed generative AI taking over rote work, such as note-taking or patient intake, thereby increasing their efficiency and freeing up more time to focus on interpersonal work with their clients.

\begin{listquote}
\item ``I wish AI could help us with note-taking in any way. That's the part of our job that I hate the most. It's so laborious... it takes up so much of our time and if there's anything that could support us with that, that would be amazing. I don't know how it would do that, but if it could, I would cry for joy.''~\aff{M1}
\end{listquote}

Further, some greatly valued the ability to scale and handle high volumes of work. For example, \pid{B6} had recently sent out several hundred thousand emails over a three month period, and imagined generative AI could increase consistency and speed.

\subsubsection{Generative AI is a useful tool but requires human review: ``Maybe a helpmate but certainly not in control.''}
\label{sec:narrative-review}

Consistent with their orientation to the use of generative AI for busywork, participants dominantly oriented to generative AI as a tool for human workers. For example, some specified that generative ``AI tools should assist'' employees, but not ``do'' their work. Many emphasized generative AI should not go beyond tool status and perform certain kinds of knowledge work, \eg generative AI should not be used for decision-making, setting strategic direction, or forming human connections.



Participants in all groups were concerned about generative AI's potential to make mistakes or produce undesirable output. For example, participants in the advertising and business groups shared concern that generated content might violate brand standards or copyright, and lawyers spoke of the need to attest to the accuracy of legal documents.
One participant recounted a news story in which an attorney submitted a legal brief created with ChatGPT, which included ``hallucinated'' cases:

\begin{listquote}
\item \affpre{L7}: I haven't used it at all. But I'm kinda wary of introducing new tech generally. Especially as a litigator, anytime you submit anything to a court, you sign it under what we call Rule 11. You're representing it's true and accurate and you’ve got a good faith argument for whatever you're doing. And I think you may have heard about that [case] where a guy wrote his brief using AI and then the AI generated six cases, citing opinions that didn't even exist apparently. Which to me is mind boggling. \newline
\affpre{L6}: I would've won so many cases if I could have done that... \newline
\affpre{L7}: That's right. Exactly. And then this guy, the attorney that did it, he's kind of... a national joke.
\end{listquote}

Accordingly, generative AI was viewed as potentially risky, and the overwhelming sentiment in all groups was that humans would need to check most or all of generative AI's output to ensure its quality. In some fields, participants stipulated that this check must be performed by a qualified professional such as an attorney.

\begin{listquote}
\item ``The AI would produce something, and then we would say yay or nay.''~\aff{J5}
\item ``[Generative AI code] should never be something that goes to production without human testing and review.''~\aff{S7}
\item ``A lot of this problematic stuff that we're hypothesizing is coming out of this idea that AI could be this standalone [advertising] team doing every single thing from start to finish and releasing this ad without consent of anyone or something like that. Which I think is still pretty sci-fi...''~\aff{A3}
\item ``I feel like you're always gonna need some type of human oversight, right? No matter how good the technology gets, I just don't see humans being obsolete <laugh>. Like how's that possible?''~\aff{A5}
\end{listquote}

\subsubsection{Existing industry governance structures are amenable to overseeing generative AI: ``It would just be part of what we've already been doing.''}

Participants spoke of new roles, responsibilities, and qualifications to enforce human oversight and monitor generative AI's use and output in knowledge work.

\begin{listquote}
\item ``Training like getting a driver's license. Once you get the ability and understand the rules you can use [generative AI].''~\aff{A7 (Policy)}
\end{listquote}

Participants pointed out that many such oversight functions, while new, integrate well in existing industry review structures and policies. For example, lawyers review the work of paralegals or junior associates, business communications undergo legal and compliance approvals, and software development has code review. Participants explained:

\begin{listquote}
\item ``There's oversight right now against a brand voice, tone, style, consistency. Anything that is generated has to go through an approval list... [There are] checks and balances in place. So I would feel really good about AI helping contribute to that message if it went against those checks and balances.''~\aff{B4}
\item ``Most of our districts already have user agreements which talk about plagiarism, hate speech and those sorts of things. I think it would just fit right into most of our districts’ user agreements.''~\aff{E5}
\end{listquote}

\subsubsection{Certain jobs will be downsized, but do not expect broad disruption: ``A lot of entry level positions will be eliminated,'' but ``I don't think therapists are gonna go away.''}
\label{sec:narrative-limited}

Participants in most industries anticipate that as generative AI streamlines formulaic work, certain jobs will likely be eliminated. They tended to think of the job risk as narrowly scoped to specific entry-level jobs or jobs with many rote tasks. For example, they ruefully highlighted jobs such as help desk, legal assistant, or product photography as being at risk.

\begin{listquote}
\item ``Commercial work I think is done <laugh>. Commercial photoshoots cost hundreds of thousands of dollars, lighting, makeup, all that stuff. [And] now you can just type it in... instead of paying for production, AI could make it all up. They won't need 99\% of roles on set. Don't need models, photographer, makeup, set designer, etc. Getting rid of the majority of roles is terrible.''~\aff{J6; J6 (Change Card)}
\end{listquote}

Many participants expected that even if workers in entry-level or similar roles in their industry were affected, they personally were somewhat unlikely to experience significant impact, feeling confident, for example, that generative AI is unlikely to be skilled enough to replace human professionals within the timeframe of their own careers. While some did acknowledge the potential for industry-wide shifts, particularly in software development, advertising, or business, sentiment such as the following was generally the exception rather than the rule:

\begin{listquote}
\item ``As a communicator, it's a little intimidating how good the bot writers can be, and I know our Bangalore teams are looking at hiring bots or putting them in place rather than hiring humans. So, there's an excitement factor in terms of cost savings as an executive, but also an intimidation factor as a communicator to see that a whole industry of journalists or people who major in [communications] could be replaced to a degree. So, I think specific jobs, organizations, entire industries, the world will be affected by that.''~\aff{B4}
\item ``Writers could start to become obsolete in journalism [because of generative AI]... We could potentially be pushed to the back burner and not relied upon whatsoever to produce content. To be frank, I'm very afraid and believe that I may have to change career paths soon.''~\aff{J5 (Change Card)}
\end{listquote}

At the same time, participants were somewhat uncertain about both their own projections as well as broader narratives. To bolster their reasoning, they sometimes drew analogies to other resources such as the internet, Google Search, or industry-specific databases, such as LexisNexis in the legal profession.
\begin{listquote}
\item ``I see certain things as tools that replaced old tools. This is not really a new tool. I think it's just an enhancement of something we already have. For example... Lexis.''~\aff{L7}
\end{listquote}

Broadly, participants tended to take a tempered view of likely changes in their respective industries, relative to the \Hype/ \Narrative/ or even the more moderate \Academic/ \Narrative/. While some participants were aware of media claims, they approached them with skepticism, although they sometimes found the claims concerning if not convincing. In some cases, they literally discounted these narratives by proposing adjustments to reported statistics. Overall, participants in most industries appeared to feel other forces are more disruptive than generative AI. We discuss some of these in Section~\ref{sec:forces} and we discuss industry-specific expectations further in Section~\ref{sec:industry}.

\begin{listquote}
\item ``I've seen it time and time again where everybody [in media companies] runs to the same goal post. Like the pivot to video is the most famous one. It's exhausting as a journalist and we watch all these people get laid off and then a year later they say, `Never mind, it wasn't the answer we thought it was, back to where we started.' <\emph{J2} interjects: [Remember] podcasting? <laugh>> Yeah, exactly. We've all seen it. So I don't know how ChatGPT shakes out in this whole picture, but I do think it's probably not as insane as some people are making it in terms of the tidal changes that it'll make to our industry. But it is probably like 60\% of the way there.''~\aff{J3}
\item
\affpre{L4}: Goldman Sachs says that AI will take over three hundred million jobs away in the coming years.\footnote[7]{Participants appear to be referring to a Goldman Sachs report released in March 2023~\cite{briggs2023} whose findings were also highlighted in a McKinsey report released in June 2023~\cite{chui2023}. The report states that 300 million full time jobs are at risk of automation, globally. The media covered both reports, sometimes sensationally (see critique:~\cite{omaar2023}), and coverage was often unclear about the geographic scope of the statistic.} \newline
\affpre{L7}: Three hundred million? \newline
\affpre{L4}: Three hundred million. \newline
\affpre{Moderator}: Do you find that plausible? \newline
\affpre{L7}: So all of the US will be laid off. \newline
\affpre{L4}: Yeah. Three hundred million... \newline
\affpre{L2}: And some of the companies like McKinsey who have studied the prognostications of those kinds of things find they're wrong way more than they're right. But the scary part is even if it's fifty million people, it is significant with ripple effects throughout everything. You know, it's amazing. They usually aren't right, completely. But the reality is that you get very scared by that headline... Talk about anxiety. `Hello, am I in the three hundred million who's not gonna be employed anymore?' Tough. Tough. \newline
\affpre{L4}: It was concerning when I read it.
\end{listquote}

\subsubsection{Need to evolve and adapt with generative AI: ``Let’s evolve with the times, right?''}

While some participants were saddened by the prospect that many would lose their jobs (particularly those in entry-level positions), others felt these changes were a natural evolution of knowledge work, which could bring new opportunities.


\begin{listquote}
\item ``Change is a constant despite humans naturally tending to try and resist it. I am all for change and constant adaptation as long as the baselines of independent thinking and learning is still held as an important skill to possess and have.''~\aff{L5 (Change Card)}
\item \affpre{A2}: [Generative AI] could eventually replace headshot photography or photo shoots for certain industries... \newline
\affpre{A8}: Yes. I wanna wait until I can say, `[Generate it] in the vein of Annie Leibovitz,' you know? ... \newline
\affpre{A2}: Gonna get there. That's the thing. It's only gonna get better. So how do you deal with that? What happens to all those creators, all those jobs? ...  I'm hopeful that there will always be a space for creative people... [but] there will be I think a big chunk of those jobs that just kind of go away. Like the simple headshot photographers or the person who does passport photos, things like that. I don't know. At this point it seems a little bleak and ambiguous, but I could see it happening... I feel like the working jobs, the creative jobs, the smaller ones will go away and then it'll just be more top heavy companies where they're overseeing the technology. \newline
\affpre{A5}: Isn't that just evolution? We went from no machines to machinery and people lost millions of jobs. Is it just not a part of evolving? I feel like there probably are still some human aspects that could go into AI... Is there a human there to say, `Oh, this isn't right.'? Maybe like quality control.
\end{listquote}

Some also emphasized that knowledge workers need to adapt and ``reformat their skillset'' in order to leverage this new tool, and stay relevant and not get left behind.
As one concrete example, participants spoke about how upcoming changes would require ``prompt engineering'' skills to elicit improved outputs from generative AI systems. 

\begin{listquote}
\item ``[To \pid{B3}] So you just graduated? I would learn to adapt with it and use it rather than ignore it... I guess it's going from a horse to a car, it's inevitable... Rather than ignoring it, just learn to drive... How [can you] do your job more efficiently and effectively by using the technology to your advantage?''~\aff{B4}
\item ``People will adjust. We are all going to move on... I do meet a lot of people, a lot of technologists, my colleagues, coworkers, they get scared by it. But then I tell myself that if you're scared by the AI itself, then maybe you as a person are not evolving fast enough... I think it gives us a new challenge and we should embrace it. We should accept it and then evolve faster, adopt and adjust as well.''~\aff{S1}
\item ``I think that if you're in the industry, keeping up with these tools and using them as an asset to enhance your work will make you irreplaceable.''~\aff{B3 (Change Card)}
\end{listquote}
\subsection{Social Forces: Deskilling, Dehumanization, Disconnection, and Disinformation}
\label{sec:forces}

We asked participants to describe important changes coming in their industry over the next one to two years. While some of these changes were driven by generative AI, many were not. For example, mental health professionals expect broader legalization of psychedelics will be transformative in their field. In some cases, participants expected that generative AI would interact with other changes to bring about particular impacts on their industry. Most prominently, participants spoke of four existing global and national forces that framed their expectations of generative AI's impact. Specifically, they anticipated generative AI would amplify the following issues that shape how knowledge workers do their work in their industries: deskilling, dehumanization, disconnection, and disinformation. These issues can be described as social forces~\cite{white2019,cox1987production}.


\subsubsection{Deskilling: The ``Uberfication'' of Knowledge Work}
\label{sec:deskilling}

Independent of generative AI, several forces are actively shifting revenue and employment in certain knowledge industries. For example, the emergence of remote services, such as BetterHelp and LegalZoom, was accelerated on both the supply and demand side by COVID-19, as both providers and clients often desired virtual sessions. While such services provide clients easier access to mental health, legal, or other advice and may reach a larger number of people at a more affordable price, participants argued they provide lower quality service and undermine demand for highly trained professionals. \pid{M6} described how telehealth tech startups pose a greater employment threat than generative AI:

\begin{listquote}
\item ``Things like BetterHelp are the biggest threat to our job security/livelihood. Because I think it's an Uberfication of mental health care. 
~\aff{M6}
\end{listquote}

\noindent
The ``Uberfication'' of work refers to a broader shift in the political economy of how work is ``arranged through the use of digital technologies...to create `platform capitalism'{''} \cite[p. 61]{simburger2016taxi}. A key element of these labor conditions is supplanting employees with self-employed workers \cite{gloss2016}. Participants, especially lawyers and journalists, reflected on a significant shift towards contract, freelance, or ``permalance'' positions (in which workers work in an extended freelance capacity with an employer). Driven by bursty demand for expertise as well as cost-saving measures, these positions are precarious and do not offer the same benefits as full-time employment \cite{wiest2018}.

Generative AI may interact with these extant labor trends to further deskill industries or reduce opportunities for highly trained professionals. For example, some participants speculated that non-permanent employment would be easier to replace with generative AI than full-time positions, since few employment protections would be in place. Some also pointed out that generative AI could facilitate inexpensive online services by proposing text strings that less-trained human workers could pass on to clients. Further, participants in fields such as law and mental health expressed concern that clients might sometimes use generative AI for advice rather rather than incurring the cost of consulting human professionals, undermining professional expertise and revenue while also yielding poor outcomes for clients.

Participants spoke of how these trends might personally affect them. For example,
while previous concerns about the gig economy have often focused on blue collar work, participants drew connections between the gig economy, generative AI, and white collar work. \pid{J2} has worked as a labor journalist, previously writing about the impact of technology on blue-collar workers. With the rise of generative AI over the past year, his concerns have become more immediate and personal.

\begin{listquote}
\item ``As a worker myself... I wasn't too worried about [the digitization of the workspace] until now because, you know, they're digitizing a factory or whatever... now it's actually starting to impact my industry. And all those questions that were arising about factory workers or Uber drivers or whatever, they're starting to be asked about us [journalists] now, to apply to us as well.''~\aff{J2}
\end{listquote}

\noindent
Participants also discussed how these forces can affect their wages and opportunities, \eg \pid{M6} felt these changes could reduce his wages or undermine his investment in a specialized career, while also reducing quality of service:

\begin{listquote}
\item ``I'm up to my neck in student loans still. Part of me is like, `Well, it's great if BetterHelp can be affordable for people.' But what about my job security, my ability to pay off my student loans? ... And then the idea of generative AI doing chat... I do have concerns that this could be used in a way that would undermine our job security, but also I think actually not provide the level of care for the consumer as well.''~\aff{M6}
\end{listquote}

\noindent
\pid{S7} shared similar concerns that generative AI might make her hard-won skills less valuable and set back her career:

\begin{listquote}
\item ``I didn't really know that I as a girl with a non-traditional background could be an engineer. And I also saw it as a way to make a lot more money than I was making. It was definitely a way for me to double, triple my salary and have access to a life that I could only dream of. And so it is kind of a bummer, even though it feels like a very first world problem, to be like, `Oh, now I'm back to potentially in a few years not having a very valuable skillset or a skillset that everyone can have.' ... It is a little disheartening to be like, `Oh, finally ahead. I'm finally at this place I wanted it to be,' and now I have to play catch up again.''~\aff{S7}
\end{listquote}

\noindent
Participants also expressed concern that, given its fitness for menial tasks, generative AI might eliminate many entry-level positions and therefore remove pathways into more senior roles:

\begin{listquote}
\item ``I would not be in my job, in the role I have today, if I didn't start as an SDR [sales development representative]. It makes me sad to think these entry level roles may no longer be necessary... [It's] upsetting because I think everyone has to start somewhere, and if you don't have that basic knowledge, how do you grow from there?''~\aff{S6; S6 (Change Card)}
\item ``A lot of help desk positions can be eliminated... I’m not too happy about that. I got my start on help desk. I didn't go to school for computers at all. So taking a lot of those jobs away would be really sad cause they're just a really easy entry into a good field.''~\aff{S4}
\end{listquote}

These issues around deskilling, menial work, the gig economy, and more widespread job loss all arise in the context of broader economic uncertainty in the US. Participants, at times, speculated on how these changes to jobs could potentially be addressed through social welfare approaches like universal basic income, but were not confident governments or employers would meaningfully address these labor changes, or other ethical issues related to generative AI, due to their position operating within a capitalist environment:

\begin{listquote}
\item 
\affpre{S7}: I just feel like under capitalism there can be no good AI... I'm still figuring out how I truly feel about AI. I don't think it's inherently evil or bad, but I think when you're talking about using it at a company or corporation, the people at the top are always concerned about making a profit and cutting out jobs and whatnot. So I just am like, the end stage of this, no matter how good it can be, it just always feels really icky and bad. \newline
\affpre{S5}: Ideally it's like the Star Trek universe. \newline
\affpre{S7}: Yeah. \newline
\affpre{S5}: I'd be into that <S7 laughs>, nobody has to work. We’re all like happy. \newline
\affpre{S7}: That would be great. But then there's no one at the top getting all the profit. \newline
\affpre{S5}: Yeah.
\end{listquote}

%
%
\subsubsection{Dehumanization: ``Whose job will it be to find out how to incorporate human nature into the AI?''}

Replacing human workers with algorithms raises concerns about dehumanization in which the task loses a characteristic of how humans interact with each other \cite{fritts2021ai}, becoming more sterile and impersonal \cite{shortliffe1993}.
Many participants expressed concern that use of generative AI might in various ways lead to a loss of humanity. 
Several participants offered a literal mechanism for retaining humanity in the use of generative AI, proposing a specific quota, such as ``at least 80\% of words, photos, everything on our site is created by human minds'' or ``60\% [of the job] needs to be conducted by humans.''
\label{sec:dehumanization}


One concern was that they felt generative AI does not have the capacity to perform interpersonal work. For example, business communications professionals spoke of the importance of personal touch and authentic human communication, and mental health professionals voiced that generative AI can not establish human-to-human rapport, which is required for therapy to be effective.
Another concern was that participants feel joy in performing tasks themselves, an emotional experience not provided by using generative AI. The introduction of algorithmic technologies into organizations in dehumanizing ways erodes workers' sense of autonomy \cite{sewell2015out, newlands2021algorithmic} and overall job satisfaction \cite{lagios2021explaining}. Participants echoed these insights in the context of the introduction of generative AI into their industries. 

\begin{listquote}
\item ``From the photography viewpoint for me, they're replacing actual picture taking with just generating in AI, and the whole reason I like photography is I like holding a camera. I like framing it. I like the interaction with the person and all that. And generative AI just completely gets rid of all that. The human connection...''~\aff{J6}
\item ``There are folks who get their joy and their sense of meaning from writing code and that's kind of their thing... these are people that spent years mastering this trade. And I think that's where it gets sad.''~\aff{S2}
\item 
\affpre{Moderator}: So should [an undergraduate] go get a computer science degree where they’re going to do a lot of math and code, or should they go to some prompt engineering boot camp? \newline
... \newline
\affpre{S2}: Well, it’s the same question in the context of art school. Should someone go and learn how to paint and all the intricacies involved there? Or is it futile because a computer can spit out that painting for you? I don't think there's a right answer... If someone wants to go to school for computer science, because... learning and tinkering is fun, they should do it and don't let the robots stop them.
\end{listquote}

Participants emphasized the value of human production and creativity. Their comments were reminiscent of the Arts \& Crafts movement, which challenged the integrity of mass produced objects, associating them with dehumanization and a decline in production standards~\cite{penick2019}. The Arts \& Crafts movement valued artisanal production and human craft, even though it was less efficient than the mass production that became common in the industrial age. Similarly, participants spoke about the value of making things by hand, and the high quality and meaning of human-created objects. At the same time, some worried the public’s aesthetic standards could shift over time to less human content.

\begin{listquote}
\item ``I think that there's potential for real things outside of screens, like actual photos in an album. Like who makes those anymore? But I think things like that would become almost like sacred and similar to... the dumb phone movement straying away from smartphones and screens. I think that there's gonna be at some point a push towards that. Like the things you could physically touch and make with your hands... Wanting the personal touch of knowing it took time and intention and purpose and thoughtfulness to create something. I think there's real beauty in that. And I guess I don't really see the beauty in AI.''~\aff{A1}
\item ``I think a lot of skilled labor and artisan style jobs are going to come back into fashion and it'll be very hard for AI to replace them. My boyfriend went to furniture school recently and so he's a carpenter and a fine furniture maker. And I was just kind of musing about how right now I'm the breadwinner, but I think soon he might be the breadwinner because there's something about something that's made by hand, something that is of quality. We live in this fast fashion age where things are just mass produced. And I think there's a real desire for something that is made with care that will last, that is creative... It takes a lot of personal skill and a lot of physical labor to make a table… and  I have a hard time thinking that generative AI is really gonna be able to replace that.''~\aff{S7}
\item ``I believe when AI created content (articles, videos, infographics, etc.) begin to seep into our feeds, we will eventually accept its strange aesthetic (which we call `robotic', `uncanny', etc.) as normal. This could create an entire new aesthetic of digital media as we know it. I'm concerned.''~\aff{J8 (Change Card)}
\end{listquote}

Some participants emphasized that people must retain critical thinking skills as well as the ability to do the type of work that is being offloaded onto generative AI, and that people must not become lazy. At the same time, some recognized a delicate line between being lazy and being efficient, therefore saving time for more meaningful activities.

\begin{listquote}
\item ``I hope my daughter never discovers [ChatGPT]... I don't want it to take away creativity. I want my daughter to think of her own thoughts. Like when she's doing her studies. That's really important to me.''~\aff{L3}
\item ``I think you need to actively not be lazy, because it's so helpful. You need to say, `Okay, I'm not gonna just let this responsibility fall onto this machine. I'm gonna still really play the biggest role here...'{''}~\aff{A1}
\item 
\affpre{A7}: I assume I'd be pretty lazy right now if I was in high school and I would just try to take advantage of [generative AI] as much as I could... You can literally not learn for the rest of your life if you don't want to now. \newline
...  \newline
\affpre{A6}: Is it lazy or is it clever? Are you learning the long way or are you streamlining the process?
\end{listquote}
\subsubsection{Disconnection: ``You have to be there.''}

Feelings of disconnection can manifest in knowledge work, \eg new tech-mediated working arrangements can increase feelings of worker isolation \cite{vega2000isolation} and broader social trends in perceived isolation can exert external pressures on knowledge work practice \cite{erickson2016}. Participants spoke about increased social disconnection, such as the ``loneliness epidemic'' \cite{epidemic2023} stemming in part from COVID-19 and the corresponding social isolation and increase in remote/virtual work and schooling, as well as other factors, such as escalating phone and social media use or even addiction.
They were concerned that generative AI would contribute to further disconnection from physical and social experiences. For example, mental health professionals worried that therapeutic uses of generative AI could have the opposite of intended effects and exacerbate loneliness, even as patients turned to it for comfort:

\begin{listquote}
\item ``I hear people talking about the uses of AI [for mental health] in the chatbot model, and to a great extent, I'm wildly against it. Because I think that a big part of the reason we have the increase in people needing mental health services is because of disconnection. And I think that while it might really have helped your client in that moment to have a bot to chat to, it's still not a replacement for human interaction. And I think that that disconnection is a lot of what is driving this [mental health] crisis.''~\aff{M6}
\end{listquote}

Our participants also described how good journalism requires being ``on the ground'' and how the continuing rise in freelance and remote work threatens the quality of news production, a trend that could be further exacerbated by the use of generative AI. \pid{J8} characterized the future of journalism with generative AI as ``just more recycled and disconnected from reality, allowing people more to be separated from what they're doing instead of fully immersed in it.'' Participants also expressed concern about content becoming increasingly disconnected from reality over time as new generative AIs are trained on layer upon layer of recycled, increasingly non-human content.

\begin{listquote}
\item ``As journalists, when we cover things, we are currently on the ground doing reporting, talking to human beings and things like that. And more and more it's becoming like we're talking to people over Zoom or we're not being flown out to places to cover things anymore and we're hiring freelancers... We get a bunch of footage from overseas and then we have to make something from it because we've hired freelancers there instead of being on the ground doing it. So it's this gradual, very gradual disconnection from what we're covering and what we're thinking and the actual reality of the world.''~\aff{J8}
\item ``[News aggregators are] a homogenization of all the stories. We lose the individual voices of the people who are witnessing or describing, you know? ... [As a reporter] you still need to experience it, you need to taste it, you need to see it, you need to feel it in order to transcribe those sensations of the human experience to your readers.''~\aff{J1}
\item ``I think it's a lot of dumbing down and recycling that's happening. I'm worried about the long-term effects of recycling things that have been recycled already with text and images and what that will do to ideas.''~\aff{J8}
\end{listquote}
\subsubsection{Disinformation: ``I think they call them hallucinations... it puts it out so authoritatively.''}

Disinformation is the intentional spread of false, inaccurate, or misleading information designed to intentionally cause public harm \cite{HLEG2018}, and its production is often motivated by ``ideology, money, and/or status and attention'' \cite[p. 27]{marwick2017media}.
Participants expressed concern about the role generative AI may play in disinformation, and more broadly in the production of low quality content, particularly in the broader context of US discussion of polarization, culture wars, and media bias. One concern was that since generative AI typically incorporates internet data in its training, its quality would be undermined by existing low-quality content on the internet:

\begin{listquote}
\item ``It can't just pull it from the internet as it is right now because there's so much misinformation pre-AI.''~\aff{J7}
\item ``I [am] really concerned about AI learning things all over the internet, just chaotic. But if there's an AI that collects everything from [an authoritative source]... I'll be really excited about it.''~\aff{A4}
\end{listquote}

Participants further expressed concern that generative AI would itself proliferate, or would be leveraged to produce, additional low-quality content. They were worried about misinformation and disinformation currently in the media and online, and saw generative AI as a tool that could make it even easier and faster to produce harmful content. They anticipate this will have significant negative social and political repercussions:

\begin{listquote}
\item ``[I fear] propaganda would be so undistinguishable that no one's gonna know what's real... We're not gonna be able to determine what's real and what's fake and it could be the doom of us.''~\aff{A6}
\item ``It does seem like a major issue with deep fakes, especially in the realm of pornography, which is what keeps the internet running. There's already issues with revenge porn, and the notion that you can do this with anybody's face, anybody's likeness... The worst human urges <laugh> and desires are gonna be generated by the generative AI.''~\aff{J7}
\item ``Generative AI just comes at a bad time for us as Americans... and for us as journalists. We have such a problem in sensitivity around words like truth and falsehood and what is the truth and objective truth versus subjective truth.''~\aff{J3}
\end{listquote}


Participants were also concerned that media upheaval, caused by wide-ranging factors such as economic or political instability, is currently driving news aggregation and media concentration, thereby eroding the quality of information available to the public and ``dumbing down'', homogenizing, or politicizing content. They were concerned generative AI could exacerbate these issues. \pid{J2} spoke of media trends that started in Europe:

\begin{listquote}
\item ``[A change] that is not caused by generative AI, but that will be made worse probably by it is the further media concentration. That's caused by billionaires buying media left and right, agglomerating them and altering and then sometimes increasing their editorial lines, to their benefit... [as] a tool for [their] political agenda... I believe generative AI might make it even worse because all of a sudden these people will be able to buy and concentrate media. If anyone has any ethics issues, well, you know, the door's right there. And generative AI can do it for even cheaper anyway... just buy a newspaper, fire the entire newsroom... we'll just run it by generative AI.''~\aff{J2}
\end{listquote}

\subsubsection{Industry Perspectives} 
\label{sec:industry}

Above, we drew themes across the industries we studied. Here we provide additional detail for each industry.
To illustrate the most salient points and capture the unique character of each discussion, we created composites consistent with the content, language, and tone of responses we received from each group~\cite{creswell2018,willis2019}, shown in Table~\ref{table:industry_composites}. These composites illustrate nuance in how themes play out across industries; for example, which roles participants think are most likely to be affected in their specific industry.

\begin{table*}[th!]
  \centering
  \begin{center}
\begin{tabular}{ p{8.2em} p{44.5em} }
 
 \toprule
 \textbf{Mental Health} \newline\small
 \tagForc{deskilling}
 \tagForc{dehumanization}
 \tagForc{disconnection} 
 & \small{We work on one of society's most important problems: mental health is a national crisis, exacerbated by loneliness. There is a huge shortage of qualified mental health professionals, and \TopThree{therapy requires human-to-human connection which a computer cannot provide}, so we feel \notTopThree{our jobs are extremely secure against generative AI}. However, \textbf{we worry about telehealth services that provide low quality service but still undermine our job security}. Generative AI may be a \notTopThree{useful tool for administrative work}, simulated talk-therapy training, or as a stopgap when people can’t immediately access a human therapist.} \\
 
 \midrule
 \textbf{Education} \newline\small
 \small
 \tagNarr{human review} 
 \tagForc{dehumanization}
 \tagForc{disconnection} 
 & \small{\notTopThree{We don't expect generative AI to affect our work deeply.} Students already have many ways of cheating, and generative AI is just a newer, better way that we  deal with by \TopThree{reviewing} their work. We're worried that generative AI \TopThree{may make students lazy} if it's not used well. We're also sensitive to \TopThree{overuse of technology and social isolation} because of remote schooling during the pandemic. We’re in the middle of restructuring much of our curriculum due to standards-based grading. This gives us more latitude to design new activities, so maybe there are opportunities to work generative AI into our lessons.} \\
 
 \midrule
 \textbf{Law} \newline\small
 \tagNarr{menial tasks} 
 \tagNarr{tool} 
 \tagNarr{human review} 
 & \small{Generative AI is a \TopThree{tool that can perform formulaic legal tasks} like drafting or research, \TopThree{provided its work is reviewed by qualified attorneys}, \notTopThree{replacing entry-level and support positions} and contributing to job loss. We’ve heard \notTopThree{press stories about false citations}, which makes us wary of using it, even with human oversight. \notTopThree{Generative AI is unlikely to be skilled enough to replace human professionals}, and it should never make decisions, or act as lawyer, judge, or jury. However, we anticipate clients will begin to use generative AI tools to do some legal work for themselves, with poor results.} \\
 
 \midrule
 \textbf{Journalism} \newline\small
 \tagForc{deskilling}
 \tagForc{dehumanization}
 \tagForc{disinformation} 
 & \small{We work in a noble profession and \notTopThree{we are already embattled on many fronts}, facing misinformation, news aggregators, decreased revenue streams, and more. Generative AI will \TopThree{exacerbate ongoing precarious/freelance employment}, poor wages, job loss, and the erosion of journalism as a field. Generative AI \TopThree{makes it easier for low-quality providers to produce mis/disinformation}, with harmful effects on society. Generative AI \TopThree{may replace some of the work we most enjoy}, like writing or photojournalism.} \\

 \midrule
 \textbf{Business \newline Communications} \newline\small
 \tagNarr{tool} \tagNarr{human review} \tagNarr{need to adapt} &
 \small{Generative AI can do a lot of our work, and it will inevitably be adopted in our industry. \TopThree{Companies and employees need to adapt, learn, and ``grow with AI’’} in order to stay relevant and protect job security. \notTopThree{Much of our writing is formulaic}, and generative AI is a \TopThree{great tool to produce early drafts} of routine, high-volume, and/or disposable communications like email notifications. Generative AI \TopThree{will need a lot of human oversight and guidance}, especially to make sure it meets compliance and legal standards and is consistent with our brand voice. We can leverage our existing approval structures to check its output.} \\

 

 \midrule
 \textbf{Advertising} \newline\small
 \tagNarr{tool} \tagNarr{human review} \tagNarr{lose certain jobs} & 
 \small{Generative AI is well suited to many tasks in the advertising industry. Sadly, \TopThree{certain jobs will go away}, most immediately those related to product photography and video production, with layout and copywriting under heavy threat. However, creative oversight will remain in the human purview, and generative AI \TopThree{will be an exciting brainstorming tool}. \TopThree{A human check will be required} for most generative AI output, especially to make sure it doesn’t violate brand standards or copyright. In fact, new teams may be created to do these checks. Our industry promotes and embraces change, but we also value human craft and \notTopThree{we often challenge the quality of digitally produced content and experiences}.} \\ 
 
 \midrule
 \textbf{Software \newline Development} \newline\small
 \tagNarr{need to adapt}
 \tagForc{deskilling}
 \tagForc{dehumanization}
 & \small{Generative AI is likely to drive change across most aspects of software development. This includes, for example, \notTopThree{the automation of menial and boring tasks}, \notTopThree{the elimination of many entry-level roles} (which will \TopThree{limit paths for new people to enter the field}), and the ability of AI to reshape low-code technology solutions and even write production-ready code within serious software companies. We feel uncertainty about the future and \TopThree{pressure to adapt} to new advances in generative AI. Some of us worry that \TopThree{the time and education we invested in a lucrative career might be nullified}, setting us back to where we started. Some people get into software engineering for the money, but some of us are motivated by our love of tinkering and problem solving, and generative AI \TopThree{may take over a lot of the work that brings us joy}. } \\ 

\bottomrule
\end{tabular}
\end{center}
  \caption{Composites illustrating the most salient points of participants' expectations regarding the impact of generative AI on their industry, with industries ordered top to bottom by least to greatest expected impact. Under the name of each industry we list the top three most salient themes (\Tool/ \Narrative/--\textcolor{royalazure}{blue}; social forces--\textcolor{brickred}{red}). Neither the top three themes nor the composites comprehensively represent all points, \eg although deskilling was discussed extensively in Law, it was not one of the top three.}
  \label{table:industry_composites}
\end{table*}
\section{Discussion}
In this section, we build on our participants' insights to surface key HCI research questions at the intersection of generative AI and knowledge work. Rather than focusing on current user practice or design recommendations for early versions of these systems, we seek to frame larger questions about how generative AI's impact may be understood and shaped. Some of this work falls squarely in the purview of HCI, but much of it likely requires broader collaboration with other academic disciplines as well as stakeholders such as knowledge workers, policy makers, civil society, and more.

\subsection{Research Challenges: Human-in-the-Loop}
\label{sec:hitl}


Participants overwhelmingly favored a human-in-the-loop (HITL) approach (\ie adding human review or oversight as described in Section~\ref{sec:narrative-review}) as a necessary and sufficient remediation for many problems with generative AI. For example, they expected that in their industries, human reviewers could check all of generative AI's output and correct any inaccuracies or other quality issues. Legal scholars report that regulators have a similar inclination to address a ``panoply'' of concerns about AI with what they refer to as a ``slap a human in it'' approach~\cite{crootof2023}.

However, the HITL literature points out serious, unsolved practical problems that were not apparent to participants. For example, humans make many errors when reviewing algorithmic output and often override algorithms in detrimental ways; therefore, human oversight policies can provide a false sense of security rather than improving outcomes overall~\cite{bucinca2021, green2022}. Further, research has shown that effectively configuring human-AI coordination is extremely difficult, so handoffs are often poorly designed and yield harmful results~\cite{crootof2023, elish2019, green2019}. Additionally, scholars argue that HITL approaches disproportionately hold humans accountable, even when in practice they have very little influence or do not have appropriate skills or time to review, leaving reviewers to shoulder the blame for technical or structural failures~\cite{awad2020, crootof2023, douer2020, elish2019, green2019, wagner2019}.
This suggests two main research questions:

\rqf{\#1. How do we raise awareness of the limitations of HITL?}
\rqft{
We believe it is important to exercise caution when applying HITL solutions. Those who are making decisions about mitigations for generative AI (\eg regulators, decision makers in knowledge industries) should be aware of the current limitations of HITL so they are not overly optimistic about its potential to remediate problems.}

\rqf{\#2. How do we make better HITL systems?}
\rqft{Despite its current limitations, many opportunities exist to improve HITL's effectiveness. The advent of generative AI particularly highlights the need for HITL approaches that will work well at scale for review of generated text, images, and video. Also, cognitive forcing interventions (such as asking workers to generate certain preliminary content before being shown generative AI's output) can engage analytical rather than heuristic thinking~\cite{bucinca2021}. Further, reviewers could receive specialized training in critically reviewing generative AI's output; these review skills might ultimately be at least as valuable as other skills like prompt engineering. Overall, while improved solutions are unlikely to alleviate all concerns with HITL~\cite{green2019}, additional research and innovative design and development could lead to substantially better outcomes.}
\subsection{Research Challenges: Knowledge Worker Expectations of Impact}




As described in Section~\ref{sec:narrative-limited}, we observed a significant gap between participants' expectations of how generative AI might change their field versus broader narratives of disruption offered by the media, technologists, and academics\textemdash participants generally took a more limited view of potential impact. This suggests the following research questions:

\newpage
\rqf{\#3 Why do some knowledge workers feel they will be largely unaffected by generative AI?} 
\rqft{Are there certain workers who feel more immune to changes from generative AI? Is this due to a failure of imagination or lack of awareness, or do knowledge workers have particular understandings of their industries that are not being taken into account in other narratives? 
}

\rqf{\#4 How do we raise awareness among knowledge workers about the \Academic/ \Narrative/?} 
\rqft{Improved understanding of how and why generative AI is being positioned as capable of competently performing work at human expert levels would prepare knowledge workers to more meaningfully provide stakeholder input. Further, is it the case that knowledge workers underestimate the scope of what generative AI can do and what it may be used for, and how it may transform not just specific tasks or jobs but industries more generally? If so, what interventions and transparency artifacts might be most meaningful? For example, AI systems are largely not yet specialized; perhaps developing and sharing industry-specific demos would be helpful.}

\subsection{Research Challenges: Deskilling}

As technologies shift how work is performed, the necessary skill sets of existing and new professions also shift \cite{acemoglu2011skills}, creating a paradox of concurrent unemployment and labor shortages \cite{autor2015there}. Consistent with this, participants expressed concern that generative AI might negatively affect the value and development of their skills, as described in Section~\ref{sec:deskilling}. This suggests the following research questions:


\rqf{\#5 How can we help knowledge workers adapt to potential, yet uncertain, wide-ranging transformation?}
\rqft{What training would benefit knowledge workers, to help them reskill for likely changes? Beyond prompt engineering, training for human review of AI output seems promising, as does development of critical thinking skills and the ability to manage complex, higher-level use cases. Additionally, conventional national policy responses to labor market changes often rely on traditional, secondary education \cite{collins2018rethinking}, which is often inadequate against the pace of technological change \cite{stephany2021}. Reskilling within individual organizations is a more cost-effective and worker-centric strategy to evolve with technological changes in knowledge work \cite{hu2021}. Future work could explore workforce transition including how to re- or up-skill workers in ways that minimize precarity.

Further, researchers have discussed how human labor is used to cover current gaps in the capabilities of automated systems, and how humans’ successful performance of that often invisible work can ultimately lead to the replacement of human labor with automated systems, \eg by demonstrating that business models are viable or by training systems to perform work previously done by humans~\cite{catanzariti2021,fox2023,iantorno2022,vertesi2020}. How can we explore protections, opportunities, and reward structures for those workers, often from marginalized groups, who enable and/or are affected by generative AI automation?}


\rqf{\#6 During a potential transition to generative AI, how can we better design for more equitable distribution and valuation of knowledge work?} 
\rqft{Will changes such as the potential erosion of entry-level positions further entrench the existing racial and gendered hierarchies in the workplace~\cite{atanasoski2019}? To what extent does realizing a liberatory or emancipatory future involve counteracting the devaluing and deskilling of work, so that AI-mediated systems do not reinforce patterns of oppression that have justified historical exploitation along racial lines~\cite{benjamin2019}? How might we better approach design or pursue other avenues to advance justice in this context~\cite{irani2016}?
And in industries that rely on development of foundational skills, how might the most talented experts arise if they do not work through entry-level tasks? What systems might support human development of necessary skills, even if generative AI can perform those tasks?}
\subsection{Research Challenges: Dehumanization}

As a research community primarily studying the interface between humans and computer systems, understanding where these systems are uncomfortably and unacceptably encroaching on people's inherent sense of humanity should be a key research agenda. To explore dehumanization, we suggest two avenues of inquiry to address the serious concerns raised by our participants as described in Section~\ref{sec:dehumanization}:

\rqf{\#7 How can we learn about and protect tasks that bring people joy and meaning, but generative AI can do as well or better than humans?} 
\rqft{We should understand not just, as many economists focus on (\eg~\cite{wef2023,albanesi2023}), which tasks can be replaced effectively by AI systems, but also which tasks inherently bring joy and reinforce humanity. 
By beginning to document these tasks that are meaningful but in some cases can be done as well, or even at higher quality or more efficiently, by AI systems, we can begin to design ways to protect these tasks, or at least make reasoned decisions about when to automate them. This may involve HCI interventions to design systems that highlight or reserve certain work or decisions for humans, or broader questions of regulation to protect certain kinds of work across society. While many utopian visions involve humans who are freed from work by AI and take up new, creative, artistic endeavors, losing the craft, joy, and humanity that exists in current work and problem-solving may have significant downsides, including perpetuating and further entrenching existing inequalities~\cite{atanasoski2019}.
Rather than ``liberating'' individuals, generative AI could merely shift the burden of dehumanized labor onto different groups or create new forms of devalued labor. Moreover, in performing tasks that require creativity or intellectual rigor, it could diminish the value and uniqueness of human contributions in these areas, reinforcing a narrow and exclusionary definition of what it means to be human.}

\newpage
\rqf{\#8 How can we promote critical thinking and also prevent people from becoming lazy?}
\rqft{Separate from the joy and meaning inherent in certain tasks, our participants also found innate value in performing reasoning and critical thinking. Participants were concerned that AI could reduce or remove tasks that challenge humans or force them to solve problems, thereby leading to a lazy society. While considerable work within the HCI community has explored how human and AI systems can collaborate for higher quality and more creative outcomes~\cite{suh2021,verheijden2023,louie2020,hwang2022,chung2022}, this question encourages research into how these collaborations can be designed to support knowledge workers~\cite{gero2023}, as well as the design of systems that directly engage people in problem solving rather than simply providing answers.}
\subsection{Research Challenges: Guardrails}


Responsible AI has gained significant traction, from published principles to more actionable and/or measurable strategies. While some issues like explainability and fairness move closer to having concrete remediations in system design, other broad social issues require additional work beyond engineering and system design, including harms assessments, stakeholder engagements, and increased focus on community outcomes. Considering the social forces prioritized by participants as described in Section~\ref{sec:forces}, current responsibility metrics are most relevant to disinformation, and have less to offer regarding deskilling, dehumanization, and disconnection. This leads us to the following questions:

\rqf{\#9 How can we design responsible AI approaches to generative AI that embrace complex global dynamics?} 
\rqft{How can responsible AI more fully and holistically consider impact and harm?
Findings from our study highlight the ways in which the impacts of generative AI must be considered with respect to social forces that intersect with conditions of its development, deployment, and use. Thus, an approach to harm or impact analysis that stops at model evaluations (\eg analyzing outputs with respect to pre-determined benchmarks) will be insufficient.}

\rqf{\#10 How can responsible AI assessments more fully incorporate stakeholder input?}
\rqft{Responsible AI product assessments often center on anticipating harms from launching a product~\cite{boyarskaya2020overcoming,brey2012anticipatory} through consideration of how their contextual use may engender harms to users and communities~\cite{alkhatib2021}. Many responsible AI interventions orient towards developer-facing interventions, such as developing AI principles \cite{fjeld2020principled}, educating practitioners to foster ethical awareness~\cite{lange2023engaging}, and moderating generative AI systems through training data mitigations, in-model controls (e.g., reinforcement learning), and safety classifiers that gate outputs \cite{hao2023safety}. However, limited attention has focused on assessing communities' desired forms and uses of AI. Meaningfully engaging communities of practice~\cite{wenger2002} in exploration of these questions, such as through community-based participatory research~\cite{hayes2011, cooper2022}, can inform development of practices that scaffold community engagement into responsible AI practice. 

Similarly, in a labor context, complicated questions arise regarding how to assess the impact of AI systems that may reduce or drastically change specific jobs. While companies may often consider the needs of an overall business, how should they consider the needs of individual workers, particularly those whose jobs may be cut or changed, and are there ways to directly engage them for input before developing these AI solutions? And how can we educate or support decision-makers in engaging stakeholders?}
\section{Conclusions}

The historical context in which technology appears influences how it is ultimately adopted and used. Generative AI has become more available and visible during the confluence of a global pandemic, economic uncertainty, and more. Many knowledge workers in our study situated generative AI in this context, highlighting generative AI as exacerbating the following four social forces: deskilling, dehumanization, disconnection, and disinformation. In other words, rather than seeing generative AI as an independent disruptor of their work or their industry, they positioned generative AI as extending and exacerbating existing forces. As described in the discussion, this framing raises important new questions and opportunities for exploring and shaping the future impact of generative AI on knowledge workers and their industries.

\begin{acks}
We thank our participants for generously sharing their insights and expertise. We thank our partner team at Gemic, including T.J. Foley, Roger Galvez, David Ginsborg, Rebekah Park, and Rohini Shah, for their support in the field. We are grateful to our colleagues at Google, including Marian Croak, Jen Gennai, Angela McKay, Anoop Sinha, and Ashley Walker, for supporting this work.
\end{acks}


\bibliographystyle{ACM-Reference-Format}
\bibliography{ourbib}


\begin{thebibliography}{193}


\ifx \showCODEN    \undefined \def \showCODEN     #1{\unskip}     \fi
\ifx \showDOI      \undefined \def \showDOI       #1{#1}\fi
\ifx \showISBNx    \undefined \def \showISBNx     #1{\unskip}     \fi
\ifx \showISBNxiii \undefined \def \showISBNxiii  #1{\unskip}     \fi
\ifx \showISSN     \undefined \def \showISSN      #1{\unskip}     \fi
\ifx \showLCCN     \undefined \def \showLCCN      #1{\unskip}     \fi
\ifx \shownote     \undefined \def \shownote      #1{#1}          \fi
\ifx \showarticletitle \undefined \def \showarticletitle #1{#1}   \fi
\ifx \showURL      \undefined \def \showURL       {\relax}        \fi
\providecommand\bibfield[2]{#2}
\providecommand\bibinfo[2]{#2}
\providecommand\natexlab[1]{#1}
\providecommand\showeprint[2][]{arXiv:#2}

\bibitem[Acemoglu and Autor(2011)]%
        {acemoglu2011skills}
\bibfield{author}{\bibinfo{person}{Daron Acemoglu} {and} \bibinfo{person}{David
  Autor}.} \bibinfo{year}{2011}\natexlab{}.
\newblock \showarticletitle{Skills, tasks and technologies: Implications for
  employment and earnings}.
\newblock In \bibinfo{booktitle}{\emph{Handbook of Labor Economics}},
  \bibfield{editor}{\bibinfo{person}{David Card} {and} \bibinfo{person}{Orley
  Ashenfelter}} (Eds.). Vol.~\bibinfo{volume}{4}.
  \bibinfo{publisher}{Elsevier}, \bibinfo{pages}{1043--1171}.
\newblock
\urldef\tempurl%
\url{https://doi.org/10.1016/S0169-7218(11)02410-5}
\showDOI{\tempurl}


\bibitem[AI(2023)]%
        {bard}
\bibfield{author}{\bibinfo{person}{Google AI}.}
  \bibinfo{year}{2023}\natexlab{}.
\newblock \bibinfo{title}{Bard}.
\newblock
\newblock
\urldef\tempurl%
\url{https://ai.google/research/projects/bard}
\showURL{%
\tempurl}


\bibitem[AI(2022)]%
        {stablediffusion}
\bibfield{author}{\bibinfo{person}{Stability AI}.}
  \bibinfo{year}{2022}\natexlab{}.
\newblock \bibinfo{title}{Stable Diffusion}.
\newblock
\newblock
\urldef\tempurl%
\url{https://stablediffusionweb.com/}
\showURL{%
\tempurl}


\bibitem[Albanesi et~al\mbox{.}(2023)]%
        {albanesi2023}
\bibfield{author}{\bibinfo{person}{Stefania Albanesi},
  \bibinfo{person}{Ant{\'o}nio~Dias da Silva}, \bibinfo{person}{Juan~F.
  Jimeno}, \bibinfo{person}{Ana Lamo}, {and} \bibinfo{person}{Alena Wabitsch}.}
  \bibinfo{year}{2023}\natexlab{}.
\newblock \bibinfo{booktitle}{\emph{New Technologies and Jobs in Europe}}.
\newblock \bibinfo{type}{Working Paper} 31357. \bibinfo{institution}{National
  Bureau of Economic Research}.
\newblock
\urldef\tempurl%
\url{https://doi.org/10.3386/w31357}
\showDOI{\tempurl}


\bibitem[Ali et~al\mbox{.}(2021)]%
        {ali2021}
\bibfield{author}{\bibinfo{person}{Safinah Ali}, \bibinfo{person}{Daniella
  DiPaola}, \bibinfo{person}{Irene Lee}, \bibinfo{person}{Jenna Hong}, {and}
  \bibinfo{person}{Cynthia Breazeal}.} \bibinfo{year}{2021}\natexlab{}.
\newblock \showarticletitle{Exploring Generative Models with Middle School
  Students}. In \bibinfo{booktitle}{\emph{Proceedings of the 2021 CHI
  Conference on Human Factors in Computing Systems}} (Yokohama, Japan)
  \emph{(\bibinfo{series}{CHI '21})}. \bibinfo{publisher}{Association for
  Computing Machinery}, \bibinfo{address}{New York, NY, USA}, Article
  \bibinfo{articleno}{678}, \bibinfo{numpages}{13}~pages.
\newblock
\showISBNx{9781450380966}
\urldef\tempurl%
\url{https://doi.org/10.1145/3411764.3445226}
\showDOI{\tempurl}


\bibitem[Alkhatib(2021)]%
        {alkhatib2021}
\bibfield{author}{\bibinfo{person}{Ali Alkhatib}.}
  \bibinfo{year}{2021}\natexlab{}.
\newblock \showarticletitle{To Live in Their Utopia: Why Algorithmic Systems
  Create Absurd Outcomes}. In \bibinfo{booktitle}{\emph{Proceedings of the 2021
  CHI Conference on Human Factors in Computing Systems}} (Yokohama, Japan)
  \emph{(\bibinfo{series}{CHI '21})}. \bibinfo{publisher}{Association for
  Computing Machinery}, \bibinfo{address}{New York, NY, USA}, Article
  \bibinfo{articleno}{95}, \bibinfo{numpages}{9}~pages.
\newblock
\showISBNx{9781450380966}
\urldef\tempurl%
\url{https://doi.org/10.1145/3411764.3445740}
\showDOI{\tempurl}


\bibitem[Arakawa et~al\mbox{.}(2023)]%
        {arakawa2023}
\bibfield{author}{\bibinfo{person}{Riku Arakawa}, \bibinfo{person}{Hiromu
  Yakura}, {and} \bibinfo{person}{Masataka Goto}.}
  \bibinfo{year}{2023}\natexlab{}.
\newblock \showarticletitle{CatAlyst: Domain-Extensible Intervention for
  Preventing Task Procrastination Using Large Generative Models}. In
  \bibinfo{booktitle}{\emph{Proceedings of the 2023 CHI Conference on Human
  Factors in Computing Systems}} (Hamburg, Germany) \emph{(\bibinfo{series}{CHI
  '23})}. \bibinfo{publisher}{Association for Computing Machinery},
  \bibinfo{address}{New York, NY, USA}, Article \bibinfo{articleno}{157},
  \bibinfo{numpages}{19}~pages.
\newblock
\showISBNx{9781450394215}
\urldef\tempurl%
\url{https://doi.org/10.1145/3544548.3581133}
\showDOI{\tempurl}


\bibitem[{ARM | Northstar}(2017)]%
        {arm2017}
\bibfield{author}{\bibinfo{person}{{ARM | Northstar}}.}
  \bibinfo{year}{2017}\natexlab{}.
\newblock \bibinfo{title}{{AI} Today, {AI} Tomorrow. {A}wareness and
  Anticipation of {AI}: A Global Perspective}.
\newblock
\newblock


\bibitem[{ARM | Northstar}(2020)]%
        {arm2020}
\bibfield{author}{\bibinfo{person}{{ARM | Northstar}}.}
  \bibinfo{year}{2020}\natexlab{}.
\newblock \bibinfo{title}{{AI} Today, {AI} Tomorrow: The {A}rm 2020 Global {AI}
  Survey}.
\newblock
\newblock


\bibitem[Arnold et~al\mbox{.}(2018)]%
        {arnold2018sentiment}
\bibfield{author}{\bibinfo{person}{Kenneth~C. Arnold}, \bibinfo{person}{Krysta
  Chauncey}, {and} \bibinfo{person}{Krzysztof~Z. Gajos}.}
  \bibinfo{year}{2018}\natexlab{}.
\newblock \showarticletitle{Sentiment Bias in Predictive Text Recommendations
  Results in Biased Writing}. In \bibinfo{booktitle}{\emph{Proceedings of the
  44th Graphics Interface Conference}}. \bibinfo{publisher}{Association for
  Computing Machinery}, \bibinfo{address}{Toronto, Canada},
  \bibinfo{pages}{42--49}.
\newblock


\bibitem[Arnold et~al\mbox{.}(2020)]%
        {arnold2020}
\bibfield{author}{\bibinfo{person}{Kenneth~C. Arnold}, \bibinfo{person}{Krysta
  Chauncey}, {and} \bibinfo{person}{Krzysztof~Z. Gajos}.}
  \bibinfo{year}{2020}\natexlab{}.
\newblock \showarticletitle{Predictive Text Encourages Predictable Writing}. In
  \bibinfo{booktitle}{\emph{Proceedings of the 25th International Conference on
  Intelligent User Interfaces}} (Cagliari, Italy) \emph{(\bibinfo{series}{IUI
  '20})}. \bibinfo{publisher}{Association for Computing Machinery},
  \bibinfo{address}{New York, NY, USA}, \bibinfo{pages}{128–138}.
\newblock
\showISBNx{9781450371186}
\urldef\tempurl%
\url{https://doi.org/10.1145/3377325.3377523}
\showDOI{\tempurl}


\bibitem[Atanasoski and Vora(2019)]%
        {atanasoski2019}
\bibfield{author}{\bibinfo{person}{Neda Atanasoski} {and}
  \bibinfo{person}{Kalindi Vora}.} \bibinfo{year}{2019}\natexlab{}.
\newblock \bibinfo{booktitle}{\emph{Surrogate Humanity: Race, Robots, and the
  Politics of Technological Futures}}.
\newblock \bibinfo{publisher}{Duke University Press}, \bibinfo{address}{Durham
  and London}.
\newblock


\bibitem[Autor(2015)]%
        {autor2015there}
\bibfield{author}{\bibinfo{person}{David~H. Autor}.}
  \bibinfo{year}{2015}\natexlab{}.
\newblock \showarticletitle{Why are there still so many jobs? The history and
  future of workplace automation}.
\newblock \bibinfo{journal}{\emph{Journal of Economic Perspectives}}
  \bibinfo{volume}{29}, \bibinfo{number}{3} (\bibinfo{year}{2015}),
  \bibinfo{pages}{3--30}.
\newblock


\bibitem[Awad et~al\mbox{.}(2020)]%
        {awad2020}
\bibfield{author}{\bibinfo{person}{Edmond Awad}, \bibinfo{person}{Sydney
  Levine}, \bibinfo{person}{Max Kleiman-Weiner}, \bibinfo{person}{Sohan
  Dsouza}, \bibinfo{person}{Joshua~B. Tenenbaum}, \bibinfo{person}{Azim
  Shariff}, \bibinfo{person}{Jean-Fran{\c{c}}ois Bonnefon}, {and}
  \bibinfo{person}{Iyad Rahwan}.} \bibinfo{year}{2020}\natexlab{}.
\newblock \showarticletitle{Drivers are blamed more than their automated cars
  when both make mistakes}.
\newblock \bibinfo{journal}{\emph{Nature Human Behaviour}} \bibinfo{volume}{4},
  \bibinfo{number}{2} (\bibinfo{year}{2020}), \bibinfo{pages}{134--143}.
\newblock


\bibitem[Bao et~al\mbox{.}(2022)]%
        {bao2022}
\bibfield{author}{\bibinfo{person}{Luye Bao}, \bibinfo{person}{Nicole~M.
  Krause}, \bibinfo{person}{Mikhaila~N. Calice}, \bibinfo{person}{Dietram~A.
  Scheufele}, \bibinfo{person}{Christopher~D. Wirz}, \bibinfo{person}{Dominique
  Brossard}, \bibinfo{person}{Todd~P. Newman}, {and}
  \bibinfo{person}{Michael~A. Xenos}.} \bibinfo{year}{2022}\natexlab{}.
\newblock \showarticletitle{Whose AI? How different publics think about AI and
  its social impacts}.
\newblock \bibinfo{journal}{\emph{Computers in Human Behavior}}
  \bibinfo{volume}{130} (\bibinfo{year}{2022}), \bibinfo{pages}{107182}.
\newblock


\bibitem[Bardzell(2010)]%
        {bardzell2010}
\bibfield{author}{\bibinfo{person}{Shaowen Bardzell}.}
  \bibinfo{year}{2010}\natexlab{}.
\newblock \showarticletitle{Feminist HCI: Taking Stock and Outlining an Agenda
  for Design}. In \bibinfo{booktitle}{\emph{Proceedings of the SIGCHI
  Conference on Human Factors in Computing Systems}} (Atlanta, Georgia, USA)
  \emph{(\bibinfo{series}{CHI '10})}. \bibinfo{publisher}{Association for
  Computing Machinery}, \bibinfo{address}{New York, NY, USA},
  \bibinfo{pages}{1301–1310}.
\newblock
\showISBNx{9781605589299}
\urldef\tempurl%
\url{https://doi.org/10.1145/1753326.1753521}
\showDOI{\tempurl}


\bibitem[Barrabi(2023)]%
        {barrabi2023}
\bibfield{author}{\bibinfo{person}{Thomas Barrabi}.}
  \bibinfo{year}{2023}\natexlab{}.
\newblock \bibinfo{title}{AI will create a ‘serious number of losers’ in
  job market, DeepMind co-founder warns}.
\newblock \bibinfo{howpublished}{New York Post}.
\newblock
\urldef\tempurl%
\url{https://nypost.com/2023/05/10/ai-will-create-a-serious-number-of-losers-in-job-market-deepmind-co-founder-warns/}
\showURL{%
\tempurl}


\bibitem[Bartholomew and Mehta(2023)]%
        {bartholomew2023}
\bibfield{author}{\bibinfo{person}{Jem Bartholomew} {and}
  \bibinfo{person}{Dhrumil Mehta}.} \bibinfo{year}{2023}\natexlab{}.
\newblock \bibinfo{title}{The Business of Artificial Intelligence}.
\newblock \bibinfo{howpublished}{Columbia Journalism Review}.
\newblock
\urldef\tempurl%
\url{https://www.cjr.org/tow_center/media-coverage-chatgpt.php}
\showURL{%
\tempurl}


\bibitem[Beignon et~al\mbox{.}(2020)]%
        {beignon2020}
\bibfield{author}{\bibinfo{person}{Ana\"{e}lle Beignon},
  \bibinfo{person}{Emeline Brul\'{e}}, \bibinfo{person}{Jean-Baptiste Joatton},
  {and} \bibinfo{person}{Aur\'{e}lien Tabard}.}
  \bibinfo{year}{2020}\natexlab{}.
\newblock \showarticletitle{Tricky Design Probes: Triggering Reflection on
  Design Research Methods in Service Design}. In
  \bibinfo{booktitle}{\emph{Proceedings of the 2020 ACM Designing Interactive
  Systems Conference}} (Eindhoven, Netherlands) \emph{(\bibinfo{series}{DIS
  '20})}. \bibinfo{publisher}{Association for Computing Machinery},
  \bibinfo{address}{New York, NY, USA}, \bibinfo{pages}{1647–1660}.
\newblock
\showISBNx{9781450369749}
\urldef\tempurl%
\url{https://doi.org/10.1145/3357236.3395572}
\showDOI{\tempurl}


\bibitem[Bell(1976)]%
        {bellpostindustrial}
\bibfield{author}{\bibinfo{person}{Daniel Bell}.}
  \bibinfo{year}{1976}\natexlab{}.
\newblock \bibinfo{booktitle}{\emph{The Coming of Post-Industrial Society: A
  Venture in Social Forecasting}}.
\newblock \bibinfo{publisher}{Basic Books}, \bibinfo{address}{New York, NY}.
\newblock


\bibitem[Benjamin(2019)]%
        {benjamin2019}
\bibfield{author}{\bibinfo{person}{Ruha Benjamin}.}
  \bibinfo{year}{2019}\natexlab{}.
\newblock \bibinfo{booktitle}{\emph{Race After Technology: Abolitionist Tools
  for the New Jim Code}}.
\newblock \bibinfo{publisher}{Polity}, \bibinfo{address}{Cambridge, UK and
  Medford, MA, USA}.
\newblock


\bibitem[Bhat et~al\mbox{.}(2023)]%
        {bhat2023}
\bibfield{author}{\bibinfo{person}{Advait Bhat}, \bibinfo{person}{Saaket
  Agashe}, \bibinfo{person}{Parth Oberoi}, \bibinfo{person}{Niharika Mohile},
  \bibinfo{person}{Ravi Jangir}, {and} \bibinfo{person}{Anirudha Joshi}.}
  \bibinfo{year}{2023}\natexlab{}.
\newblock \showarticletitle{Interacting with Next-Phrase Suggestions: How
  Suggestion Systems Aid and Influence the Cognitive Processes of Writing}. In
  \bibinfo{booktitle}{\emph{Proceedings of the 28th International Conference on
  Intelligent User Interfaces}} (Sydney, NSW, Australia)
  \emph{(\bibinfo{series}{IUI '23})}. \bibinfo{publisher}{Association for
  Computing Machinery}, \bibinfo{address}{New York, NY, USA},
  \bibinfo{pages}{436–452}.
\newblock
\showISBNx{9798400701061}
\urldef\tempurl%
\url{https://doi.org/10.1145/3581641.3584060}
\showDOI{\tempurl}


\bibitem[Birks et~al\mbox{.}(2008)]%
        {birks2008}
\bibfield{author}{\bibinfo{person}{Melanie Birks}, \bibinfo{person}{Ysanne
  Chapman}, {and} \bibinfo{person}{Karen Francis}.}
  \bibinfo{year}{2008}\natexlab{}.
\newblock \showarticletitle{Memoing in qualitative research: Probing data and
  processes}.
\newblock \bibinfo{journal}{\emph{Journal of Research in Nursing}}
  \bibinfo{volume}{13}, \bibinfo{number}{1} (\bibinfo{year}{2008}),
  \bibinfo{pages}{68--75}.
\newblock
\urldef\tempurl%
\url{https://doi.org/10.1177/1744987107081254}
\showDOI{\tempurl}


\bibitem[{Blumberg Capital}(2019)]%
        {blumberg2019}
\bibfield{author}{\bibinfo{person}{{Blumberg Capital}}.}
  \bibinfo{year}{2019}\natexlab{}.
\newblock \bibinfo{title}{Artificial Intelligence in 2019: Getting Past the
  Adoption Tipping Point}.
\newblock
\newblock


\bibitem[Boehner et~al\mbox{.}(2007)]%
        {boehner2007}
\bibfield{author}{\bibinfo{person}{Kirsten Boehner}, \bibinfo{person}{Janet
  Vertesi}, \bibinfo{person}{Phoebe Sengers}, {and} \bibinfo{person}{Paul
  Dourish}.} \bibinfo{year}{2007}\natexlab{}.
\newblock \showarticletitle{How HCI Interprets the Probes}. In
  \bibinfo{booktitle}{\emph{Proceedings of the SIGCHI Conference on Human
  Factors in Computing Systems}} (San Jose, California, USA)
  \emph{(\bibinfo{series}{CHI '07})}. \bibinfo{publisher}{Association for
  Computing Machinery}, \bibinfo{address}{New York, NY, USA},
  \bibinfo{pages}{1077–1086}.
\newblock
\showISBNx{9781595935939}
\urldef\tempurl%
\url{https://doi.org/10.1145/1240624.1240789}
\showDOI{\tempurl}


\bibitem[Boyarskaya et~al\mbox{.}(2020)]%
        {boyarskaya2020overcoming}
\bibfield{author}{\bibinfo{person}{Margarita Boyarskaya},
  \bibinfo{person}{Alexandra Olteanu}, {and} \bibinfo{person}{Kate Crawford}.}
  \bibinfo{year}{2020}\natexlab{}.
\newblock \bibinfo{title}{Overcoming Failures of Imagination in AI Infused
  System Development and Deployment}.
\newblock
\newblock
\showeprint[arxiv]{2011.13416}~[cs.CY]


\bibitem[Braun and Clarke(2019)]%
        {braun2019}
\bibfield{author}{\bibinfo{person}{Virginia Braun} {and}
  \bibinfo{person}{Victoria Clarke}.} \bibinfo{year}{2019}\natexlab{}.
\newblock \showarticletitle{Reflecting on reflexive thematic analysis}.
\newblock \bibinfo{journal}{\emph{Qualitative Research in Sport, Exercise and
  Health}} \bibinfo{volume}{11}, \bibinfo{number}{4} (\bibinfo{year}{2019}),
  \bibinfo{pages}{589--597}.
\newblock
\urldef\tempurl%
\url{https://doi.org/10.1080/2159676X.2019.1628806}
\showDOI{\tempurl}


\bibitem[Braun and Clarke(2021)]%
        {braun2020}
\bibfield{author}{\bibinfo{person}{Virginia Braun} {and}
  \bibinfo{person}{Victoria Clarke}.} \bibinfo{year}{2021}\natexlab{}.
\newblock \showarticletitle{One size fits all? What counts as quality practice
  in (reflexive) thematic analysis?}
\newblock \bibinfo{journal}{\emph{Qualitative Research in Psychology}}
  \bibinfo{volume}{18}, \bibinfo{number}{3} (\bibinfo{year}{2021}),
  \bibinfo{pages}{328--352}.
\newblock
\urldef\tempurl%
\url{https://doi.org/10.1080/14780887.2020.1769238}
\showDOI{\tempurl}


\bibitem[Braun and Clarke(2022)]%
        {braun2022}
\bibfield{author}{\bibinfo{person}{Virginia Braun} {and}
  \bibinfo{person}{Victoria Clarke}.} \bibinfo{year}{2022}\natexlab{}.
\newblock \bibinfo{booktitle}{\emph{Thematic Analysis: A Practical Guide}}.
\newblock \bibinfo{publisher}{SAGE}, \bibinfo{address}{Thousand Oaks, CA}.
\newblock


\bibitem[Bray et~al\mbox{.}(2022)]%
        {bray2022}
\bibfield{author}{\bibinfo{person}{Kirsten~E. Bray}, \bibinfo{person}{Christina
  Harrington}, \bibinfo{person}{Andrea~G. Parker}, \bibinfo{person}{N'Deye
  Diakhate}, {and} \bibinfo{person}{Jennifer Roberts}.}
  \bibinfo{year}{2022}\natexlab{}.
\newblock \showarticletitle{Radical Futures: Supporting Community-Led Design
  Engagements through an Afrofuturist Speculative Design Toolkit}. In
  \bibinfo{booktitle}{\emph{Proceedings of the 2022 CHI Conference on Human
  Factors in Computing Systems}} (New Orleans, LA, USA)
  \emph{(\bibinfo{series}{CHI '22})}. \bibinfo{publisher}{Association for
  Computing Machinery}, \bibinfo{address}{New York, NY, USA}, Article
  \bibinfo{articleno}{452}, \bibinfo{numpages}{13}~pages.
\newblock
\showISBNx{9781450391573}
\urldef\tempurl%
\url{https://doi.org/10.1145/3491102.3501945}
\showDOI{\tempurl}


\bibitem[Brey(2012)]%
        {brey2012anticipatory}
\bibfield{author}{\bibinfo{person}{Philip~A.E. Brey}.}
  \bibinfo{year}{2012}\natexlab{}.
\newblock \showarticletitle{Anticipatory ethics for emerging technologies}.
\newblock \bibinfo{journal}{\emph{NanoEthics}} \bibinfo{volume}{6},
  \bibinfo{number}{1} (\bibinfo{year}{2012}), \bibinfo{pages}{1--13}.
\newblock


\bibitem[Briggs and Kodnani(2023)]%
        {briggs2023}
\bibfield{author}{\bibinfo{person}{Joseph Briggs} {and} \bibinfo{person}{Devesh
  Kodnani}.} \bibinfo{year}{2023}\natexlab{}.
\newblock \bibinfo{booktitle}{\emph{The Potentially Large Effects of Artificial
  Intelligence on Economic Growth}}.
\newblock \bibinfo{type}{Global Economics Analyst}.
  \bibinfo{institution}{Goldman Sachs}.
\newblock


\bibitem[Brown et~al\mbox{.}(2020)]%
        {brown2020language}
\bibfield{author}{\bibinfo{person}{Tom Brown}, \bibinfo{person}{Benjamin Mann},
  \bibinfo{person}{Nick Ryder}, \bibinfo{person}{Melanie Subbiah},
  \bibinfo{person}{Jared~D. Kaplan}, \bibinfo{person}{Prafulla Dhariwal},
  \bibinfo{person}{Arvind Neelakantan}, \bibinfo{person}{Pranav Shyam},
  \bibinfo{person}{Girish Sastry}, \bibinfo{person}{Amanda Askell},
  \bibinfo{person}{Sandhini Agarwal}, \bibinfo{person}{Ariel Herbert-Voss},
  \bibinfo{person}{Gretchen Krueger}, \bibinfo{person}{Tom Henighan},
  \bibinfo{person}{Rewon Child}, \bibinfo{person}{Aditya Ramesh},
  \bibinfo{person}{Daniel Ziegler}, \bibinfo{person}{Jeffrey Wu},
  \bibinfo{person}{Clemens Winter}, \bibinfo{person}{Chris Hesse},
  \bibinfo{person}{Mark Chen}, \bibinfo{person}{Eric Sigler},
  \bibinfo{person}{Mateusz Litwin}, \bibinfo{person}{Scott Gray},
  \bibinfo{person}{Benjamin Chess}, \bibinfo{person}{Jack Clark},
  \bibinfo{person}{Christopher Berner}, \bibinfo{person}{Sam McCandlish},
  \bibinfo{person}{Alec Radford}, \bibinfo{person}{Ilya Sutskever}, {and}
  \bibinfo{person}{Dario Amodei}.} \bibinfo{year}{2020}\natexlab{}.
\newblock \showarticletitle{Language Models are Few-Shot Learners}. In
  \bibinfo{booktitle}{\emph{Advances in Neural Information Processing
  Systems}}, \bibfield{editor}{\bibinfo{person}{H.~Larochelle},
  \bibinfo{person}{M.~Ranzato}, \bibinfo{person}{R.~Hadsell},
  \bibinfo{person}{M.F. Balcan}, {and} \bibinfo{person}{H.~Lin}} (Eds.),
  Vol.~\bibinfo{volume}{33}. \bibinfo{publisher}{Curran Associates, Inc.},
  \bibinfo{address}{Vancouver, Canada}, \bibinfo{pages}{1877--1901}.
\newblock
\urldef\tempurl%
\url{https://proceedings.neurips.cc/paper_files/paper/2020/file/1457c0d6bfcb4967418bfb8ac142f64a-Paper.pdf}
\showURL{%
\tempurl}


\bibitem[Brynjolfsson et~al\mbox{.}(2023)]%
        {brynjolfsson2023}
\bibfield{author}{\bibinfo{person}{Erik Brynjolfsson},
  \bibinfo{person}{Danielle Li}, {and} \bibinfo{person}{Lindsey~R. Raymond}.}
  \bibinfo{year}{2023}\natexlab{}.
\newblock \bibinfo{booktitle}{\emph{Generative AI at work}}.
\newblock \bibinfo{type}{Working Paper} 31161. \bibinfo{institution}{National
  Bureau of Economic Research}.
\newblock
\urldef\tempurl%
\url{https://doi.org/10.3386/w31161}
\showDOI{\tempurl}


\bibitem[Bu\c{c}inca et~al\mbox{.}(2021)]%
        {bucinca2021}
\bibfield{author}{\bibinfo{person}{Zana Bu\c{c}inca},
  \bibinfo{person}{Maja~Barbara Malaya}, {and} \bibinfo{person}{Krzysztof~Z.
  Gajos}.} \bibinfo{year}{2021}\natexlab{}.
\newblock \showarticletitle{To Trust or to Think: Cognitive Forcing Functions
  Can Reduce Overreliance on AI in AI-Assisted Decision-Making}.
\newblock \bibinfo{journal}{\emph{Proceedings of the ACM on Human-Computer
  Interaction}} \bibinfo{volume}{5}, \bibinfo{number}{CSCW1}, Article
  \bibinfo{articleno}{188} (\bibinfo{date}{April} \bibinfo{year}{2021}),
  \bibinfo{numpages}{21}~pages.
\newblock
\urldef\tempurl%
\url{https://doi.org/10.1145/3449287}
\showDOI{\tempurl}


\bibitem[Castells(2011)]%
        {castells2011rise}
\bibfield{author}{\bibinfo{person}{Manuel Castells}.}
  \bibinfo{year}{2011}\natexlab{}.
\newblock \bibinfo{booktitle}{\emph{The Rise of the Network Society}}.
\newblock \bibinfo{publisher}{John Wiley \& Sons}, \bibinfo{address}{Hoboken,
  NJ}.
\newblock


\bibitem[Catanzariti et~al\mbox{.}(2021)]%
        {catanzariti2021}
\bibfield{author}{\bibinfo{person}{Benedetta Catanzariti},
  \bibinfo{person}{Srravya Chandhiramowuli}, \bibinfo{person}{Suha Mohamed},
  \bibinfo{person}{Sarayu Natarajan}, \bibinfo{person}{Shantanu Prabhat},
  \bibinfo{person}{Noopur Raval}, \bibinfo{person}{Alex~S. Taylor}, {and}
  \bibinfo{person}{Ding Wang}.} \bibinfo{year}{2021}\natexlab{}.
\newblock \showarticletitle{The Global Labours of AI and Data Intensive
  Systems}. In \bibinfo{booktitle}{\emph{Companion Publication of the 2021
  Conference on Computer Supported Cooperative Work and Social Computing}}
  (Virtual Event, USA) \emph{(\bibinfo{series}{CSCW '21 Companion})}.
  \bibinfo{publisher}{Association for Computing Machinery},
  \bibinfo{address}{New York, NY, USA}, \bibinfo{pages}{319–322}.
\newblock
\showISBNx{9781450384797}
\urldef\tempurl%
\url{https://doi.org/10.1145/3462204.3481725}
\showDOI{\tempurl}


\bibitem[Cave et~al\mbox{.}(2019)]%
        {cave2019}
\bibfield{author}{\bibinfo{person}{Stephen Cave}, \bibinfo{person}{Kate
  Coughlan}, {and} \bibinfo{person}{Kanta Dihal}.}
  \bibinfo{year}{2019}\natexlab{}.
\newblock \showarticletitle{``{S}cary {R}obots'': Examining Public Responses to
  {AI}}. In \bibinfo{booktitle}{\emph{Proceedings of the 2019 AAAI/ACM
  Conference on AI, Ethics, and Society}} (Honolulu, HI, USA)
  \emph{(\bibinfo{series}{AIES '19})}. \bibinfo{publisher}{Association for
  Computing Machinery}, \bibinfo{address}{New York, NY, USA},
  \bibinfo{pages}{331–337}.
\newblock
\showISBNx{9781450363242}
\urldef\tempurl%
\url{https://doi.org/10.1145/3306618.3314232}
\showDOI{\tempurl}


\bibitem[Chomsky(2023)]%
        {chomsky2023}
\bibfield{author}{\bibinfo{person}{Noam Chomsky}.}
  \bibinfo{year}{2023}\natexlab{}.
\newblock \showarticletitle{Opinion Guest Essay: The False Promise of ChatGPT}.
\newblock \bibinfo{journal}{\emph{The New York Times}} (\bibinfo{date}{8 March}
  \bibinfo{year}{2023}).
\newblock
\urldef\tempurl%
\url{https://www.nytimes.com/2023/03/08/opinion/noam-chomsky-chatgpt-ai.html}
\showURL{%
\tempurl}


\bibitem[Chui et~al\mbox{.}(2023)]%
        {chui2023}
\bibfield{author}{\bibinfo{person}{Michael Chui}, \bibinfo{person}{Eric Hazan},
  \bibinfo{person}{Roger Roberts}, \bibinfo{person}{Alex Singla},
  \bibinfo{person}{Kate Smaje}, \bibinfo{person}{Alex Sukharevsky},
  \bibinfo{person}{Lareina Yee}, {and} \bibinfo{person}{Rodney Zemmel}.}
  \bibinfo{year}{2023}\natexlab{}.
\newblock \showarticletitle{The economic potential of generative AI: The next
  productivity frontier}.
\newblock \bibinfo{journal}{\emph{McKinsey \& Company}} (\bibinfo{date}{14
  June} \bibinfo{year}{2023}).
\newblock
\urldef\tempurl%
\url{https://www.mckinsey.com/capabilities/mckinsey-digital/our-insights/the-economic-potential-of-generative-ai-the-next-productivity-frontier}
\showURL{%
\tempurl}


\bibitem[Chung et~al\mbox{.}(2022)]%
        {chung2022}
\bibfield{author}{\bibinfo{person}{John Joon~Young Chung},
  \bibinfo{person}{Wooseok Kim}, \bibinfo{person}{Kang~Min Yoo},
  \bibinfo{person}{Hwaran Lee}, \bibinfo{person}{Eytan Adar}, {and}
  \bibinfo{person}{Minsuk Chang}.} \bibinfo{year}{2022}\natexlab{}.
\newblock \showarticletitle{TaleBrush: Sketching Stories with Generative
  Pretrained Language Models}. In \bibinfo{booktitle}{\emph{Proceedings of the
  2022 CHI Conference on Human Factors in Computing Systems}} (New Orleans, LA,
  USA) \emph{(\bibinfo{series}{CHI '22})}. \bibinfo{publisher}{Association for
  Computing Machinery}, \bibinfo{address}{New York, NY, USA}, Article
  \bibinfo{articleno}{209}, \bibinfo{numpages}{19}~pages.
\newblock
\showISBNx{9781450391573}
\urldef\tempurl%
\url{https://doi.org/10.1145/3491102.3501819}
\showDOI{\tempurl}


\bibitem[Clark et~al\mbox{.}(2018)]%
        {clark2018}
\bibfield{author}{\bibinfo{person}{Elizabeth Clark},
  \bibinfo{person}{Anne~Spencer Ross}, \bibinfo{person}{Chenhao Tan},
  \bibinfo{person}{Yangfeng Ji}, {and} \bibinfo{person}{Noah~A. Smith}.}
  \bibinfo{year}{2018}\natexlab{}.
\newblock \showarticletitle{Creative Writing with a Machine in the Loop: Case
  Studies on Slogans and Stories}. In \bibinfo{booktitle}{\emph{23rd
  International Conference on Intelligent User Interfaces}} (Tokyo, Japan)
  \emph{(\bibinfo{series}{IUI '18})}. \bibinfo{publisher}{Association for
  Computing Machinery}, \bibinfo{address}{New York, NY, USA},
  \bibinfo{pages}{329–340}.
\newblock
\showISBNx{9781450349451}
\urldef\tempurl%
\url{https://doi.org/10.1145/3172944.3172983}
\showDOI{\tempurl}


\bibitem[Collins and Halverson(2018)]%
        {collins2018rethinking}
\bibfield{author}{\bibinfo{person}{Allan Collins} {and}
  \bibinfo{person}{Richard Halverson}.} \bibinfo{year}{2018}\natexlab{}.
\newblock \bibinfo{booktitle}{\emph{Rethinking education in the age of
  technology: The digital revolution and schooling in America}}.
\newblock \bibinfo{publisher}{Teachers College Press},
  \bibinfo{address}{London}.
\newblock


\bibitem[Cooper et~al\mbox{.}(2022)]%
        {cooper2022}
\bibfield{author}{\bibinfo{person}{Ned Cooper}, \bibinfo{person}{Tiffanie
  Horne}, \bibinfo{person}{Gillian~R. Hayes}, \bibinfo{person}{Courtney
  Heldreth}, \bibinfo{person}{Michal Lahav}, \bibinfo{person}{Jess Holbrook},
  {and} \bibinfo{person}{Lauren Wilcox}.} \bibinfo{year}{2022}\natexlab{}.
\newblock \showarticletitle{A Systematic Review and Thematic Analysis of
  Community-Collaborative Approaches to Computing Research}. In
  \bibinfo{booktitle}{\emph{Proceedings of the 2022 CHI Conference on Human
  Factors in Computing Systems}} (New Orleans, LA, USA)
  \emph{(\bibinfo{series}{CHI '22})}. \bibinfo{publisher}{Association for
  Computing Machinery}, \bibinfo{address}{New York, NY, USA}, Article
  \bibinfo{articleno}{73}, \bibinfo{numpages}{18}~pages.
\newblock
\showISBNx{9781450391573}
\urldef\tempurl%
\url{https://doi.org/10.1145/3491102.3517716}
\showDOI{\tempurl}


\bibitem[Corfield and Field(2022)]%
        {corfield2022}
\bibfield{author}{\bibinfo{person}{Gareth Corfield} {and}
  \bibinfo{person}{Matthew Field}.} \bibinfo{year}{2022}\natexlab{}.
\newblock \showarticletitle{Meet ChatGPT, the scarily intelligent robot who can
  do your job better than you}.
\newblock \bibinfo{journal}{\emph{The Telegraph}} (\bibinfo{date}{6 Dec.}
  \bibinfo{year}{2022}).
\newblock


\bibitem[Cox(1987)]%
        {cox1987production}
\bibfield{author}{\bibinfo{person}{Robert~W. Cox}.}
  \bibinfo{year}{1987}\natexlab{}.
\newblock \bibinfo{booktitle}{\emph{Production, power, and world order: Social
  forces in the making of history}}. Vol.~\bibinfo{volume}{1}.
\newblock \bibinfo{publisher}{Columbia University Press}, \bibinfo{address}{New
  York}.
\newblock


\bibitem[Creswell and Poth(2018)]%
        {creswell2018}
\bibfield{author}{\bibinfo{person}{John~W. Creswell} {and}
  \bibinfo{person}{Cheryl~N. Poth}.} \bibinfo{year}{2018}\natexlab{}.
\newblock \bibinfo{booktitle}{\emph{Qualitative Inquiry and Research Design:
  Choosing among Five Approaches} (\bibinfo{edition}{fourth} ed.)}.
\newblock \bibinfo{publisher}{Sage Publications}, \bibinfo{address}{Thousand
  Oaks, CA}.
\newblock


\bibitem[Crootof et~al\mbox{.}(2023)]%
        {crootof2023}
\bibfield{author}{\bibinfo{person}{Rebecca Crootof}, \bibinfo{person}{Margot~E.
  Kaminski}, {and} \bibinfo{person}{W.~Nicholson Price~II}.}
  \bibinfo{year}{2023}\natexlab{}.
\newblock \showarticletitle{Humans in the Loop}.
\newblock \bibinfo{journal}{\emph{Vanderbilt Law Review}}  \bibinfo{volume}{76}
  (\bibinfo{year}{2023}), \bibinfo{pages}{429--510}.
\newblock


\bibitem[Dang et~al\mbox{.}(2023)]%
        {dang2023}
\bibfield{author}{\bibinfo{person}{Hai Dang}, \bibinfo{person}{Sven Goller},
  \bibinfo{person}{Florian Lehmann}, {and} \bibinfo{person}{Daniel Buschek}.}
  \bibinfo{year}{2023}\natexlab{}.
\newblock \showarticletitle{Choice Over Control: How Users Write with Large
  Language Models Using Diegetic and Non-Diegetic Prompting}. In
  \bibinfo{booktitle}{\emph{Proceedings of the 2023 CHI Conference on Human
  Factors in Computing Systems}} (Hamburg, Germany) \emph{(\bibinfo{series}{CHI
  '23})}. \bibinfo{publisher}{Association for Computing Machinery},
  \bibinfo{address}{New York, NY, USA}, Article \bibinfo{articleno}{408},
  \bibinfo{numpages}{17}~pages.
\newblock
\showISBNx{9781450394215}
\urldef\tempurl%
\url{https://doi.org/10.1145/3544548.3580969}
\showDOI{\tempurl}


\bibitem[Davenport and Mittal(2022)]%
        {davenport2022}
\bibfield{author}{\bibinfo{person}{Thomas~H. Davenport} {and}
  \bibinfo{person}{Nitin Mittal}.} \bibinfo{year}{2022}\natexlab{}.
\newblock \showarticletitle{How Generative AI is Changing Creative Work}.
\newblock \bibinfo{journal}{\emph{Harvard Business Review}} (\bibinfo{date}{14
  November} \bibinfo{year}{2022}).
\newblock


\bibitem[Davis et~al\mbox{.}(2023)]%
        {davis2023}
\bibfield{author}{\bibinfo{person}{Richard~Lee Davis}, \bibinfo{person}{Thiemo
  Wambsganss}, \bibinfo{person}{Wei Jiang}, \bibinfo{person}{Kevin~Gonyop Kim},
  \bibinfo{person}{Tanja K\"{a}ser}, {and} \bibinfo{person}{Pierre
  Dillenbourg}.} \bibinfo{year}{2023}\natexlab{}.
\newblock \showarticletitle{Fashioning the Future: Unlocking the Creative
  Potential of Deep Generative Models for Design Space Exploration}. In
  \bibinfo{booktitle}{\emph{Extended Abstracts of the 2023 CHI Conference on
  Human Factors in Computing Systems}} (Hamburg, Germany)
  \emph{(\bibinfo{series}{CHI EA '23})}. \bibinfo{publisher}{Association for
  Computing Machinery}, \bibinfo{address}{New York, NY, USA}, Article
  \bibinfo{articleno}{136}, \bibinfo{numpages}{9}~pages.
\newblock
\showISBNx{9781450394222}
\urldef\tempurl%
\url{https://doi.org/10.1145/3544549.3585644}
\showDOI{\tempurl}


\bibitem[DeLisi and Howley(2023)]%
        {gartner2023}
\bibfield{author}{\bibinfo{person}{Meghan~Rimol DeLisi} {and}
  \bibinfo{person}{Catherine Howley}.} \bibinfo{year}{2023}\natexlab{}.
\newblock \bibinfo{title}{Gartner Places Generative AI on the Peak of Inflated
  Expectations on the 2023 Hype Cycle for Emerging Technologies}.
\newblock \bibinfo{howpublished}{Gartner}.
\newblock
\urldef\tempurl%
\url{https://www.gartner.com/en/newsroom/press-releases/2023-08-16-gartner-places-generative-ai-on-the-peak-of-inflated-expectations-on-the-2023-hype-cycle-for-emerging-technologies}
\showURL{%
\tempurl}


\bibitem[Dell'Acqua et~al\mbox{.}(2023)]%
        {dell2023}
\bibfield{author}{\bibinfo{person}{Fabrizio Dell'Acqua},
  \bibinfo{person}{Edward McFowland}, \bibinfo{person}{Ethan~R. Mollick},
  \bibinfo{person}{Hila Lifshitz-Assaf}, \bibinfo{person}{Katherine Kellogg},
  \bibinfo{person}{Saran Rajendran}, \bibinfo{person}{Lisa Krayer},
  \bibinfo{person}{Fran{\c{c}}ois Candelon}, {and} \bibinfo{person}{Karim~R.
  Lakhani}.} \bibinfo{year}{2023}\natexlab{}.
\newblock \showarticletitle{{Navigating the Jagged Technological Frontier:
  Field Experimental Evidence of the Effects of AI on Knowledge Worker
  Productivity and Quality}}.
\newblock \bibinfo{journal}{\emph{Harvard Business School Technology \&
  Operations Management Unit Working Paper}} \bibinfo{number}{24-013}
  (\bibinfo{year}{2023}).
\newblock
\urldef\tempurl%
\url{https://doi.org/10.2139/ssrn.4573321}
\showDOI{\tempurl}


\bibitem[Dell'Era and Landoni(2014)]%
        {dell2014}
\bibfield{author}{\bibinfo{person}{Claudio Dell'Era} {and}
  \bibinfo{person}{Paolo Landoni}.} \bibinfo{year}{2014}\natexlab{}.
\newblock \showarticletitle{Living Lab: A methodology between user-centred
  design and participatory design}.
\newblock \bibinfo{journal}{\emph{Creativity and Innovation Management}}
  \bibinfo{volume}{23}, \bibinfo{number}{2} (\bibinfo{year}{2014}),
  \bibinfo{pages}{137--154}.
\newblock
\urldef\tempurl%
\url{https://doi.org/10.1111/caim.12061}
\showDOI{\tempurl}


\bibitem[Dosovitskiy et~al\mbox{.}(2020)]%
        {dosovitskiy2020image}
\bibfield{author}{\bibinfo{person}{Alexey Dosovitskiy}, \bibinfo{person}{Lucas
  Beyer}, \bibinfo{person}{Alexander Kolesnikov}, \bibinfo{person}{Dirk
  Weissenborn}, \bibinfo{person}{Xiaohua Zhai}, \bibinfo{person}{Thomas
  Unterthiner}, \bibinfo{person}{Mostafa Dehghani}, \bibinfo{person}{Matthias
  Minderer}, \bibinfo{person}{Georg Heigold}, \bibinfo{person}{Sylvain Gelly},
  \bibinfo{person}{Jakob Uszkoreit}, {and} \bibinfo{person}{Neil Houlsby}.}
  \bibinfo{year}{2020}\natexlab{}.
\newblock \bibinfo{title}{An Image is Worth 16x16 Words: Transformers for Image
  Recognition at Scale}.
\newblock
\newblock
\showeprint[arxiv]{2010.11929}~[cs.CV]


\bibitem[Douer and Meyer(2020)]%
        {douer2020}
\bibfield{author}{\bibinfo{person}{Nir Douer} {and} \bibinfo{person}{Joachim
  Meyer}.} \bibinfo{year}{2020}\natexlab{}.
\newblock \showarticletitle{The Responsibility Quantification Model of Human
  Interaction With Automation}.
\newblock \bibinfo{journal}{\emph{IEEE Transactions on Automation Science and
  Engineering}} \bibinfo{volume}{17}, \bibinfo{number}{2}
  (\bibinfo{year}{2020}), \bibinfo{pages}{1044--1060}.
\newblock
\urldef\tempurl%
\url{https://doi.org/10.1109/TASE.2020.2965466}
\showDOI{\tempurl}


\bibitem[Dove(1998)]%
        {dove1998}
\bibfield{author}{\bibinfo{person}{Rick Dove}.}
  \bibinfo{year}{1998}\natexlab{}.
\newblock \showarticletitle{The Knowledge Worker}.
\newblock \bibinfo{journal}{\emph{Automotive Manufacturing \& Promotion}}
  \bibinfo{volume}{110}, \bibinfo{number}{6} (\bibinfo{year}{1998}),
  \bibinfo{pages}{26--28}.
\newblock


\bibitem[Drucker(1959)]%
        {druckerlandmarks}
\bibfield{author}{\bibinfo{person}{Peter Drucker}.}
  \bibinfo{year}{1959}\natexlab{}.
\newblock \bibinfo{booktitle}{\emph{The Landmarks of Tomorrow}}.
\newblock \bibinfo{publisher}{Heineman}, \bibinfo{address}{New York, NY}.
\newblock


\bibitem[Drucker(1999)]%
        {druckerknowledge}
\bibfield{author}{\bibinfo{person}{Peter~F. Drucker}.}
  \bibinfo{year}{1999}\natexlab{}.
\newblock \showarticletitle{Knowledge-Worker Productivity: The Biggest
  Challenge}.
\newblock \bibinfo{journal}{\emph{California Management Review}}
  \bibinfo{volume}{41}, \bibinfo{number}{2} (\bibinfo{year}{1999}),
  \bibinfo{pages}{79--94}.
\newblock
\urldef\tempurl%
\url{https://doi.org/10.2307/41165987}
\showDOI{\tempurl}


\bibitem[Edelman(2019)]%
        {edelman2019}
\bibfield{author}{\bibinfo{person}{Edelman}.} \bibinfo{year}{2019}\natexlab{}.
\newblock \bibinfo{title}{2019 {E}delman {AI} Survey}.
\newblock
\newblock


\bibitem[Elish(2019)]%
        {elish2019}
\bibfield{author}{\bibinfo{person}{Madeleine~Clare Elish}.}
  \bibinfo{year}{2019}\natexlab{}.
\newblock \showarticletitle{Moral crumple zones: Cautionary tales in
  human-robot interaction}.
\newblock \bibinfo{journal}{\emph{Engaging Science, Technology, and Society}}
  \bibinfo{volume}{5} (\bibinfo{year}{2019}), \bibinfo{pages}{40--60}.
\newblock


\bibitem[Eloundou et~al\mbox{.}(2023)]%
        {eloundou2023}
\bibfield{author}{\bibinfo{person}{Tyna Eloundou}, \bibinfo{person}{Sam
  Manning}, \bibinfo{person}{Pamela Mishkin}, {and} \bibinfo{person}{Daniel
  Rock}.} \bibinfo{year}{2023}\natexlab{}.
\newblock \bibinfo{title}{GPTs are GPTs: An Early Look at the Labor Market
  Impact Potential of Large Language Models}.
\newblock
\newblock
\showeprint[arxiv]{2303.10130}~[econ.GN]


\bibitem[Erickson and Jarrahi(2016)]%
        {erickson2016}
\bibfield{author}{\bibinfo{person}{Ingrid Erickson} {and}
  \bibinfo{person}{Mohammad~Hossein Jarrahi}.} \bibinfo{year}{2016}\natexlab{}.
\newblock \showarticletitle{Infrastructuring and the Challenge of Dynamic Seams
  in Mobile Knowledge Work}. In \bibinfo{booktitle}{\emph{Proceedings of the
  19th ACM Conference on Computer-Supported Cooperative Work \& Social
  Computing}} (San Francisco, California, USA) \emph{(\bibinfo{series}{CSCW
  '16})}. \bibinfo{publisher}{Association for Computing Machinery},
  \bibinfo{address}{New York, NY, USA}, \bibinfo{pages}{1323–1336}.
\newblock
\showISBNx{9781450335928}
\urldef\tempurl%
\url{https://doi.org/10.1145/2818048.2820015}
\showDOI{\tempurl}


\bibitem[Felten et~al\mbox{.}(2023)]%
        {felten2023}
\bibfield{author}{\bibinfo{person}{Ed Felten}, \bibinfo{person}{Manav Raj},
  {and} \bibinfo{person}{Robert Seamans}.} \bibinfo{year}{2023}\natexlab{}.
\newblock \bibinfo{title}{How will Language Modelers like ChatGPT Affect
  Occupations and Industries?}
\newblock
\newblock
\showeprint[arxiv]{2303.01157}~[econ.GN]


\bibitem[Fjeld et~al\mbox{.}(2020)]%
        {fjeld2020principled}
\bibfield{author}{\bibinfo{person}{Jessica Fjeld}, \bibinfo{person}{Nele
  Achten}, \bibinfo{person}{Hannah Hilligoss}, \bibinfo{person}{Adam Nagy},
  {and} \bibinfo{person}{Madhulika Srikumar}.} \bibinfo{year}{2020}\natexlab{}.
\newblock \showarticletitle{Principled artificial intelligence: Mapping
  consensus in ethical and rights-based approaches to principles for AI}.
\newblock \bibinfo{journal}{\emph{Berkman Klein Center Research Publication}}
  (\bibinfo{year}{2020}).
\newblock


\bibitem[Fox et~al\mbox{.}(2015)]%
        {fox2015}
\bibfield{author}{\bibinfo{person}{Sarah Fox}, \bibinfo{person}{Rachel~Rose
  Ulgado}, {and} \bibinfo{person}{Daniela Rosner}.}
  \bibinfo{year}{2015}\natexlab{}.
\newblock \showarticletitle{Hacking Culture, Not Devices: Access and
  Recognition in Feminist Hackerspaces}. In
  \bibinfo{booktitle}{\emph{Proceedings of the 18th ACM Conference on Computer
  Supported Cooperative Work \& Social Computing}} (Vancouver, BC, Canada)
  \emph{(\bibinfo{series}{CSCW '15})}. \bibinfo{publisher}{Association for
  Computing Machinery}, \bibinfo{address}{New York, NY, USA},
  \bibinfo{pages}{56–68}.
\newblock
\showISBNx{9781450329224}
\urldef\tempurl%
\url{https://doi.org/10.1145/2675133.2675223}
\showDOI{\tempurl}


\bibitem[Fox et~al\mbox{.}(2023)]%
        {fox2023}
\bibfield{author}{\bibinfo{person}{Sarah~E. Fox}, \bibinfo{person}{Samantha
  Shorey}, \bibinfo{person}{Esther~Y. Kang}, \bibinfo{person}{Dominique
  Montiel~Valle}, {and} \bibinfo{person}{Estefania Rodriguez}.}
  \bibinfo{year}{2023}\natexlab{}.
\newblock \showarticletitle{Patchwork: The Hidden, Human Labor of AI
  Integration within Essential Work}.
\newblock \bibinfo{journal}{\emph{Proceedings of the ACM on Human-Computer
  Interaction}} \bibinfo{volume}{7}, \bibinfo{number}{CSCW1}, Article
  \bibinfo{articleno}{81} (\bibinfo{date}{April} \bibinfo{year}{2023}),
  \bibinfo{numpages}{20}~pages.
\newblock
\urldef\tempurl%
\url{https://doi.org/10.1145/3579514}
\showDOI{\tempurl}


\bibitem[Fritts and Cabrera(2021)]%
        {fritts2021ai}
\bibfield{author}{\bibinfo{person}{Megan Fritts} {and} \bibinfo{person}{Frank
  Cabrera}.} \bibinfo{year}{2021}\natexlab{}.
\newblock \showarticletitle{AI recruitment algorithms and the dehumanization
  problem}.
\newblock \bibinfo{journal}{\emph{Ethics and Information Technology}}
  \bibinfo{volume}{23} (\bibinfo{year}{2021}), \bibinfo{pages}{791--801}.
\newblock


\bibitem[Funk et~al\mbox{.}(2020)]%
        {funk2020}
\bibfield{author}{\bibinfo{person}{Cary Funk}, \bibinfo{person}{Alec Tyson},
  \bibinfo{person}{Brian Kennedy}, {and} \bibinfo{person}{Courtney Johnson}.}
  \bibinfo{year}{2020}\natexlab{}.
\newblock \showarticletitle{Science and Scientists Held in High Esteem Across
  Global Publics}.
\newblock \bibinfo{journal}{\emph{Pew Research Center}}
  (\bibinfo{date}{September} \bibinfo{year}{2020}).
\newblock


\bibitem[Galbraith(2007)]%
        {galbraith2007new}
\bibfield{author}{\bibinfo{person}{John~Kenneth Galbraith}.}
  \bibinfo{year}{2007}\natexlab{}.
\newblock \bibinfo{booktitle}{\emph{The New Industrial State}}.
  Vol.~\bibinfo{volume}{9}.
\newblock \bibinfo{publisher}{Princeton University Press},
  \bibinfo{address}{Princeton, NJ}.
\newblock


\bibitem[Gaver et~al\mbox{.}(1999)]%
        {gaver1999}
\bibfield{author}{\bibinfo{person}{Bill Gaver}, \bibinfo{person}{Tony Dunne},
  {and} \bibinfo{person}{Elena Pacenti}.} \bibinfo{year}{1999}\natexlab{}.
\newblock \showarticletitle{Design: Cultural Probes}.
\newblock \bibinfo{journal}{\emph{Interactions}} \bibinfo{volume}{6},
  \bibinfo{number}{1} (\bibinfo{date}{Jan.} \bibinfo{year}{1999}),
  \bibinfo{pages}{21–29}.
\newblock
\showISSN{1072-5520}
\urldef\tempurl%
\url{https://doi.org/10.1145/291224.291235}
\showDOI{\tempurl}


\bibitem[George et~al\mbox{.}(2023)]%
        {shaji2023}
\bibfield{author}{\bibinfo{person}{A.~Shaji George},
  \bibinfo{person}{A.S.~Hovan George}, {and} \bibinfo{person}{A.S.~Gabrio
  Martin}.} \bibinfo{year}{2023}\natexlab{}.
\newblock \showarticletitle{A Review of ChatGPT AI’s Impact on Several
  Business Sectors}.
\newblock \bibinfo{journal}{\emph{Partners Universal International Innovation
  Journal}} \bibinfo{volume}{1}, \bibinfo{number}{1} (\bibinfo{date}{February}
  \bibinfo{year}{2023}), \bibinfo{pages}{9–23}.
\newblock
\urldef\tempurl%
\url{https://doi.org/10.5281/zenodo.7644359}
\showDOI{\tempurl}


\bibitem[Gero et~al\mbox{.}(2023)]%
        {gero2023}
\bibfield{author}{\bibinfo{person}{Katy~Ilonka Gero}, \bibinfo{person}{Tao
  Long}, {and} \bibinfo{person}{Lydia~B. Chilton}.}
  \bibinfo{year}{2023}\natexlab{}.
\newblock \showarticletitle{Social Dynamics of AI Support in Creative Writing}.
  In \bibinfo{booktitle}{\emph{Proceedings of the 2023 CHI Conference on Human
  Factors in Computing Systems}} (Hamburg, Germany) \emph{(\bibinfo{series}{CHI
  '23})}. \bibinfo{publisher}{Association for Computing Machinery},
  \bibinfo{address}{New York, NY, USA}, Article \bibinfo{articleno}{245},
  \bibinfo{numpages}{15}~pages.
\newblock
\showISBNx{9781450394215}
\urldef\tempurl%
\url{https://doi.org/10.1145/3544548.3580782}
\showDOI{\tempurl}


\bibitem[Gl\"{o}ss et~al\mbox{.}(2016)]%
        {gloss2016}
\bibfield{author}{\bibinfo{person}{Mareike Gl\"{o}ss}, \bibinfo{person}{Moira
  McGregor}, {and} \bibinfo{person}{Barry Brown}.}
  \bibinfo{year}{2016}\natexlab{}.
\newblock \showarticletitle{Designing for Labour: Uber and the On-Demand Mobile
  Workforce}. In \bibinfo{booktitle}{\emph{Proceedings of the 2016 CHI
  Conference on Human Factors in Computing Systems}} (San Jose, California,
  USA) \emph{(\bibinfo{series}{CHI '16})}. \bibinfo{publisher}{Association for
  Computing Machinery}, \bibinfo{address}{New York, NY, USA},
  \bibinfo{pages}{1632–1643}.
\newblock
\showISBNx{9781450333627}
\urldef\tempurl%
\url{https://doi.org/10.1145/2858036.2858476}
\showDOI{\tempurl}


\bibitem[Goodfellow et~al\mbox{.}(2014)]%
        {goodfellow2014generative}
\bibfield{author}{\bibinfo{person}{Ian Goodfellow}, \bibinfo{person}{Jean
  Pouget-Abadie}, \bibinfo{person}{Mehdi Mirza}, \bibinfo{person}{Bing Xu},
  \bibinfo{person}{David Warde-Farley}, \bibinfo{person}{Sherjil Ozair},
  \bibinfo{person}{Aaron Courville}, {and} \bibinfo{person}{Yoshua Bengio}.}
  \bibinfo{year}{2014}\natexlab{}.
\newblock \showarticletitle{Generative Adversarial Nets}. In
  \bibinfo{booktitle}{\emph{Advances in Neural Information Processing
  Systems}}, \bibfield{editor}{\bibinfo{person}{Z.~Ghahramani},
  \bibinfo{person}{M.~Welling}, \bibinfo{person}{C.~Cortes},
  \bibinfo{person}{N.~Lawrence}, {and} \bibinfo{person}{K.Q. Weinberger}}
  (Eds.), Vol.~\bibinfo{volume}{27}. \bibinfo{publisher}{Curran Associates,
  Inc.}, \bibinfo{address}{Montreal, Canada}.
\newblock
\urldef\tempurl%
\url{https://proceedings.neurips.cc/paper_files/paper/2014/file/5ca3e9b122f61f8f06494c97b1afccf3-Paper.pdf}
\showURL{%
\tempurl}


\bibitem[Graham and Rouncefield(2008)]%
        {graham2008}
\bibfield{author}{\bibinfo{person}{Connor Graham} {and} \bibinfo{person}{Mark
  Rouncefield}.} \bibinfo{year}{2008}\natexlab{}.
\newblock \showarticletitle{Probes and Participation}. In
  \bibinfo{booktitle}{\emph{Proceedings of the Tenth Anniversary Conference on
  Participatory Design 2008}} (Bloomington, Indiana)
  \emph{(\bibinfo{series}{PDC '08})}. \bibinfo{publisher}{Indiana University},
  \bibinfo{address}{USA}, \bibinfo{pages}{194–197}.
\newblock
\showISBNx{9780981856100}


\bibitem[Green(2022)]%
        {green2022}
\bibfield{author}{\bibinfo{person}{Ben Green}.}
  \bibinfo{year}{2022}\natexlab{}.
\newblock \showarticletitle{The flaws of policies requiring human oversight of
  government algorithms}.
\newblock \bibinfo{journal}{\emph{Computer Law \& Security Review}}
  \bibinfo{volume}{45} (\bibinfo{year}{2022}), \bibinfo{pages}{105681}.
\newblock
\urldef\tempurl%
\url{https://doi.org/10.1016/j.clsr.2022.105681}
\showDOI{\tempurl}


\bibitem[Green and Chen(2019)]%
        {green2019}
\bibfield{author}{\bibinfo{person}{Ben Green} {and} \bibinfo{person}{Yiling
  Chen}.} \bibinfo{year}{2019}\natexlab{}.
\newblock \showarticletitle{The Principles and Limits of Algorithm-in-the-Loop
  Decision Making}.
\newblock \bibinfo{journal}{\emph{Proceedings of the ACM on Human-Computer
  Interaction}} \bibinfo{volume}{3}, \bibinfo{number}{CSCW}, Article
  \bibinfo{articleno}{50} (\bibinfo{year}{2019}), \bibinfo{numpages}{24}~pages.
\newblock
\urldef\tempurl%
\url{https://doi.org/10.1145/3359152}
\showDOI{\tempurl}


\bibitem[Guzdial et~al\mbox{.}(2019)]%
        {guzdial2019}
\bibfield{author}{\bibinfo{person}{Matthew Guzdial}, \bibinfo{person}{Nicholas
  Liao}, \bibinfo{person}{Jonathan Chen}, \bibinfo{person}{Shao-Yu Chen},
  \bibinfo{person}{Shukan Shah}, \bibinfo{person}{Vishwa Shah},
  \bibinfo{person}{Joshua Reno}, \bibinfo{person}{Gillian Smith}, {and}
  \bibinfo{person}{Mark~O. Riedl}.} \bibinfo{year}{2019}\natexlab{}.
\newblock \showarticletitle{Friend, Collaborator, Student, Manager: How Design
  of an AI-Driven Game Level Editor Affects Creators}. In
  \bibinfo{booktitle}{\emph{Proceedings of the 2019 CHI Conference on Human
  Factors in Computing Systems}} (Glasgow, Scotland UK)
  \emph{(\bibinfo{series}{CHI '19})}. \bibinfo{publisher}{Association for
  Computing Machinery}, \bibinfo{address}{New York, NY, USA},
  \bibinfo{pages}{1–13}.
\newblock
\showISBNx{9781450359702}
\urldef\tempurl%
\url{https://doi.org/10.1145/3290605.3300854}
\showDOI{\tempurl}


\bibitem[Hao et~al\mbox{.}(2023)]%
        {hao2023safety}
\bibfield{author}{\bibinfo{person}{Susan Hao}, \bibinfo{person}{Piyush Kumar},
  \bibinfo{person}{Sarah Laszlo}, \bibinfo{person}{Shivani Poddar},
  \bibinfo{person}{Bhaktipriya Radharapu}, {and} \bibinfo{person}{Renee
  Shelby}.} \bibinfo{year}{2023}\natexlab{}.
\newblock \bibinfo{title}{Safety and Fairness for Content Moderation in
  Generative Models}.
\newblock
\newblock
\showeprint[arxiv]{2306.06135}~[cs.LG]


\bibitem[Harrington et~al\mbox{.}(2019)]%
        {harrington2019}
\bibfield{author}{\bibinfo{person}{Christina~N. Harrington},
  \bibinfo{person}{Katya Borgos-Rodriguez}, {and} \bibinfo{person}{Anne~Marie
  Piper}.} \bibinfo{year}{2019}\natexlab{}.
\newblock \showarticletitle{Engaging Low-Income African American Older Adults
  in Health Discussions through Community-Based Design Workshops}. In
  \bibinfo{booktitle}{\emph{Proceedings of the 2019 CHI Conference on Human
  Factors in Computing Systems}} (Glasgow, Scotland UK)
  \emph{(\bibinfo{series}{CHI '19})}. \bibinfo{publisher}{Association for
  Computing Machinery}, \bibinfo{address}{New York, NY, USA},
  \bibinfo{pages}{1–15}.
\newblock
\urldef\tempurl%
\url{https://doi.org/10.1145/3290605.3300823}
\showDOI{\tempurl}


\bibitem[Hayes(2011)]%
        {hayes2011}
\bibfield{author}{\bibinfo{person}{Gillian~R. Hayes}.}
  \bibinfo{year}{2011}\natexlab{}.
\newblock \showarticletitle{The Relationship of Action Research to
  Human-Computer Interaction}.
\newblock \bibinfo{journal}{\emph{ACM Transactions on Computer-Human
  Interaction}} \bibinfo{volume}{18}, \bibinfo{number}{3}, Article
  \bibinfo{articleno}{15} (\bibinfo{date}{Aug.} \bibinfo{year}{2011}),
  \bibinfo{numpages}{20}~pages.
\newblock
\showISSN{1073-0516}
\urldef\tempurl%
\url{https://doi.org/10.1145/1993060.1993065}
\showDOI{\tempurl}


\bibitem[Heerwagen et~al\mbox{.}(2004)]%
        {heerwagencollaborative}
\bibfield{author}{\bibinfo{person}{Judith~H. Heerwagen}, \bibinfo{person}{Kevin
  Kampschroer}, \bibinfo{person}{Kevin~M. Powell}, {and}
  \bibinfo{person}{Vivian Loftness}.} \bibinfo{year}{2004}\natexlab{}.
\newblock \showarticletitle{Collaborative knowledge work environments}.
\newblock \bibinfo{journal}{\emph{Building Research \& Information}}
  \bibinfo{volume}{32}, \bibinfo{number}{6} (\bibinfo{year}{2004}),
  \bibinfo{pages}{510--528}.
\newblock
\urldef\tempurl%
\url{https://doi.org/10.1080/09613210412331313025}
\showDOI{\tempurl}


\bibitem[Heritage(2022)]%
        {heritage2022}
\bibfield{author}{\bibinfo{person}{Stewart Heritage}.}
  \bibinfo{year}{2022}\natexlab{}.
\newblock \showarticletitle{Could ChatGPT write my book --- and feed my kids?}
\newblock \bibinfo{journal}{\emph{The Sunday Times}} (\bibinfo{date}{11 Dec.}
  \bibinfo{year}{2022}).
\newblock
\urldef\tempurl%
\url{https://www.thetimes.co.uk/article/could-chatgpt-write-my-book-and-feed-my-kids-7972vx0xp}
\showURL{%
\tempurl}


\bibitem[HLEG(2018)]%
        {HLEG2018}
\bibfield{author}{\bibinfo{person}{HLEG}.} \bibinfo{year}{2018}\natexlab{}.
\newblock \bibinfo{title}{A multi-dimensional approach to disinformation:
  Report of the independent High level Group on fake news and online
  disinformation}.
\newblock
\newblock
\urldef\tempurl%
\url{https://data.europa.eu/doi/10.2759/739290}
\showURL{%
\tempurl}


\bibitem[Hu et~al\mbox{.}(2021)]%
        {hu2021}
\bibfield{author}{\bibinfo{person}{Han Hu}, \bibinfo{person}{Quentin Jadoul},
  {and} \bibinfo{person}{Angelika Reich}.} \bibinfo{year}{2021}\natexlab{}.
\newblock \showarticletitle{How banks can build their future
  workforce---today}.
\newblock \bibinfo{journal}{\emph{McKinsey \& Company}} (\bibinfo{date}{17
  Aug.} \bibinfo{year}{2021}).
\newblock
\urldef\tempurl%
\url{https://www.mckinsey.com/industries/financial-services/our-insights/how-banks-can-build-their-future-workforce-today}
\showURL{%
\tempurl}


\bibitem[Hwang(2022)]%
        {hwang2022}
\bibfield{author}{\bibinfo{person}{Angel Hsing-Chi Hwang}.}
  \bibinfo{year}{2022}\natexlab{}.
\newblock \showarticletitle{Too Late to Be Creative? AI-Empowered Tools in
  Creative Processes}. In \bibinfo{booktitle}{\emph{Extended Abstracts of the
  2022 CHI Conference on Human Factors in Computing Systems}} (New Orleans, LA,
  USA) \emph{(\bibinfo{series}{CHI EA '22})}. \bibinfo{publisher}{Association
  for Computing Machinery}, \bibinfo{address}{New York, NY, USA}, Article
  \bibinfo{articleno}{38}, \bibinfo{numpages}{9}~pages.
\newblock
\showISBNx{9781450391566}
\urldef\tempurl%
\url{https://doi.org/10.1145/3491101.3503549}
\showDOI{\tempurl}


\bibitem[Iantorno et~al\mbox{.}(2022)]%
        {iantorno2022}
\bibfield{author}{\bibinfo{person}{Mathew Iantorno}, \bibinfo{person}{Olivia
  Doggett}, \bibinfo{person}{Priyank Chandra}, \bibinfo{person}{Julie
  Yujie~Chen}, \bibinfo{person}{Rosemary Steup}, \bibinfo{person}{Noopur
  Raval}, \bibinfo{person}{Vera Khovanskaya}, \bibinfo{person}{Laura Lam},
  \bibinfo{person}{Anubha Singh}, \bibinfo{person}{Sarah Rotz}, {and}
  \bibinfo{person}{Matt Ratto}.} \bibinfo{year}{2022}\natexlab{}.
\newblock \showarticletitle{Outsourcing Artificial Intelligence: Responding to
  the Reassertion of the Human Element into Automation}. In
  \bibinfo{booktitle}{\emph{Extended Abstracts of the 2022 CHI Conference on
  Human Factors in Computing Systems}} (New Orleans, LA, USA)
  \emph{(\bibinfo{series}{CHI EA '22})}. \bibinfo{publisher}{Association for
  Computing Machinery}, \bibinfo{address}{New York, NY, USA}, Article
  \bibinfo{articleno}{103}, \bibinfo{numpages}{5}~pages.
\newblock
\showISBNx{9781450391566}
\urldef\tempurl%
\url{https://doi.org/10.1145/3491101.3503720}
\showDOI{\tempurl}


\bibitem[Inie et~al\mbox{.}(2023)]%
        {inie2023}
\bibfield{author}{\bibinfo{person}{Nanna Inie}, \bibinfo{person}{Jeanette
  Falk}, {and} \bibinfo{person}{Steve Tanimoto}.}
  \bibinfo{year}{2023}\natexlab{}.
\newblock \showarticletitle{Designing Participatory AI: Creative
  Professionals’ Worries and Expectations about Generative AI}. In
  \bibinfo{booktitle}{\emph{Extended Abstracts of the 2023 CHI Conference on
  Human Factors in Computing Systems}} (Hamburg, Germany)
  \emph{(\bibinfo{series}{CHI EA '23})}. \bibinfo{publisher}{Association for
  Computing Machinery}, \bibinfo{address}{New York, NY, USA}, Article
  \bibinfo{articleno}{82}, \bibinfo{numpages}{8}~pages.
\newblock
\showISBNx{9781450394222}
\urldef\tempurl%
\url{https://doi.org/10.1145/3544549.3585657}
\showDOI{\tempurl}


\bibitem[Irani and Silberman(2016)]%
        {irani2016}
\bibfield{author}{\bibinfo{person}{Lilly~C. Irani} {and}
  \bibinfo{person}{M.~Six Silberman}.} \bibinfo{year}{2016}\natexlab{}.
\newblock \showarticletitle{Stories We Tell About Labor: Turkopticon and the
  Trouble with ``Design''}. In \bibinfo{booktitle}{\emph{Proceedings of the
  2016 CHI Conference on Human Factors in Computing Systems}} (San Jose,
  California, USA) \emph{(\bibinfo{series}{CHI '16})}.
  \bibinfo{publisher}{Association for Computing Machinery},
  \bibinfo{address}{New York, NY, USA}, \bibinfo{pages}{4573–4586}.
\newblock
\showISBNx{9781450333627}
\urldef\tempurl%
\url{https://doi.org/10.1145/2858036.2858592}
\showDOI{\tempurl}


\bibitem[Jakesch et~al\mbox{.}(2023)]%
        {jakesch2023}
\bibfield{author}{\bibinfo{person}{Maurice Jakesch}, \bibinfo{person}{Advait
  Bhat}, \bibinfo{person}{Daniel Buschek}, \bibinfo{person}{Lior Zalmanson},
  {and} \bibinfo{person}{Mor Naaman}.} \bibinfo{year}{2023}\natexlab{}.
\newblock \showarticletitle{Co-Writing with Opinionated Language Models Affects
  Users’ Views}. In \bibinfo{booktitle}{\emph{Proceedings of the 2023 CHI
  Conference on Human Factors in Computing Systems}} (Hamburg, Germany)
  \emph{(\bibinfo{series}{CHI '23})}. \bibinfo{publisher}{Association for
  Computing Machinery}, \bibinfo{address}{New York, NY, USA}, Article
  \bibinfo{articleno}{111}, \bibinfo{numpages}{15}~pages.
\newblock
\showISBNx{9781450394215}
\urldef\tempurl%
\url{https://doi.org/10.1145/3544548.3581196}
\showDOI{\tempurl}


\bibitem[Janz et~al\mbox{.}(1997)]%
        {janzknowledge}
\bibfield{author}{\bibinfo{person}{Brian~D. Janz}, \bibinfo{person}{Jason~A.
  Colquitt}, {and} \bibinfo{person}{Raymond~A. Noe}.}
  \bibinfo{year}{1997}\natexlab{}.
\newblock \showarticletitle{Knowledge Worker Team Effectiveness: The Role of
  Autonomy, Interdependence, Team Development, and Contextual Support
  Variables}.
\newblock \bibinfo{journal}{\emph{Personnel Psychology}} \bibinfo{volume}{50},
  \bibinfo{number}{4} (\bibinfo{year}{1997}), \bibinfo{pages}{877--904}.
\newblock
\urldef\tempurl%
\url{https://doi.org/10.1111/j.1744-6570.1997.tb01486.x}
\showDOI{\tempurl}


\bibitem[Jaupi(2023)]%
        {jaupi2023}
\bibfield{author}{\bibinfo{person}{Jona Jaupi}.}
  \bibinfo{year}{2023}\natexlab{}.
\newblock \showarticletitle{YOU'VE BEEN REPLACED: I’m going to lose my job to
  an AI – ChatGPT does an hour of my work in seconds}.
\newblock \bibinfo{journal}{\emph{The U.S. Sun}} (\bibinfo{date}{24 Jan.}
  \bibinfo{year}{2023}).
\newblock
\urldef\tempurl%
\url{https://www.the-sun.com/tech/7211994/lose-job-ai-chatgpt-work-hour-seconds/}
\showURL{%
\tempurl}


\bibitem[Jesuthasan(2023)]%
        {jesuthasan2023}
\bibfield{author}{\bibinfo{person}{Ravin Jesuthasan}.}
  \bibinfo{year}{2023}\natexlab{}.
\newblock \showarticletitle{Cutting Through The Hype Cycle Of Generative AI}.
\newblock \bibinfo{journal}{\emph{Forbes}} (\bibinfo{date}{19 Aug.}
  \bibinfo{year}{2023}).
\newblock
\urldef\tempurl%
\url{https://www.forbes.com/sites/ravinjesuthasan/2023/08/19/cutting-through-the-hype-cycle-of-generative-ai}
\showURL{%
\tempurl}


\bibitem[Jones et~al\mbox{.}(2023)]%
        {jones2023}
\bibfield{author}{\bibinfo{person}{Bronwyn Jones}, \bibinfo{person}{Ewa Luger},
  {and} \bibinfo{person}{Rhianne Jones}.} \bibinfo{year}{2023}\natexlab{}.
\newblock \bibinfo{booktitle}{\emph{Generative AI \& journalism: A rapid
  risk-based review}}.
\newblock \bibinfo{type}{Edinburgh Research Explorer}.
  \bibinfo{institution}{University of Edinburgh}.
\newblock


\bibitem[Joshi(2023)]%
        {joshi2023}
\bibfield{author}{\bibinfo{person}{Bhautik Joshi}.}
  \bibinfo{year}{2023}\natexlab{}.
\newblock \showarticletitle{Is AI Going to Replace Creative Professionals?}
\newblock \bibinfo{journal}{\emph{Interactions}} \bibinfo{volume}{30},
  \bibinfo{number}{5} (\bibinfo{date}{Aug.} \bibinfo{year}{2023}),
  \bibinfo{pages}{24–29}.
\newblock
\showISSN{1072-5520}
\urldef\tempurl%
\url{https://doi.org/10.1145/3610529}
\showDOI{\tempurl}


\bibitem[Kapania et~al\mbox{.}(2022)]%
        {kapania2022}
\bibfield{author}{\bibinfo{person}{Shivani Kapania}, \bibinfo{person}{Oliver
  Siy}, \bibinfo{person}{Gabe Clapper}, \bibinfo{person}{Azhagu~Meena SP},
  {and} \bibinfo{person}{Nithya Sambasivan}.} \bibinfo{year}{2022}\natexlab{}.
\newblock \showarticletitle{{``Because AI is 100\% Right and Safe'': User
  Attitudes and Sources of AI Authority in India}}. In
  \bibinfo{booktitle}{\emph{Proceedings of the 2022 {CHI} {Conference} on
  {Human} {Factors} in {Computing} {Systems}}} (New Orleans, LA, USA)
  \emph{(\bibinfo{series}{{CHI} '22})}. \bibinfo{publisher}{ACM},
  \bibinfo{address}{New York, NY, USA}, Article \bibinfo{articleno}{158},
  \bibinfo{numpages}{18}~pages.
\newblock
\urldef\tempurl%
\url{https://doi.org/10.1145/3491102.3517533}
\showDOI{\tempurl}


\bibitem[Katz(2020)]%
        {katz2020}
\bibfield{author}{\bibinfo{person}{Yarden Katz}.}
  \bibinfo{year}{2020}\natexlab{}.
\newblock \bibinfo{booktitle}{\emph{Artificial Whiteness: Politics and Ideology
  in Artificial Intelligence}}.
\newblock \bibinfo{publisher}{Columbia University Press}, \bibinfo{address}{New
  York, NY}.
\newblock


\bibitem[Kazemitabaar et~al\mbox{.}(2023)]%
        {kazemitabaar2023}
\bibfield{author}{\bibinfo{person}{Majeed Kazemitabaar},
  \bibinfo{person}{Justin Chow}, \bibinfo{person}{Carl Ka~To Ma},
  \bibinfo{person}{Barbara~J. Ericson}, \bibinfo{person}{David Weintrop}, {and}
  \bibinfo{person}{Tovi Grossman}.} \bibinfo{year}{2023}\natexlab{}.
\newblock \showarticletitle{Studying the Effect of AI Code Generators on
  Supporting Novice Learners in Introductory Programming}. In
  \bibinfo{booktitle}{\emph{Proceedings of the 2023 CHI Conference on Human
  Factors in Computing Systems}} (Hamburg, Germany) \emph{(\bibinfo{series}{CHI
  '23})}. \bibinfo{publisher}{Association for Computing Machinery},
  \bibinfo{address}{New York, NY, USA}, Article \bibinfo{articleno}{455},
  \bibinfo{numpages}{23}~pages.
\newblock
\showISBNx{9781450394215}
\urldef\tempurl%
\url{https://doi.org/10.1145/3544548.3580919}
\showDOI{\tempurl}


\bibitem[Kelley et~al\mbox{.}(2023)]%
        {kelley2023}
\bibfield{author}{\bibinfo{person}{Patrick~Gage Kelley},
  \bibinfo{person}{Celestina Cornejo}, \bibinfo{person}{Lisa Hayes},
  \bibinfo{person}{Ellie~Shuo Jin}, \bibinfo{person}{Aaron Sedley},
  \bibinfo{person}{Kurt Thomas}, \bibinfo{person}{Yongwei Yang}, {and}
  \bibinfo{person}{Allison Woodruff}.} \bibinfo{year}{2023}\natexlab{}.
\newblock \showarticletitle{"There will be less privacy, of course": How and
  why people in 10 countries expect {AI} will affect privacy in the future}. In
  \bibinfo{booktitle}{\emph{Nineteenth Symposium on Usable Privacy and Security
  (SOUPS 2023)}}. \bibinfo{publisher}{USENIX Association},
  \bibinfo{address}{Anaheim, CA}, \bibinfo{pages}{579--603}.
\newblock
\showISBNx{978-1-939133-36-6}
\urldef\tempurl%
\url{https://www.usenix.org/conference/soups2023/presentation/kelley}
\showURL{%
\tempurl}


\bibitem[Kelley et~al\mbox{.}(2021b)]%
        {kelley2021}
\bibfield{author}{\bibinfo{person}{Patrick~Gage Kelley},
  \bibinfo{person}{Yongwei Yang}, \bibinfo{person}{Courtney Heldreth},
  \bibinfo{person}{Christopher Moessner}, \bibinfo{person}{Aaron Sedley},
  \bibinfo{person}{Andreas Kramm}, \bibinfo{person}{David~T. Newman}, {and}
  \bibinfo{person}{Allison Woodruff}.} \bibinfo{year}{2021}\natexlab{b}.
\newblock \showarticletitle{Exciting, Useful, Worrying, Futuristic: Public
  Perception of Artificial Intelligence in 8 Countries}. In
  \bibinfo{booktitle}{\emph{Proceedings of the 2021 AAAI/ACM Conference on AI,
  Ethics, and Society}} (Virtual Event, USA) \emph{(\bibinfo{series}{AIES
  '21})}. \bibinfo{publisher}{Association for Computing Machinery},
  \bibinfo{address}{New York, NY, USA}, \bibinfo{pages}{627–637}.
\newblock
\showISBNx{9781450384735}
\urldef\tempurl%
\url{https://doi.org/10.1145/3461702.3462605}
\showDOI{\tempurl}


\bibitem[Kelley et~al\mbox{.}(2021a)]%
        {kelley2021b}
\bibfield{author}{\bibinfo{person}{Patrick~Gage Kelley},
  \bibinfo{person}{Yongwei Yang}, \bibinfo{person}{Courtney Heldreth},
  \bibinfo{person}{Christopher Moessner}, \bibinfo{person}{Aaron~M. Sedley},
  {and} \bibinfo{person}{Allison Woodruff}.} \bibinfo{year}{2021}\natexlab{a}.
\newblock \showarticletitle{``{M}ixture of amazement at the potential of this
  technology and concern about possible pitfalls'': Public sentiment towards
  {AI} in 15 countries}.
\newblock \bibinfo{journal}{\emph{Bulletin of the IEEE Computer Society
  Technical Committee on Data Engineering}} \bibinfo{volume}{44},
  \bibinfo{number}{4} (\bibinfo{year}{2021}), \bibinfo{pages}{28--46}.
\newblock


\bibitem[Kingma and Welling(2013)]%
        {kingma2013auto}
\bibfield{author}{\bibinfo{person}{Diederik~P. Kingma} {and}
  \bibinfo{person}{Max Welling}.} \bibinfo{year}{2013}\natexlab{}.
\newblock \bibinfo{title}{Auto-encoding Variational Bayes}.
\newblock
\newblock
\showeprint[arxiv]{1312.6114}~[stat.ML]


\bibitem[Knill and Young(1997)]%
        {knill1997hidden}
\bibfield{author}{\bibinfo{person}{K. Knill} {and} \bibinfo{person}{S. Young}.}
  \bibinfo{year}{1997}\natexlab{}.
\newblock \bibinfo{booktitle}{\emph{Hidden Markov Models in Speech and Language
  Processing}}.
\newblock \bibinfo{publisher}{Springer Netherlands},
  \bibinfo{address}{Dordrecht}, \bibinfo{pages}{27--68}.
\newblock
\showISBNx{978-94-017-1183-8}
\urldef\tempurl%
\url{https://doi.org/10.1007/978-94-017-1183-8_2}
\showDOI{\tempurl}


\bibitem[Krizhevsky et~al\mbox{.}(2012)]%
        {krizhevsky2012imagenet}
\bibfield{author}{\bibinfo{person}{Alex Krizhevsky}, \bibinfo{person}{Ilya
  Sutskever}, {and} \bibinfo{person}{Geoffrey~E Hinton}.}
  \bibinfo{year}{2012}\natexlab{}.
\newblock \showarticletitle{ImageNet Classification with Deep Convolutional
  Neural Networks}. In \bibinfo{booktitle}{\emph{Advances in Neural Information
  Processing Systems}}, \bibfield{editor}{\bibinfo{person}{F.~Pereira},
  \bibinfo{person}{C.J. Burges}, \bibinfo{person}{L.~Bottou}, {and}
  \bibinfo{person}{K.Q. Weinberger}} (Eds.), Vol.~\bibinfo{volume}{25}.
  \bibinfo{publisher}{Curran Associates, Inc.}, \bibinfo{address}{Lake Tahoe,
  NV, USA}.
\newblock
\urldef\tempurl%
\url{https://proceedings.neurips.cc/paper_files/paper/2012/file/c399862d3b9d6b76c8436e924a68c45b-Paper.pdf}
\showURL{%
\tempurl}


\bibitem[Kumar and Karusala(2019)]%
        {kumar2019}
\bibfield{author}{\bibinfo{person}{Neha Kumar} {and} \bibinfo{person}{Naveena
  Karusala}.} \bibinfo{year}{2019}\natexlab{}.
\newblock \showarticletitle{Intersectional Computing}.
\newblock \bibinfo{journal}{\emph{Interactions}} \bibinfo{volume}{26},
  \bibinfo{number}{2} (\bibinfo{date}{Feb.} \bibinfo{year}{2019}),
  \bibinfo{pages}{50–54}.
\newblock
\showISSN{1072-5520}
\urldef\tempurl%
\url{https://doi.org/10.1145/3305360}
\showDOI{\tempurl}


\bibitem[Lagios et~al\mbox{.}(2021)]%
        {lagios2021explaining}
\bibfield{author}{\bibinfo{person}{Constantin Lagios}, \bibinfo{person}{Gaetane
  Caesens}, \bibinfo{person}{Nathan Nguyen}, {and} \bibinfo{person}{Florence
  Stinglhamber}.} \bibinfo{year}{2021}\natexlab{}.
\newblock \showarticletitle{Explaining the negative consequences of
  organizational dehumanization}.
\newblock \bibinfo{journal}{\emph{Journal of Personnel Psychology}}
  \bibinfo{volume}{21}, \bibinfo{number}{2} (\bibinfo{year}{2021}),
  \bibinfo{pages}{86--93}.
\newblock


\bibitem[Lange et~al\mbox{.}(2023)]%
        {lange2023engaging}
\bibfield{author}{\bibinfo{person}{Benjamin Lange}, \bibinfo{person}{Amanda
  McCroskery}, \bibinfo{person}{Ben Zevenbergen}, \bibinfo{person}{Geoff
  Keeling}, \bibinfo{person}{Sandra Blascovich}, \bibinfo{person}{Kyle
  Pedersen}, \bibinfo{person}{Alison Lentz}, {and}
  \bibinfo{person}{Blaise~Ag{\"u}era y Arcas}.}
  \bibinfo{year}{2023}\natexlab{}.
\newblock \bibinfo{title}{Engaging Google Teams Through Moral Imagination: A
  Bottom-Up Approach for Responsible Innovation and Ethical Culture Change in
  Technology Companies}.
\newblock
\newblock
\showeprint[arxiv]{2306.06901}~[cs.CY]


\bibitem[Le~Dantec and Fox(2015)]%
        {ledantec2015}
\bibfield{author}{\bibinfo{person}{Christopher~A. Le~Dantec} {and}
  \bibinfo{person}{Sarah Fox}.} \bibinfo{year}{2015}\natexlab{}.
\newblock \showarticletitle{Strangers at the Gate: Gaining Access, Building
  Rapport, and Co-Constructing Community-Based Research}. In
  \bibinfo{booktitle}{\emph{Proceedings of the 18th ACM Conference on Computer
  Supported Cooperative Work \& Social Computing}} (Vancouver, BC, Canada)
  \emph{(\bibinfo{series}{CSCW '15})}. \bibinfo{publisher}{Association for
  Computing Machinery}, \bibinfo{address}{New York, NY, USA},
  \bibinfo{pages}{1348–1358}.
\newblock
\urldef\tempurl%
\url{https://doi.org/10.1145/2675133.2675147}
\showDOI{\tempurl}


\bibitem[Lewis et~al\mbox{.}(2019)]%
        {lewis2019bart}
\bibfield{author}{\bibinfo{person}{Mike Lewis}, \bibinfo{person}{Yinhan Liu},
  \bibinfo{person}{Naman Goyal}, \bibinfo{person}{Marjan Ghazvininejad},
  \bibinfo{person}{Abdelrahman Mohamed}, \bibinfo{person}{Omer Levy},
  \bibinfo{person}{Ves Stoyanov}, {and} \bibinfo{person}{Luke Zettlemoyer}.}
  \bibinfo{year}{2019}\natexlab{}.
\newblock \bibinfo{title}{BART: Denoising sequence-to-sequence pre-training for
  natural language generation, translation, and comprehension}.
\newblock
\newblock
\showeprint[arxiv]{1910.13461}~[cs.CL]


\bibitem[Light(2011)]%
        {light2011}
\bibfield{author}{\bibinfo{person}{Ann Light}.}
  \bibinfo{year}{2011}\natexlab{}.
\newblock \showarticletitle{HCI as Heterodoxy: Technologies of Identity and the
  Queering of Interaction with Computers}.
\newblock \bibinfo{journal}{\emph{Interacting with Computers}}
  \bibinfo{volume}{23}, \bibinfo{number}{5} (\bibinfo{date}{Sept.}
  \bibinfo{year}{2011}), \bibinfo{pages}{430–438}.
\newblock
\showISSN{0953-5438}
\urldef\tempurl%
\url{https://doi.org/10.1016/j.intcom.2011.02.002}
\showDOI{\tempurl}


\bibitem[Liu et~al\mbox{.}(2021)]%
        {liu2021swin}
\bibfield{author}{\bibinfo{person}{Ze Liu}, \bibinfo{person}{Yutong Lin},
  \bibinfo{person}{Yue Cao}, \bibinfo{person}{Han Hu}, \bibinfo{person}{Yixuan
  Wei}, \bibinfo{person}{Zheng Zhang}, \bibinfo{person}{Stephen Lin}, {and}
  \bibinfo{person}{Baining Guo}.} \bibinfo{year}{2021}\natexlab{}.
\newblock \showarticletitle{Swin transformer: Hierarchical vision transformer
  using shifted windows}. In \bibinfo{booktitle}{\emph{Proceedings of the
  IEEE/CVF International Conference on Computer Vision}} (Virtual Event, USA).
  \bibinfo{pages}{10012--10022}.
\newblock


\bibitem[{Lloyd’s Register Foundation}(2020)]%
        {lloyds2020}
\bibfield{author}{\bibinfo{person}{{Lloyd’s Register Foundation}}.}
  \bibinfo{year}{2020}\natexlab{}.
\newblock \bibinfo{title}{World {R}isk {P}oll {R}eport 2019}.
\newblock
\newblock


\bibitem[Lock(2022)]%
        {lock2022}
\bibfield{author}{\bibinfo{person}{Samantha Lock}.}
  \bibinfo{year}{2022}\natexlab{}.
\newblock \showarticletitle{What is AI chatbot phenomenon ChatGPT and could it
  replace humans?}
\newblock \bibinfo{journal}{\emph{The Guardian}} (\bibinfo{date}{5 Dec.}
  \bibinfo{year}{2022}).
\newblock
\urldef\tempurl%
\url{https://www.theguardian.com/technology/2022/dec/05/what-is-ai-chatbot-phenomenon-chatgpt-and-could-it-replace-humans}
\showURL{%
\tempurl}


\bibitem[Louie et~al\mbox{.}(2020)]%
        {louie2020}
\bibfield{author}{\bibinfo{person}{Ryan Louie}, \bibinfo{person}{Andy Coenen},
  \bibinfo{person}{Cheng~Zhi Huang}, \bibinfo{person}{Michael Terry}, {and}
  \bibinfo{person}{Carrie~J. Cai}.} \bibinfo{year}{2020}\natexlab{}.
\newblock \showarticletitle{Novice-AI Music Co-Creation via AI-Steering Tools
  for Deep Generative Models}. In \bibinfo{booktitle}{\emph{Proceedings of the
  2020 CHI Conference on Human Factors in Computing Systems}} (Honolulu, HI,
  USA) \emph{(\bibinfo{series}{CHI '20})}. \bibinfo{publisher}{Association for
  Computing Machinery}, \bibinfo{address}{New York, NY, USA},
  \bibinfo{pages}{1–13}.
\newblock
\showISBNx{9781450367080}
\urldef\tempurl%
\url{https://doi.org/10.1145/3313831.3376739}
\showDOI{\tempurl}


\bibitem[Lu et~al\mbox{.}(2023)]%
        {lu2023}
\bibfield{author}{\bibinfo{person}{Xinyi Lu}, \bibinfo{person}{Simin Fan},
  \bibinfo{person}{Jessica Houghton}, \bibinfo{person}{Lu Wang}, {and}
  \bibinfo{person}{Xu Wang}.} \bibinfo{year}{2023}\natexlab{}.
\newblock \showarticletitle{ReadingQuizMaker: A Human-NLP Collaborative System
  That Supports Instructors to Design High-Quality Reading Quiz Questions}. In
  \bibinfo{booktitle}{\emph{Proceedings of the 2023 CHI Conference on Human
  Factors in Computing Systems}} (Hamburg, Germany) \emph{(\bibinfo{series}{CHI
  '23})}. \bibinfo{publisher}{Association for Computing Machinery},
  \bibinfo{address}{New York, NY, USA}, Article \bibinfo{articleno}{454},
  \bibinfo{numpages}{18}~pages.
\newblock
\showISBNx{9781450394215}
\urldef\tempurl%
\url{https://doi.org/10.1145/3544548.3580957}
\showDOI{\tempurl}


\bibitem[Marcus(2019)]%
        {marcus2019}
\bibfield{author}{\bibinfo{person}{Gary Marcus}.}
  \bibinfo{year}{2019}\natexlab{}.
\newblock \showarticletitle{An Epidemic of AI Misinformation}.
\newblock \bibinfo{journal}{\emph{The Gradient}} (\bibinfo{date}{30 Nov.}
  \bibinfo{year}{2019}).
\newblock
\urldef\tempurl%
\url{https://thegradient.pub/an-epidemic-of-ai-misinformation/}
\showURL{%
\tempurl}


\bibitem[Marr(2023)]%
        {marr2023}
\bibfield{author}{\bibinfo{person}{Bernard Marr}.}
  \bibinfo{year}{2023}\natexlab{}.
\newblock \showarticletitle{A Short History Of ChatGPT: How We Got To Where We
  Are Today}.
\newblock \bibinfo{journal}{\emph{Forbes}} (\bibinfo{date}{19 May}
  \bibinfo{year}{2023}).
\newblock
\urldef\tempurl%
\url{https://www.forbes.com/sites/bernardmarr/2023/05/19/a-short-history-of-chatgpt-how-we-got-to-where-we-are-today/?sh=286b1544674f}
\showURL{%
\tempurl}


\bibitem[Marwick and Lewis(2017)]%
        {marwick2017media}
\bibfield{author}{\bibinfo{person}{Alice~E. Marwick} {and}
  \bibinfo{person}{Rebecca Lewis}.} \bibinfo{year}{2017}\natexlab{}.
\newblock \bibinfo{booktitle}{\emph{Media Manipulation and Disinformation
  Online}}.
\newblock \bibinfo{type}{Report}. \bibinfo{institution}{Data \& Society
  Research Institute}.
\newblock
\urldef\tempurl%
\url{https://datasociety.net/library/media-manipulation-and-disinfo-online}
\showURL{%
\tempurl}


\bibitem[Metz(2023)]%
        {metz2023}
\bibfield{author}{\bibinfo{person}{Cade Metz}.}
  \bibinfo{year}{2023}\natexlab{}.
\newblock \showarticletitle{`The Godfather of A.I.' Leaves Google and Warns of
  Danger Ahead}.
\newblock \bibinfo{journal}{\emph{The New York Times}} (\bibinfo{date}{1 May}
  \bibinfo{year}{2023}).
\newblock
\urldef\tempurl%
\url{https://www.nytimes.com/2023/05/01/technology/ai-google-chatbot-engineer-quits-hinton.html}
\showURL{%
\tempurl}


\bibitem[Meyer(2023)]%
        {meyer2023}
\bibfield{author}{\bibinfo{person}{David Meyer}.}
  \bibinfo{year}{2023}\natexlab{}.
\newblock \bibinfo{title}{A.I.’s threat to jobs is not hypothetical—just
  ask IBM’s boss}.
\newblock \bibinfo{howpublished}{Yahoo Finance}.
\newblock
\urldef\tempurl%
\url{https://finance.yahoo.com/news/threat-jobs-not-hypothetical-just-173432317.html}
\showURL{%
\tempurl}


\bibitem[{Midjourney, Inc.}(2022)]%
        {midjourney}
\bibfield{author}{\bibinfo{person}{{Midjourney, Inc.}}}
  \bibinfo{year}{2022}\natexlab{}.
\newblock \bibinfo{title}{Midjourney}.
\newblock
\newblock
\urldef\tempurl%
\url{https://www.midjourney.com/}
\showURL{%
\tempurl}


\bibitem[Mikolov et~al\mbox{.}(2010)]%
        {mikolov2010recurrent}
\bibfield{author}{\bibinfo{person}{Tomas Mikolov}, \bibinfo{person}{Martin
  Karafi{\'a}t}, \bibinfo{person}{Lukas Burget}, \bibinfo{person}{Jan
  Cernock{\`y}}, {and} \bibinfo{person}{Sanjeev Khudanpur}.}
  \bibinfo{year}{2010}\natexlab{}.
\newblock \showarticletitle{Recurrent neural network based language model.}. In
  \bibinfo{booktitle}{\emph{Interspeech}}. \bibinfo{publisher}{ICSA},
  \bibinfo{address}{Makuhari, Chiba, Japan}, \bibinfo{pages}{1045--1048}.
\newblock


\bibitem[Mosco(2008)]%
        {moscoknowledge}
\bibfield{author}{\bibinfo{person}{Vincent Mosco}.}
  \bibinfo{year}{2008}\natexlab{}.
\newblock \showarticletitle{{Knowledge Workers of the World! Unite?}}
\newblock \bibinfo{journal}{\emph{Communication, Culture and Critique}}
  \bibinfo{volume}{1}, \bibinfo{number}{1} (\bibinfo{date}{March}
  \bibinfo{year}{2008}), \bibinfo{pages}{105--115}.
\newblock
\showISSN{1753-9129}
\urldef\tempurl%
\url{https://doi.org/10.1111/j.1753-9137.2007.00011.x}
\showDOI{\tempurl}


\bibitem[Mozilla(2019)]%
        {mozilla2019}
\bibfield{author}{\bibinfo{person}{Mozilla}.} \bibinfo{year}{2019}\natexlab{}.
\newblock \bibinfo{title}{We Asked People Around the World How They Feel About
  Artificial Intelligence. {H}ere's What We Learned.}
\newblock
\newblock
\urldef\tempurl%
\url{https://foundation.mozilla.org/en/blog/we-asked-people-around-the-world-how-they-feel-about-artificial-intelligence-heres-what-we-learned/}
\showURL{%
\tempurl}


\bibitem[Muller et~al\mbox{.}(2022)]%
        {muller2022}
\bibfield{author}{\bibinfo{person}{Michael Muller}, \bibinfo{person}{Lydia~B.
  Chilton}, \bibinfo{person}{Anna Kantosalo}, \bibinfo{person}{Charles~Patrick
  Martin}, {and} \bibinfo{person}{Greg Walsh}.}
  \bibinfo{year}{2022}\natexlab{}.
\newblock \showarticletitle{GenAICHI: Generative AI and HCI}. In
  \bibinfo{booktitle}{\emph{Extended Abstracts of the 2022 CHI Conference on
  Human Factors in Computing Systems}} (New Orleans, LA, USA)
  \emph{(\bibinfo{series}{CHI EA '22})}. \bibinfo{publisher}{Association for
  Computing Machinery}, \bibinfo{address}{New York, NY, USA}, Article
  \bibinfo{articleno}{110}, \bibinfo{numpages}{7}~pages.
\newblock
\showISBNx{9781450391566}
\urldef\tempurl%
\url{https://doi.org/10.1145/3491101.3503719}
\showDOI{\tempurl}


\bibitem[Murphy(2022)]%
        {murphy2022probabilistic}
\bibfield{author}{\bibinfo{person}{Kevin~P. Murphy}.}
  \bibinfo{year}{2022}\natexlab{}.
\newblock \bibinfo{booktitle}{\emph{Probabilistic machine learning: an
  introduction}}.
\newblock \bibinfo{publisher}{MIT Press}, \bibinfo{address}{Boston, MA}.
\newblock


\bibitem[Neudert et~al\mbox{.}(2020)]%
        {neudert2020}
\bibfield{author}{\bibinfo{person}{Lisa-Maria Neudert}, \bibinfo{person}{Aleksi
  Knuutila}, {and} \bibinfo{person}{Philip~N. Howard}.}
  \bibinfo{year}{2020}\natexlab{}.
\newblock \bibinfo{booktitle}{\emph{Global Attitudes Towards {AI}, Machine
  Learning \& Automated Decision Making: Implications for Involving Artificial
  Intelligence in Public Service and Good Governance}}.
\newblock \bibinfo{type}{{T}echnical {R}eport}. \bibinfo{institution}{Oxford
  Internet Institute}.
\newblock


\bibitem[Newlands(2021)]%
        {newlands2021algorithmic}
\bibfield{author}{\bibinfo{person}{Gemma Newlands}.}
  \bibinfo{year}{2021}\natexlab{}.
\newblock \showarticletitle{Algorithmic surveillance in the gig economy: The
  organization of work through Lefebvrian conceived space}.
\newblock \bibinfo{journal}{\emph{Organization Studies}} \bibinfo{volume}{42},
  \bibinfo{number}{5} (\bibinfo{year}{2021}), \bibinfo{pages}{719--737}.
\newblock


\bibitem[Ng(2017)]%
        {ng2017}
\bibfield{author}{\bibinfo{person}{Andrew Ng}.}
  \bibinfo{year}{2017}\natexlab{}.
\newblock \bibinfo{title}{Andrew Ng: Artificial Intelligence is the New
  Electricity}.
\newblock \bibinfo{howpublished}{Stanford Graduate School of Business}.
\newblock
\urldef\tempurl%
\url{https://www.youtube.com/watch?v=21EiKfQYZXc}
\showURL{%
\tempurl}


\bibitem[{Northeastern University and Gallup}(2018)]%
        {northeastern2018}
\bibfield{author}{\bibinfo{person}{{Northeastern University and Gallup}}.}
  \bibinfo{year}{2018}\natexlab{}.
\newblock \bibinfo{title}{Optimism and Anxiety: Views on the Impact of
  Artificial Intelligence and Higher Education's Response}.
\newblock
\newblock


\bibitem[Noy and Zhang(2023)]%
        {noy2023}
\bibfield{author}{\bibinfo{person}{Shakked Noy} {and} \bibinfo{person}{Whitney
  Zhang}.} \bibinfo{year}{2023}\natexlab{}.
\newblock \showarticletitle{Experimental evidence on the productivity effects
  of generative artificial intelligence}.
\newblock \bibinfo{journal}{\emph{Science}} \bibinfo{volume}{381},
  \bibinfo{number}{6654} (\bibinfo{year}{2023}), \bibinfo{pages}{187--192}.
\newblock
\urldef\tempurl%
\url{https://doi.org/10.1126/science.adh2586}
\showDOI{\tempurl}


\bibitem[of~the Surgeon~General(2023)]%
        {epidemic2023}
\bibfield{author}{\bibinfo{person}{Office of~the Surgeon~General}.}
  \bibinfo{year}{2023}\natexlab{}.
\newblock \bibinfo{title}{Our Epidemic of Loneliness and Isolation: The U.S.
  Surgeon General’s Advisory on the Healing Effects of Social Connection and
  Community}.
\newblock
\newblock
\urldef\tempurl%
\url{https://www.hhs.gov/sites/default/files/surgeon-general-social-connection-advisory.pdf}
\showURL{%
\tempurl}


\bibitem[Ogbonnaya-Ogburu et~al\mbox{.}(2020)]%
        {ogbonnaya2020}
\bibfield{author}{\bibinfo{person}{Ihudiya~Finda Ogbonnaya-Ogburu},
  \bibinfo{person}{Angela~D.R. Smith}, \bibinfo{person}{Alexandra To}, {and}
  \bibinfo{person}{Kentaro Toyama}.} \bibinfo{year}{2020}\natexlab{}.
\newblock \showarticletitle{Critical Race Theory for HCI}. In
  \bibinfo{booktitle}{\emph{Proceedings of the 2020 CHI Conference on Human
  Factors in Computing Systems}} (Honolulu, HI, USA)
  \emph{(\bibinfo{series}{CHI '20})}. \bibinfo{publisher}{Association for
  Computing Machinery}, \bibinfo{address}{New York, NY, USA},
  \bibinfo{pages}{1–16}.
\newblock
\showISBNx{9781450367080}
\urldef\tempurl%
\url{https://doi.org/10.1145/3313831.3376392}
\showDOI{\tempurl}


\bibitem[Omaar(2023)]%
        {omaar2023}
\bibfield{author}{\bibinfo{person}{Hodan Omaar}.}
  \bibinfo{year}{2023}\natexlab{}.
\newblock \bibinfo{title}{Claims About Generative AI Replacing Jobs Are
  Hyperbolic and Misleading}.
\newblock
\newblock
\urldef\tempurl%
\url{https://datainnovation.org/2023/04/claims-about-generative-ai-replacing-are-hyperbolic-and-misleading/}
\showURL{%
\tempurl}


\bibitem[OpenAI(2022a)]%
        {codex}
\bibfield{author}{\bibinfo{person}{OpenAI}.} \bibinfo{year}{2022}\natexlab{a}.
\newblock \bibinfo{title}{Codex}.
\newblock
\newblock
\urldef\tempurl%
\url{https://openai.com/codex/}
\showURL{%
\tempurl}


\bibitem[OpenAI(2022b)]%
        {dalle2}
\bibfield{author}{\bibinfo{person}{OpenAI}.} \bibinfo{year}{2022}\natexlab{b}.
\newblock \bibinfo{title}{DALL-E 2}.
\newblock
\newblock
\urldef\tempurl%
\url{https://openai.com/dall-e-2}
\showURL{%
\tempurl}


\bibitem[OpenAI(2023)]%
        {chatgpt}
\bibfield{author}{\bibinfo{person}{OpenAI}.} \bibinfo{year}{2023}\natexlab{}.
\newblock \bibinfo{title}{ChatGPT}.
\newblock
\newblock
\urldef\tempurl%
\url{https://chat.openai.com/}
\showURL{%
\tempurl}


\bibitem[Penick and Long(2019)]%
        {penick2019}
\bibfield{author}{\bibinfo{person}{Monica Penick} {and}
  \bibinfo{person}{Christopher Long}.} \bibinfo{year}{2019}\natexlab{}.
\newblock \bibinfo{booktitle}{\emph{The rise of everyday design: The arts and
  crafts movement in Britain and America}}.
\newblock \bibinfo{publisher}{Yale University Press}, \bibinfo{address}{New
  Haven, Connecticut}.
\newblock


\bibitem[Poddar et~al\mbox{.}(2023)]%
        {poddar2023}
\bibfield{author}{\bibinfo{person}{Ritika Poddar}, \bibinfo{person}{Rashmi
  Sinha}, \bibinfo{person}{Mor Naaman}, {and} \bibinfo{person}{Maurice
  Jakesch}.} \bibinfo{year}{2023}\natexlab{}.
\newblock \showarticletitle{AI Writing Assistants Influence Topic Choice in
  Self-Presentation}. In \bibinfo{booktitle}{\emph{Extended Abstracts of the
  2023 CHI Conference on Human Factors in Computing Systems}} (Hamburg,
  Germany) \emph{(\bibinfo{series}{CHI EA '23})}.
  \bibinfo{publisher}{Association for Computing Machinery},
  \bibinfo{address}{New York, NY, USA}, Article \bibinfo{articleno}{29},
  \bibinfo{numpages}{6}~pages.
\newblock
\showISBNx{9781450394222}
\urldef\tempurl%
\url{https://doi.org/10.1145/3544549.3585893}
\showDOI{\tempurl}


\bibitem[Py{\"o}ri{\"a}(2005)]%
        {pyoria2005concept}
\bibfield{author}{\bibinfo{person}{Pasi Py{\"o}ri{\"a}}.}
  \bibinfo{year}{2005}\natexlab{}.
\newblock \showarticletitle{The concept of knowledge work revisited}.
\newblock \bibinfo{journal}{\emph{Journal of Knowledge Management}}
  \bibinfo{volume}{9}, \bibinfo{number}{3} (\bibinfo{year}{2005}),
  \bibinfo{pages}{116--127}.
\newblock


\bibitem[Rainie et~al\mbox{.}(2022)]%
        {rainie2022}
\bibfield{author}{\bibinfo{person}{Lee Rainie}, \bibinfo{person}{Cary Funk},
  \bibinfo{person}{Monica Anderson}, {and} \bibinfo{person}{Alec Tyson}.}
  \bibinfo{year}{2022}\natexlab{}.
\newblock \bibinfo{booktitle}{\emph{{AI} and human enhancement: {A}mericans’
  openness is tempered by a range of concerns}}.
\newblock \bibinfo{type}{{T}echnical {R}eport}. \bibinfo{institution}{Pew
  Research Center}.
\newblock


\bibitem[Ramesh et~al\mbox{.}(2022)]%
        {ramesh2022hierarchical}
\bibfield{author}{\bibinfo{person}{Aditya Ramesh}, \bibinfo{person}{Prafulla
  Dhariwal}, \bibinfo{person}{Alex Nichol}, \bibinfo{person}{Casey Chu}, {and}
  \bibinfo{person}{Mark Chen}.} \bibinfo{year}{2022}\natexlab{}.
\newblock \bibinfo{title}{Hierarchical text-conditional image generation with
  CLIP latents}.
\newblock
\newblock
\showeprint[arxiv]{2204.06125}~[cs.CV]


\bibitem[Ramesh et~al\mbox{.}(2021)]%
        {ramesh2021zero}
\bibfield{author}{\bibinfo{person}{Aditya Ramesh}, \bibinfo{person}{Mikhail
  Pavlov}, \bibinfo{person}{Gabriel Goh}, \bibinfo{person}{Scott Gray},
  \bibinfo{person}{Chelsea Voss}, \bibinfo{person}{Alec Radford},
  \bibinfo{person}{Mark Chen}, {and} \bibinfo{person}{Ilya Sutskever}.}
  \bibinfo{year}{2021}\natexlab{}.
\newblock \showarticletitle{Zero-shot Text-to-Image Generation}. In
  \bibinfo{booktitle}{\emph{International Conference on Machine Learning}}
  (Virtual Event, USA). \bibinfo{publisher}{PMLR}, \bibinfo{pages}{8821--8831}.
\newblock


\bibitem[Rankin et~al\mbox{.}(2020)]%
        {rankin2020}
\bibfield{author}{\bibinfo{person}{Yolanda~A. Rankin},
  \bibinfo{person}{Jakita~O. Thomas}, {and} \bibinfo{person}{Nicole~M.
  Joseph}.} \bibinfo{year}{2020}\natexlab{}.
\newblock \showarticletitle{Intersectionality in HCI: Lost in Translation}.
\newblock \bibinfo{journal}{\emph{Interactions}} \bibinfo{volume}{27},
  \bibinfo{number}{5} (\bibinfo{date}{Sept.} \bibinfo{year}{2020}),
  \bibinfo{pages}{68–71}.
\newblock
\showISSN{1072-5520}
\urldef\tempurl%
\url{https://doi.org/10.1145/3416498}
\showDOI{\tempurl}


\bibitem[Research(2022)]%
        {imagen}
\bibfield{author}{\bibinfo{person}{Google Research}.}
  \bibinfo{year}{2022}\natexlab{}.
\newblock \bibinfo{title}{Imagen: Text to Image Diffusion Models}.
\newblock \bibinfo{howpublished}{\url{https://imagen.research.google}}.
\newblock


\bibitem[Reynolds(2009)]%
        {reynolds2009gaussian}
\bibfield{author}{\bibinfo{person}{Douglas~A. Reynolds}.}
  \bibinfo{year}{2009}\natexlab{}.
\newblock \bibinfo{booktitle}{\emph{Gaussian Mixture Models}}.
  Vol.~\bibinfo{volume}{741}.
\newblock \bibinfo{publisher}{Springer}, \bibinfo{address}{Berlin, Germany},
  \bibinfo{pages}{659--663}.
\newblock


\bibitem[Rode(2011)]%
        {rode2011}
\bibfield{author}{\bibinfo{person}{Jennifer~A. Rode}.}
  \bibinfo{year}{2011}\natexlab{}.
\newblock \showarticletitle{A Theoretical Agenda for Feminist HCI}.
\newblock \bibinfo{journal}{\emph{Interacting with Computers}}
  \bibinfo{volume}{23}, \bibinfo{number}{5} (\bibinfo{date}{Sept.}
  \bibinfo{year}{2011}), \bibinfo{pages}{393–400}.
\newblock
\showISSN{0953-5438}
\urldef\tempurl%
\url{https://doi.org/10.1016/j.intcom.2011.04.005}
\showDOI{\tempurl}


\bibitem[Roemmele and Gordon(2015)]%
        {roemmele2015creative}
\bibfield{author}{\bibinfo{person}{Melissa Roemmele} {and}
  \bibinfo{person}{Andrew~S. Gordon}.} \bibinfo{year}{2015}\natexlab{}.
\newblock \showarticletitle{Creative help: A story writing assistant}. In
  \bibinfo{booktitle}{\emph{Interactive Storytelling: 8th International
  Conference on Interactive Digital Storytelling}}
  \emph{(\bibinfo{series}{ICIDS 2015})}. \bibinfo{publisher}{Springer},
  \bibinfo{address}{Copenhagen, Denmark}, \bibinfo{pages}{81--92}.
\newblock


\bibitem[Roemmele and Gordon(2018)]%
        {roemmele2018}
\bibfield{author}{\bibinfo{person}{Melissa Roemmele} {and}
  \bibinfo{person}{Andrew~S. Gordon}.} \bibinfo{year}{2018}\natexlab{}.
\newblock \showarticletitle{Automated Assistance for Creative Writing with an
  RNN Language Model}. In \bibinfo{booktitle}{\emph{Proceedings of the 23rd
  International Conference on Intelligent User Interfaces Companion}} (Tokyo,
  Japan) \emph{(\bibinfo{series}{IUI '18 Companion})}.
  \bibinfo{publisher}{Association for Computing Machinery},
  \bibinfo{address}{New York, NY, USA}, Article \bibinfo{articleno}{21},
  \bibinfo{numpages}{2}~pages.
\newblock
\showISBNx{9781450355711}
\urldef\tempurl%
\url{https://doi.org/10.1145/3180308.3180329}
\showDOI{\tempurl}


\bibitem[Roose(2023)]%
        {roose2023}
\bibfield{author}{\bibinfo{person}{Kevin Roose}.}
  \bibinfo{year}{2023}\natexlab{}.
\newblock \showarticletitle{Bing’s A.I. Chat: `I Want to Be Alive'}.
\newblock \bibinfo{journal}{\emph{The New York Times}} (\bibinfo{date}{16 Feb.}
  \bibinfo{year}{2023}).
\newblock
\urldef\tempurl%
\url{https://www.nytimes.com/2023/02/16/technology/bing-chatbot-transcript.html}
\showURL{%
\tempurl}


\bibitem[Rosner et~al\mbox{.}(2016)]%
        {rosner2016}
\bibfield{author}{\bibinfo{person}{Daniela~K. Rosner}, \bibinfo{person}{Saba
  Kawas}, \bibinfo{person}{Wenqi Li}, \bibinfo{person}{Nicole Tilly}, {and}
  \bibinfo{person}{Yi-Chen Sung}.} \bibinfo{year}{2016}\natexlab{}.
\newblock \showarticletitle{Out of Time, Out of Place: Reflections on Design
  Workshops as a Research Method}. In \bibinfo{booktitle}{\emph{Proceedings of
  the 19th ACM Conference on Computer-Supported Cooperative Work \& Social
  Computing}} (San Francisco, California, USA) \emph{(\bibinfo{series}{CSCW
  '16})}. \bibinfo{publisher}{Association for Computing Machinery},
  \bibinfo{address}{New York, NY, USA}, \bibinfo{pages}{1131–1141}.
\newblock
\showISBNx{9781450335928}
\urldef\tempurl%
\url{https://doi.org/10.1145/2818048.2820021}
\showDOI{\tempurl}


\bibitem[Sahai(2023)]%
        {sahai2023}
\bibfield{author}{\bibinfo{person}{Sandeep Sahai}.}
  \bibinfo{year}{2023}\natexlab{}.
\newblock \showarticletitle{Generative AI: A Big Bang Moment For FinTech}.
\newblock \bibinfo{journal}{\emph{Forbes}} (\bibinfo{date}{26 July}
  \bibinfo{year}{2023}).
\newblock
\urldef\tempurl%
\url{https://www.forbes.com/sites/forbestechcouncil/2023/07/26/generative-ai-a-big-bang-moment-for-fintech}
\showURL{%
\tempurl}


\bibitem[Sarsa et~al\mbox{.}(2022)]%
        {sarsa2022}
\bibfield{author}{\bibinfo{person}{Sami Sarsa}, \bibinfo{person}{Paul Denny},
  \bibinfo{person}{Arto Hellas}, {and} \bibinfo{person}{Juho Leinonen}.}
  \bibinfo{year}{2022}\natexlab{}.
\newblock \showarticletitle{Automatic Generation of Programming Exercises and
  Code Explanations Using Large Language Models}. In
  \bibinfo{booktitle}{\emph{Proceedings of the 2022 ACM Conference on
  International Computing Education Research - Volume 1}} (Lugano and Virtual
  Event, Switzerland) \emph{(\bibinfo{series}{ICER '22})}.
  \bibinfo{publisher}{Association for Computing Machinery},
  \bibinfo{address}{New York, NY, USA}, \bibinfo{pages}{27–43}.
\newblock
\showISBNx{9781450391948}
\urldef\tempurl%
\url{https://doi.org/10.1145/3501385.3543957}
\showDOI{\tempurl}


\bibitem[Scheuerman et~al\mbox{.}(2019)]%
        {scheuerman2019}
\bibfield{author}{\bibinfo{person}{Morgan~Klaus Scheuerman},
  \bibinfo{person}{Jacob~M. Paul}, {and} \bibinfo{person}{Jed~R. Brubaker}.}
  \bibinfo{year}{2019}\natexlab{}.
\newblock \showarticletitle{How Computers See Gender: An Evaluation of Gender
  Classification in Commercial Facial Analysis Services}.
\newblock \bibinfo{journal}{\emph{Proceedings of the ACM on Human-Computer
  Interaction}} \bibinfo{volume}{3}, \bibinfo{number}{CSCW}, Article
  \bibinfo{articleno}{144} (\bibinfo{date}{Nov.} \bibinfo{year}{2019}),
  \bibinfo{numpages}{33}~pages.
\newblock
\urldef\tempurl%
\url{https://doi.org/10.1145/3359246}
\showDOI{\tempurl}


\bibitem[Schlesinger et~al\mbox{.}(2017)]%
        {schlesinger2017}
\bibfield{author}{\bibinfo{person}{Ari Schlesinger}, \bibinfo{person}{W.~Keith
  Edwards}, {and} \bibinfo{person}{Rebecca~E. Grinter}.}
  \bibinfo{year}{2017}\natexlab{}.
\newblock \showarticletitle{Intersectional HCI: Engaging Identity through
  Gender, Race, and Class}. In \bibinfo{booktitle}{\emph{Proceedings of the
  2017 CHI Conference on Human Factors in Computing Systems}} (Denver,
  Colorado, USA) \emph{(\bibinfo{series}{CHI '17})}.
  \bibinfo{publisher}{Association for Computing Machinery},
  \bibinfo{address}{New York, NY, USA}, \bibinfo{pages}{5412–5427}.
\newblock
\showISBNx{9781450346559}
\urldef\tempurl%
\url{https://doi.org/10.1145/3025453.3025766}
\showDOI{\tempurl}


\bibitem[Schlesinger et~al\mbox{.}(2018)]%
        {schlesinger2018}
\bibfield{author}{\bibinfo{person}{Ari Schlesinger}, \bibinfo{person}{Kenton~P.
  O'Hara}, {and} \bibinfo{person}{Alex~S. Taylor}.}
  \bibinfo{year}{2018}\natexlab{}.
\newblock \showarticletitle{Let's Talk About Race: Identity, Chatbots, and AI}.
  In \bibinfo{booktitle}{\emph{Proceedings of the 2018 CHI Conference on Human
  Factors in Computing Systems}} (Montreal, Canada) \emph{(\bibinfo{series}{CHI
  '18})}. \bibinfo{publisher}{Association for Computing Machinery},
  \bibinfo{address}{New York, NY, USA}, \bibinfo{pages}{1–14}.
\newblock
\showISBNx{9781450356206}
\urldef\tempurl%
\url{https://doi.org/10.1145/3173574.3173889}
\showDOI{\tempurl}


\bibitem[Selwyn et~al\mbox{.}(2020)]%
        {selwyn2020}
\bibfield{author}{\bibinfo{person}{Neil Selwyn}, \bibinfo{person}{Beatriz~Gallo
  Cordoba}, \bibinfo{person}{Mark Andrejevic}, {and} \bibinfo{person}{Liz
  Campbell}.} \bibinfo{year}{2020}\natexlab{}.
\newblock \bibinfo{booktitle}{\emph{AI for social good: Australian public
  attitudes toward AI and society}}.
\newblock \bibinfo{type}{{T}echnical {R}eport}. \bibinfo{institution}{Monash
  Data Futures Institute}.
\newblock


\bibitem[Sewell and Taskin(2015)]%
        {sewell2015out}
\bibfield{author}{\bibinfo{person}{Graham Sewell} {and}
  \bibinfo{person}{Laurent Taskin}.} \bibinfo{year}{2015}\natexlab{}.
\newblock \showarticletitle{Out of sight, out of mind in a new world of work?
  Autonomy, control, and spatiotemporal scaling in telework}.
\newblock \bibinfo{journal}{\emph{Organization studies}} \bibinfo{volume}{36},
  \bibinfo{number}{11} (\bibinfo{year}{2015}), \bibinfo{pages}{1507--1529}.
\newblock


\bibitem[Shakeri et~al\mbox{.}(2021)]%
        {shakeri2021}
\bibfield{author}{\bibinfo{person}{Hanieh Shakeri}, \bibinfo{person}{Carman
  Neustaedter}, {and} \bibinfo{person}{Steve DiPaola}.}
  \bibinfo{year}{2021}\natexlab{}.
\newblock \showarticletitle{SAGA: Collaborative Storytelling with GPT-3}. In
  \bibinfo{booktitle}{\emph{Companion Publication of the 2021 Conference on
  Computer Supported Cooperative Work and Social Computing}} (Virtual Event,
  USA) \emph{(\bibinfo{series}{CSCW '21})}. \bibinfo{publisher}{Association for
  Computing Machinery}, \bibinfo{address}{New York, NY, USA},
  \bibinfo{pages}{163–166}.
\newblock
\showISBNx{9781450384797}
\urldef\tempurl%
\url{https://doi.org/10.1145/3462204.3481771}
\showDOI{\tempurl}


\bibitem[Shortliffe(1993)]%
        {shortliffe1993}
\bibfield{author}{\bibinfo{person}{Edward~H. Shortliffe}.}
  \bibinfo{year}{1993}\natexlab{}.
\newblock \showarticletitle{Doctors, Patients, and Computers: Will Information
  Technology Dehumanize Health-Care Delivery?}
\newblock \bibinfo{journal}{\emph{Proceedings of the American Philosophical
  Society}} \bibinfo{volume}{137}, \bibinfo{number}{3} (\bibinfo{year}{1993}),
  \bibinfo{pages}{390--398}.
\newblock
\showISSN{0003049X}
\urldef\tempurl%
\url{http://www.jstor.org/stable/986999}
\showURL{%
\tempurl}


\bibitem[Siegel(2023)]%
        {siegel2023}
\bibfield{author}{\bibinfo{person}{Eric Siegel}.}
  \bibinfo{year}{2023}\natexlab{}.
\newblock \bibinfo{title}{The AI Hype Cycle Is Distracting Companies}.
\newblock \bibinfo{howpublished}{Harvard Business Review}.
\newblock
\urldef\tempurl%
\url{https://hbr.org/2023/06/the-ai-hype-cycle-is-distracting-companies}
\showURL{%
\tempurl}


\bibitem[Simb{\"u}rger and Neary(2016)]%
        {simburger2016taxi}
\bibfield{author}{\bibinfo{person}{Elisabeth Simb{\"u}rger} {and}
  \bibinfo{person}{Mike Neary}.} \bibinfo{year}{2016}\natexlab{}.
\newblock \showarticletitle{Taxi professors: academic labour in Chile, a
  critical-practical response to the politics of worker identity}.
\newblock \bibinfo{journal}{\emph{Workplace: A Journal for Academic Labor}}
  \bibinfo{number}{28} (\bibinfo{year}{2016}), \bibinfo{pages}{48--73}.
\newblock
\urldef\tempurl%
\url{https://doi.org/10.14288/workplace.v0i28.186212}
\showDOI{\tempurl}


\bibitem[Singh et~al\mbox{.}(2022)]%
        {singh2022}
\bibfield{author}{\bibinfo{person}{Nikhil Singh}, \bibinfo{person}{Guillermo
  Bernal}, \bibinfo{person}{Daria Savchenko}, {and} \bibinfo{person}{Elena~L.
  Glassman}.} \bibinfo{year}{2022}\natexlab{}.
\newblock \showarticletitle{Where to Hide a Stolen Elephant: Leaps in Creative
  Writing with Multimodal Machine Intelligence}.
\newblock \bibinfo{journal}{\emph{ACM Transactions on Compututer-Human
  Interaction}} (\bibinfo{date}{Feb.} \bibinfo{year}{2022}),
  \bibinfo{pages}{1--53}.
\newblock
\showISSN{1073-0516}
\urldef\tempurl%
\url{https://doi.org/10.1145/3511599}
\showDOI{\tempurl}


\bibitem[Smith(2018)]%
        {smith2018}
\bibfield{author}{\bibinfo{person}{Aaron Smith}.}
  \bibinfo{year}{2018}\natexlab{}.
\newblock \showarticletitle{Public Attitudes Toward Computer Algorithms}.
\newblock \bibinfo{journal}{\emph{Pew Research Center}} (\bibinfo{date}{Nov.}
  \bibinfo{year}{2018}).
\newblock


\bibitem[Song and Ermon(2019)]%
        {song2019generative}
\bibfield{author}{\bibinfo{person}{Yang Song} {and} \bibinfo{person}{Stefano
  Ermon}.} \bibinfo{year}{2019}\natexlab{}.
\newblock \showarticletitle{Generative Modeling by Estimating Gradients of the
  Data Distribution}. In \bibinfo{booktitle}{\emph{Advances in Neural
  Information Processing Systems}},
  \bibfield{editor}{\bibinfo{person}{H.~Wallach},
  \bibinfo{person}{H.~Larochelle}, \bibinfo{person}{A.~Beygelzimer},
  \bibinfo{person}{F.~d\textquotesingle Alch\'{e}-Buc},
  \bibinfo{person}{E.~Fox}, {and} \bibinfo{person}{R.~Garnett}} (Eds.),
  Vol.~\bibinfo{volume}{32}. \bibinfo{publisher}{Curran Associates, Inc.},
  \bibinfo{address}{Vancouver, Canada}, \bibinfo{pages}{1--13}.
\newblock
\urldef\tempurl%
\url{https://proceedings.neurips.cc/paper_files/paper/2019/file/3001ef257407d5a371a96dcd947c7d93-Paper.pdf}
\showURL{%
\tempurl}


\bibitem[Spiel et~al\mbox{.}(2019)]%
        {spiel2019}
\bibfield{author}{\bibinfo{person}{Katta Spiel}, \bibinfo{person}{Os Keyes},
  \bibinfo{person}{Ashley~Marie Walker}, \bibinfo{person}{Michael~A. DeVito},
  \bibinfo{person}{Jeremy Birnholtz}, \bibinfo{person}{Emeline Brul\'{e}},
  \bibinfo{person}{Ann Light}, \bibinfo{person}{P\i{}nar Barlas},
  \bibinfo{person}{Jean Hardy}, \bibinfo{person}{Alex Ahmed},
  \bibinfo{person}{Jennifer~A. Rode}, \bibinfo{person}{Jed~R. Brubaker}, {and}
  \bibinfo{person}{Gopinaath Kannabiran}.} \bibinfo{year}{2019}\natexlab{}.
\newblock \showarticletitle{Queer(Ing) HCI: Moving Forward in Theory and
  Practice}. In \bibinfo{booktitle}{\emph{Extended Abstracts of the 2019 CHI
  Conference on Human Factors in Computing Systems}} (Glasgow, Scotland UK)
  \emph{(\bibinfo{series}{CHI EA '19})}. \bibinfo{publisher}{Association for
  Computing Machinery}, \bibinfo{address}{New York, NY, USA},
  \bibinfo{pages}{1–4}.
\newblock
\showISBNx{9781450359719}
\urldef\tempurl%
\url{https://doi.org/10.1145/3290607.3311750}
\showDOI{\tempurl}


\bibitem[Stephany(2021)]%
        {stephany2021}
\bibfield{author}{\bibinfo{person}{Fabian Stephany}.}
  \bibinfo{year}{2021}\natexlab{}.
\newblock \showarticletitle{One size does not fit all: Constructing
  complementary digital reskilling strategies using online labour market data}.
\newblock \bibinfo{journal}{\emph{Big Data \& Society}} \bibinfo{volume}{8},
  \bibinfo{number}{1} (\bibinfo{year}{2021}),
  \bibinfo{pages}{20539517211003120}.
\newblock
\urldef\tempurl%
\url{https://doi.org/10.1177/20539517211003120}
\showDOI{\tempurl}


\bibitem[Suh et~al\mbox{.}(2021)]%
        {suh2021}
\bibfield{author}{\bibinfo{person}{Minhyang~(Mia) Suh}, \bibinfo{person}{Emily
  Youngblom}, \bibinfo{person}{Michael Terry}, {and} \bibinfo{person}{Carrie~J.
  Cai}.} \bibinfo{year}{2021}\natexlab{}.
\newblock \showarticletitle{AI as Social Glue: Uncovering the Roles of Deep
  Generative AI during Social Music Composition}. In
  \bibinfo{booktitle}{\emph{Proceedings of the 2021 CHI Conference on Human
  Factors in Computing Systems}} (Yokohama, Japan) \emph{(\bibinfo{series}{CHI
  '21})}. \bibinfo{publisher}{Association for Computing Machinery},
  \bibinfo{address}{New York, NY, USA}, Article \bibinfo{articleno}{582},
  \bibinfo{numpages}{11}~pages.
\newblock
\showISBNx{9781450380966}
\urldef\tempurl%
\url{https://doi.org/10.1145/3411764.3445219}
\showDOI{\tempurl}


\bibitem[{The European Commission}(2017)]%
        {european2017}
\bibfield{author}{\bibinfo{person}{{The European Commission}}.}
  \bibinfo{year}{2017}\natexlab{}.
\newblock \bibinfo{title}{Special {E}urobarometer 460: Attitudes towards the
  impact of digitisation and automation on daily life}.
\newblock
\newblock


\bibitem[Thoppilan et~al\mbox{.}(2022)]%
        {thoppilan2022lamda}
\bibfield{author}{\bibinfo{person}{Romal Thoppilan}, \bibinfo{person}{Daniel~De
  Freitas}, \bibinfo{person}{Jamie Hall}, \bibinfo{person}{Noam Shazeer},
  \bibinfo{person}{Apoorv Kulshreshtha}, \bibinfo{person}{Heng-Tze Cheng},
  \bibinfo{person}{Alicia Jin}, \bibinfo{person}{Taylor Bos},
  \bibinfo{person}{Leslie Baker}, \bibinfo{person}{Yu Du},
  \bibinfo{person}{YaGuang Li}, \bibinfo{person}{Hongrae Lee},
  \bibinfo{person}{Huaixiu~Steven Zheng}, \bibinfo{person}{Amin Ghafouri},
  \bibinfo{person}{Marcelo Menegali}, \bibinfo{person}{Yanping Huang},
  \bibinfo{person}{Maxim Krikun}, \bibinfo{person}{Dmitry Lepikhin},
  \bibinfo{person}{James Qin}, \bibinfo{person}{Dehao Chen},
  \bibinfo{person}{Yuanzhong Xu}, \bibinfo{person}{Zhifeng Chen},
  \bibinfo{person}{Adam Roberts}, \bibinfo{person}{Maarten Bosma},
  \bibinfo{person}{Vincent Zhao}, \bibinfo{person}{Yanqi Zhou},
  \bibinfo{person}{Chung-Ching Chang}, \bibinfo{person}{Igor Krivokon},
  \bibinfo{person}{Will Rusch}, \bibinfo{person}{Marc Pickett},
  \bibinfo{person}{Pranesh Srinivasan}, \bibinfo{person}{Laichee Man},
  \bibinfo{person}{Kathleen Meier-Hellstern}, \bibinfo{person}{Meredith~Ringel
  Morris}, \bibinfo{person}{Tulsee Doshi}, \bibinfo{person}{Renelito~Delos
  Santos}, \bibinfo{person}{Toju Duke}, \bibinfo{person}{Johnny Soraker},
  \bibinfo{person}{Ben Zevenbergen}, \bibinfo{person}{Vinodkumar Prabhakaran},
  \bibinfo{person}{Mark Diaz}, \bibinfo{person}{Ben Hutchinson},
  \bibinfo{person}{Kristen Olson}, \bibinfo{person}{Alejandra Molina},
  \bibinfo{person}{Erin Hoffman-John}, \bibinfo{person}{Josh Lee},
  \bibinfo{person}{Lora Aroyo}, \bibinfo{person}{Ravi Rajakumar},
  \bibinfo{person}{Alena Butryna}, \bibinfo{person}{Matthew Lamm},
  \bibinfo{person}{Viktoriya Kuzmina}, \bibinfo{person}{Joe Fenton},
  \bibinfo{person}{Aaron Cohen}, \bibinfo{person}{Rachel Bernstein},
  \bibinfo{person}{Ray Kurzweil}, \bibinfo{person}{Blaise~Ag{\"u}era y Arcas},
  \bibinfo{person}{Claire Cui}, \bibinfo{person}{Marian Croak},
  \bibinfo{person}{Ed Chi}, {and} \bibinfo{person}{Quoc Le}.}
  \bibinfo{year}{2022}\natexlab{}.
\newblock \bibinfo{title}{LaMDA: Language Models for Dialog Applications}.
\newblock
\newblock
\showeprint[arxiv]{2201.08239}~[cs.CL]


\bibitem[Till et~al\mbox{.}(2022)]%
        {till2022}
\bibfield{author}{\bibinfo{person}{Sarina Till}, \bibinfo{person}{Jaydon
  Farao}, \bibinfo{person}{Toshka~Lauren Coleman},
  \bibinfo{person}{Londiwe~Deborah Shandu}, \bibinfo{person}{Nonkululeko
  Khuzwayo}, \bibinfo{person}{Livhuwani Muthelo},
  \bibinfo{person}{Masenyani~Oupa Mbombi}, \bibinfo{person}{Mamare Bopane},
  \bibinfo{person}{Molebogeng Motlhatlhedi}, \bibinfo{person}{Gugulethu
  Mabena}, \bibinfo{person}{Alastair Van~Heerden},
  \bibinfo{person}{Tebogo~Maria Mothiba}, \bibinfo{person}{Shane Norris},
  \bibinfo{person}{Nervo Verdezoto~Dias}, {and} \bibinfo{person}{Melissa
  Densmore}.} \bibinfo{year}{2022}\natexlab{}.
\newblock \showarticletitle{Community-Based Co-Design across Geographic
  Locations and Cultures: Methodological Lessons from Co-Design Workshops in
  South Africa}. In \bibinfo{booktitle}{\emph{Proceedings of the Participatory
  Design Conference 2022 - Volume 1}} (Newcastle upon Tyne, United Kingdom)
  \emph{(\bibinfo{series}{PDC '22})}. \bibinfo{publisher}{Association for
  Computing Machinery}, \bibinfo{address}{New York, NY, USA},
  \bibinfo{pages}{120–132}.
\newblock
\showISBNx{9781450393881}
\urldef\tempurl%
\url{https://doi.org/10.1145/3536169.3537786}
\showDOI{\tempurl}


\bibitem[Vallance(2023)]%
        {BBC-jobs}
\bibfield{author}{\bibinfo{person}{Chris Vallance}.}
  \bibinfo{year}{2023}\natexlab{}.
\newblock \showarticletitle{AI could replace equivalent of 300 million jobs}.
\newblock \bibinfo{journal}{\emph{BBC News}} (\bibinfo{date}{28 March}
  \bibinfo{year}{2023}).
\newblock
\urldef\tempurl%
\url{https://www.bbc.com/news/technology-65102150}
\showURL{%
\tempurl}


\bibitem[Varghese and Chapiro(2023)]%
        {varghese2023}
\bibfield{author}{\bibinfo{person}{Julian Varghese} {and}
  \bibinfo{person}{Julius Chapiro}.} \bibinfo{year}{2023}\natexlab{}.
\newblock \showarticletitle{ChatGPT: The transformative influence of generative
  AI on science and healthcare}.
\newblock \bibinfo{journal}{\emph{Journal of Hepatology}} (\bibinfo{date}{Aug.}
  \bibinfo{year}{2023}).
\newblock
\urldef\tempurl%
\url{https://doi.org/10.1016/j.jhep.2023.07.028}
\showDOI{\tempurl}


\bibitem[Vaswani et~al\mbox{.}(2017)]%
        {vaswani2017attention}
\bibfield{author}{\bibinfo{person}{Ashish Vaswani}, \bibinfo{person}{Noam
  Shazeer}, \bibinfo{person}{Niki Parmar}, \bibinfo{person}{Jakob Uszkoreit},
  \bibinfo{person}{Llion Jones}, \bibinfo{person}{Aidan~N. Gomez},
  \bibinfo{person}{{\L}ukasz Kaiser}, {and} \bibinfo{person}{Illia
  Polosukhin}.} \bibinfo{year}{2017}\natexlab{}.
\newblock \showarticletitle{Attention is All you Need}. In
  \bibinfo{booktitle}{\emph{Advances in Neural Information Processing
  Systems}}, \bibfield{editor}{\bibinfo{person}{I.~Guyon},
  \bibinfo{person}{U.~Von Luxburg}, \bibinfo{person}{S.~Bengio},
  \bibinfo{person}{H.~Wallach}, \bibinfo{person}{R.~Fergus},
  \bibinfo{person}{S.~Vishwanathan}, {and} \bibinfo{person}{R.~Garnett}}
  (Eds.), Vol.~\bibinfo{volume}{30}. \bibinfo{publisher}{Curran Associates,
  Inc.}, \bibinfo{address}{Long Beach, CA, USA}, \bibinfo{pages}{1--11}.
\newblock
\urldef\tempurl%
\url{https://proceedings.neurips.cc/paper_files/paper/2017/file/3f5ee243547dee91fbd053c1c4a845aa-Paper.pdf}
\showURL{%
\tempurl}


\bibitem[Vega and Brennan(2000)]%
        {vega2000isolation}
\bibfield{author}{\bibinfo{person}{Gina Vega} {and} \bibinfo{person}{Louis
  Brennan}.} \bibinfo{year}{2000}\natexlab{}.
\newblock \showarticletitle{Isolation and technology: The human disconnect}.
\newblock \bibinfo{journal}{\emph{Journal of Organizational Change Management}}
  \bibinfo{volume}{13}, \bibinfo{number}{5} (\bibinfo{year}{2000}),
  \bibinfo{pages}{468--481}.
\newblock


\bibitem[Verheijden and Funk(2023)]%
        {verheijden2023}
\bibfield{author}{\bibinfo{person}{Mathias~Peter Verheijden} {and}
  \bibinfo{person}{Mathias Funk}.} \bibinfo{year}{2023}\natexlab{}.
\newblock \showarticletitle{Collaborative Diffusion: Boosting Designerly
  Co-Creation with Generative AI}. In \bibinfo{booktitle}{\emph{Extended
  Abstracts of the 2023 CHI Conference on Human Factors in Computing Systems}}
  (Hamburg, Germany) \emph{(\bibinfo{series}{CHI EA '23})}.
  \bibinfo{publisher}{Association for Computing Machinery},
  \bibinfo{address}{New York, NY, USA}, Article \bibinfo{articleno}{73},
  \bibinfo{numpages}{8}~pages.
\newblock
\showISBNx{9781450394222}
\urldef\tempurl%
\url{https://doi.org/10.1145/3544549.3585680}
\showDOI{\tempurl}


\bibitem[Vertesi et~al\mbox{.}(2020)]%
        {vertesi2020}
\bibfield{author}{\bibinfo{person}{Janet~A. Vertesi}, \bibinfo{person}{Adam
  Goldstein}, \bibinfo{person}{Diana Enriquez}, \bibinfo{person}{Larry Liu},
  {and} \bibinfo{person}{Katherine~T. Miller}.}
  \bibinfo{year}{2020}\natexlab{}.
\newblock \showarticletitle{Pre-Automation: Insourcing and Automating the Gig
  Economy}.
\newblock \bibinfo{journal}{\emph{Sociologica}} \bibinfo{volume}{14},
  \bibinfo{number}{3} (\bibinfo{date}{Jan.} \bibinfo{year}{2020}),
  \bibinfo{pages}{167–193}.
\newblock
\urldef\tempurl%
\url{https://doi.org/10.6092/issn.1971-8853/11657}
\showDOI{\tempurl}


\bibitem[Vynck(2023)]%
        {vynck2023}
\bibfield{author}{\bibinfo{person}{Gerrit~De Vynck}.}
  \bibinfo{year}{2023}\natexlab{}.
\newblock \showarticletitle{Every start-up is an AI company now. Bubble fears
  are growing.}
\newblock \bibinfo{journal}{\emph{Washington Post}} (\bibinfo{date}{5 Aug.}
  \bibinfo{year}{2023}).
\newblock
\urldef\tempurl%
\url{https://www.washingtonpost.com/technology/2023/08/05/ai-hype-bubble-chatgpt/}
\showURL{%
\tempurl}


\bibitem[Wagner(2019)]%
        {wagner2019}
\bibfield{author}{\bibinfo{person}{Ben Wagner}.}
  \bibinfo{year}{2019}\natexlab{}.
\newblock \showarticletitle{Liable, but Not in Control? Ensuring Meaningful
  Human Agency in Automated Decision-Making Systems}.
\newblock \bibinfo{journal}{\emph{Policy \& Internet}} \bibinfo{volume}{11},
  \bibinfo{number}{1} (\bibinfo{year}{2019}), \bibinfo{pages}{104--122}.
\newblock


\bibitem[Weisz et~al\mbox{.}(2021)]%
        {weisz2021}
\bibfield{author}{\bibinfo{person}{Justin~D. Weisz}, \bibinfo{person}{Michael
  Muller}, \bibinfo{person}{Stephanie Houde}, \bibinfo{person}{John Richards},
  \bibinfo{person}{Steven~I. Ross}, \bibinfo{person}{Fernando Martinez},
  \bibinfo{person}{Mayank Agarwal}, {and} \bibinfo{person}{Kartik
  Talamadupula}.} \bibinfo{year}{2021}\natexlab{}.
\newblock \showarticletitle{Perfection Not Required? Human-AI Partnerships in
  Code Translation}. In \bibinfo{booktitle}{\emph{26th International Conference
  on Intelligent User Interfaces}} (College Station, TX, USA)
  \emph{(\bibinfo{series}{IUI '21})}. \bibinfo{publisher}{Association for
  Computing Machinery}, \bibinfo{address}{New York, NY, USA},
  \bibinfo{pages}{402–412}.
\newblock
\showISBNx{9781450380171}
\urldef\tempurl%
\url{https://doi.org/10.1145/3397481.3450656}
\showDOI{\tempurl}


\bibitem[Welsh(2022)]%
        {welsh2023}
\bibfield{author}{\bibinfo{person}{Matt Welsh}.}
  \bibinfo{year}{2022}\natexlab{}.
\newblock \showarticletitle{The End of Programming}.
\newblock \bibinfo{journal}{\emph{Commun. ACM}} \bibinfo{volume}{66},
  \bibinfo{number}{1} (\bibinfo{date}{Dec.} \bibinfo{year}{2022}),
  \bibinfo{pages}{34–35}.
\newblock
\urldef\tempurl%
\url{https://doi.org/10.1145/3570220}
\showDOI{\tempurl}


\bibitem[Wenger et~al\mbox{.}(2002)]%
        {wenger2002}
\bibfield{author}{\bibinfo{person}{Etienne Wenger}, \bibinfo{person}{Richard
  McDermott}, {and} \bibinfo{person}{William~M. Snyder}.}
  \bibinfo{year}{2002}\natexlab{}.
\newblock \bibinfo{booktitle}{\emph{Cultivating Communities of Practice}}.
\newblock \bibinfo{publisher}{Harvard Business School Press},
  \bibinfo{address}{Cambridge, MA, USA}.
\newblock


\bibitem[White et~al\mbox{.}(2019)]%
        {white2019}
\bibfield{author}{\bibinfo{person}{Gwen White}, \bibinfo{person}{Thilini
  Ariyachandra}, {and} \bibinfo{person}{David White}.}
  \bibinfo{year}{2019}\natexlab{}.
\newblock \showarticletitle{Big Data, Ethics, and Social Impact Theory--A
  Conceptual Framework}.
\newblock \bibinfo{journal}{\emph{Journal of Management \& Engineering
  Integration}} \bibinfo{volume}{12}, \bibinfo{number}{1}
  (\bibinfo{year}{2019}), \bibinfo{pages}{9--15}.
\newblock


\bibitem[Wiest(2018)]%
        {wiest2018}
\bibfield{author}{\bibinfo{person}{Brianna Wiest}.}
  \bibinfo{year}{2018}\natexlab{}.
\newblock \showarticletitle{`Permalancing' Is The New Self-Employment Trend
  You'll Be Seeing Everywhere}.
\newblock \bibinfo{journal}{\emph{Forbes}} (\bibinfo{date}{13 June}
  \bibinfo{year}{2018}).
\newblock
\urldef\tempurl%
\url{https://www.forbes.com/sites/briannawiest/2018/06/13/permalancing-is-the-new-self-employment-trend-youll-be-seeing-everywhere/?sh=74d0a4dca383}
\showURL{%
\tempurl}


\bibitem[Willis(2019)]%
        {willis2019}
\bibfield{author}{\bibinfo{person}{Rebecca Willis}.}
  \bibinfo{year}{2019}\natexlab{}.
\newblock \showarticletitle{The use of composite narratives to present
  interview findings}.
\newblock \bibinfo{journal}{\emph{Qualitative Research}} \bibinfo{volume}{19},
  \bibinfo{number}{4} (\bibinfo{year}{2019}), \bibinfo{pages}{471--480}.
\newblock
\urldef\tempurl%
\url{https://doi.org/10.1177/1468794118787711}
\showDOI{\tempurl}


\bibitem[Wong-Villacres et~al\mbox{.}(2018)]%
        {wong2018}
\bibfield{author}{\bibinfo{person}{Marisol Wong-Villacres},
  \bibinfo{person}{Arkadeep Kumar}, \bibinfo{person}{Aditya Vishwanath},
  \bibinfo{person}{Naveena Karusala}, \bibinfo{person}{Betsy DiSalvo}, {and}
  \bibinfo{person}{Neha Kumar}.} \bibinfo{year}{2018}\natexlab{}.
\newblock \showarticletitle{Designing for Intersections}. In
  \bibinfo{booktitle}{\emph{Proceedings of the 2018 Designing Interactive
  Systems Conference}} (Hong Kong, China) \emph{(\bibinfo{series}{DIS '18})}.
  \bibinfo{publisher}{Association for Computing Machinery},
  \bibinfo{address}{New York, NY, USA}, \bibinfo{pages}{45–58}.
\newblock
\showISBNx{9781450351980}
\urldef\tempurl%
\url{https://doi.org/10.1145/3196709.3196794}
\showDOI{\tempurl}


\bibitem[Yang et~al\mbox{.}(2022)]%
        {yang2022ai}
\bibfield{author}{\bibinfo{person}{Daijin Yang}, \bibinfo{person}{Yanpeng
  Zhou}, \bibinfo{person}{Zhiyuan Zhang}, \bibinfo{person}{Toby Jia-Jun Li},
  {and} \bibinfo{person}{Ray LC}.} \bibinfo{year}{2022}\natexlab{}.
\newblock \showarticletitle{AI as an Active Writer: Interaction strategies with
  generated text in human-AI collaborative fiction writing}. In
  \bibinfo{booktitle}{\emph{Joint Proceedings of the ACM IUI Workshops}},
  Vol.~\bibinfo{volume}{10}. CEUR-WS Team, \bibinfo{publisher}{Association for
  Computing Machinery}, \bibinfo{address}{Helsinki, Finland},
  \bibinfo{pages}{1--11}.
\newblock


\bibitem[{YouGov}(2021)]%
        {yougov2021}
\bibfield{author}{\bibinfo{person}{{YouGov}}.} \bibinfo{year}{2021}\natexlab{}.
\newblock \bibinfo{title}{International Technology Report 2021: Automation \&
  {AI}}.
\newblock
\newblock


\bibitem[Yuan et~al\mbox{.}(2022)]%
        {yuan2022}
\bibfield{author}{\bibinfo{person}{Ann Yuan}, \bibinfo{person}{Andy Coenen},
  \bibinfo{person}{Emily Reif}, {and} \bibinfo{person}{Daphne Ippolito}.}
  \bibinfo{year}{2022}\natexlab{}.
\newblock \showarticletitle{Wordcraft: Story Writing With Large Language
  Models}. In \bibinfo{booktitle}{\emph{27th International Conference on
  Intelligent User Interfaces}} (Helsinki, Finland) \emph{(\bibinfo{series}{IUI
  '22})}. \bibinfo{publisher}{Association for Computing Machinery},
  \bibinfo{address}{New York, NY, USA}, \bibinfo{pages}{841–852}.
\newblock
\showISBNx{9781450391443}
\urldef\tempurl%
\url{https://doi.org/10.1145/3490099.3511105}
\showDOI{\tempurl}


\bibitem[Zahidi(2023)]%
        {wef2023}
\bibfield{author}{\bibinfo{person}{Saadia Zahidi}.}
  \bibinfo{year}{2023}\natexlab{}.
\newblock \showarticletitle{The Future of Jobs Report 2023}.
\newblock \bibinfo{journal}{\emph{World Economic Forum}} (\bibinfo{date}{30
  April} \bibinfo{year}{2023}).
\newblock
\urldef\tempurl%
\url{https://www.weforum.org/reports/the-future-of-jobs-report-2023}
\showURL{%
\tempurl}


\bibitem[Zhang and Dafoe(2019)]%
        {zhang2019artificial}
\bibfield{author}{\bibinfo{person}{Baobao Zhang} {and} \bibinfo{person}{Allan
  Dafoe}.} \bibinfo{year}{2019}\natexlab{}.
\newblock \showarticletitle{Artificial intelligence: American attitudes and
  trends}.
\newblock \bibinfo{journal}{\emph{Available at SSRN 3312874}}
  (\bibinfo{year}{2019}).
\newblock
\urldef\tempurl%
\url{https://doi.org/10.2139/ssrn.3312874}
\showDOI{\tempurl}


\bibitem[Zheng et~al\mbox{.}(2022)]%
        {zheng2022}
\bibfield{author}{\bibinfo{person}{Chengbo Zheng}, \bibinfo{person}{Dakuo
  Wang}, \bibinfo{person}{April~Yi Wang}, {and} \bibinfo{person}{Xiaojuan Ma}.}
  \bibinfo{year}{2022}\natexlab{}.
\newblock \showarticletitle{Telling Stories from Computational Notebooks:
  AI-Assisted Presentation Slides Creation for Presenting Data Science Work}.
  In \bibinfo{booktitle}{\emph{Proceedings of the 2022 CHI Conference on Human
  Factors in Computing Systems}} (New Orleans, LA, USA)
  \emph{(\bibinfo{series}{CHI '22})}. \bibinfo{publisher}{Association for
  Computing Machinery}, \bibinfo{address}{New York, NY, USA}, Article
  \bibinfo{articleno}{53}, \bibinfo{numpages}{20}~pages.
\newblock
\showISBNx{9781450391573}
\urldef\tempurl%
\url{https://doi.org/10.1145/3491102.3517615}
\showDOI{\tempurl}


\end{thebibliography}

\newpage
\appendix

\begin{table*}[h!]
\section{Industry Details}
\vspace{2.0em}
\centering
\begin{tabular}{ p{8em} p{18em} p{24em}  }

    Industries &     Brief Overview &      Recruitment For Roles \\
    \midrule
    Advertising & \emph{Conduct promotional work which aims to sell a project or service}   &              \textbf{Central:} Art director, Creative director \newline\newline \textbf{Secondary:} Copywriter, Graphic designer, Illustrator, Animator, Social media or SEO strategist, Media planner, Account manager \\

    \midrule
    Business \newline Communications & \emph{Manage internal and external messaging to audiences such as employees, customers, stakeholders, the media, or the general public} &   \textbf{Central:} Internal communications director, Communications specialist, Administrative assistant \newline\newline \textbf{Secondary:} HR expert, Content strategist, Corporate PR  \\

        \midrule
    Education  & \emph{Teach students and contribute to their learning and development}         &               \textbf{Central:} Middle school \& high school teachers, Course instructor or professor (post-secondary) \newline\newline \textbf{Secondary:} School principal or assistant principal, Freelance tutor, Reading specialist \\
    
    \midrule
    Journalism & \emph{Produce and disseminate reports to inform society on events, people, ideas, and anything else that might be considered ``newsworthy''} &               \textbf{Central:} Newspaper reporter, Columnist, Editorial writer \newline\newline \textbf{Secondary:} Copyeditor, Fact checker, Section/Content editor, Newsroom editor/manager, Freelance reporter (print/online or broadcast) \\

    \midrule
    Law & \emph{Support and enforce established legal standards; advocate for clients} &               \textbf{Central:} Junior and senior associates \newline\newline \textbf{Secondary:} Law firm partner, Freelance attorney, Paralegal, Law clerk, Public defender  \\
    
    \midrule
    Mental Health & \emph{Support, stabilize, and improve individuals' mental health}            &               \textbf{Central:} Psychologist, Therapist, Psychiatrist \newline\newline \textbf{Secondary:} Social worker, Case manager/coordinator, Clinical counselor  \\

    \midrule
    Software \newline Development & \emph{Design, program, deploy, and maintain software}          &               \textbf{Central:} Junior and senior full-stack developers \newline\newline \textbf{Secondary:} Database engineer, Data scientist, Product manager  \\

     \bottomrule
\end{tabular}
  \vspace{0.7em}
  \caption{Overview of how we thought about each industry, including the roles we described as central and secondary shared with our recruitment partners.}
  \label{table:industry_overview}
\end{table*}

\clearpage
\section{Introduction to Generative AI}
\vspace{2.0em}
A key component of our workshops was offering participants a foundation for thinking about generative AI. To this end, early in each workshop we led an education section about 40 minutes in length. We began with a 20-minute presentation covering:

\begin{itemize}
  \item a shared definition of AI
  \item a very condensed history of AI and generative AI, focusing on key concepts like the early aims in developing AI
  \item a brief, non-technical explanation of what has changed recently with transformer models and LLMs
  \item 15 key concepts regarding characteristics, benefits, and risks of generative AI systems, to refer to throughout the workshop
\end{itemize}

Participants were encouraged to ask questions at any point during the presentation, and then we spent an additional 20 minutes on further questions and discussion. In each workshop, the presentation and Q\&A was led by one of two authors, both of whom are researchers who work in AI. We lightly customized materials to each group's industry.
 
 \vspace{1em} 
 
The definition we provided is: ``Artificial Intelligence is the ability of a computer or a machine to think or learn,'' and we provided additional color on how we think about the terms: ``computer or machine,'' ``think,'' and ``learn.''

 \vspace{1em} 

\newpage

The 15 concepts we shared include the following:

\begin{description}

\item[Bias] --- 
Generative AI tools may reflect social biases that are present in their training data

\item[Bland] ---
Generative AI often generates ``flat'' or generic text, unless explicitly directed to do otherwise

\item[Brainstorming] --- 
Generative AI tools can create outlines, lists, drafts, possible solutions, and more

\item[Emergent Properties] --- 
Generative AI models may seem to possess abilities they were not designed to have

\item[Falsehoods] ---
Generative AI can fabricate information or sources, or get facts wrong, yet seem confident and compelling

\item[Grammatical] --- 
Content generated by text-based Generative AI tools can be well-written, using good syntax and avoiding typos

\item[Identifies Tacit Structure] ---
Generative AI can uncover steps and processes which were not previously articulated

\item[Memorization/Privacy Breaches] ---
Generative AI may generate content identical to its training data

\item[Mimicry] --- 
Generative AI can be asked to mimic genre, tone, phrasing, visual style, or more

\item[Non-Deterministic] ---
Generative AI models can give responses which are variable, not consistent. This means that when users input the same or similar prompts, the system may not respond in the same way

\item[Provenance Is Unclear] --- 
Generative AI tools may not be able to reliably trace specific content back to a direct source in training data

\item[Remixes] ---
Generative AI always generates content based on its training data. It can recombine data in unique ways, but is limited to re-mixing training data

\item[Safety Not Guaranteed] ---
Generative AI tools may have built-in safety systems to attempt to prevent certain types of content or topics, but these are not infallible

\item[Scale / Speed] ---
Generative AI, like other AI and ML systems, is able to consider large amounts of data and handle many tasks, over and over again, extremely quickly

\item[Tweakable] ---
Through ``prompt engineering,'' Generative AI tools can often be influenced to generate content in a certain way

\end{description}

\clearpage

\end{document}